\documentclass[12pt]{article}
\usepackage{array}
\usepackage{tabularx}
\usepackage{multirow}
\usepackage{slashed}
\usepackage{subcaption}
\usepackage{bm}
\usepackage{youngtab}
\usepackage{graphicx}
\usepackage{enumitem}
\usepackage{authblk}
\usepackage{amsmath,amssymb,amsfonts,epsfig,cite,setspace,bigstrut,framed,eufrak}
\usepackage[all]{xy}
\usepackage{color}
\usepackage{pifont}
\usepackage{xcolor}
\usepackage{tikz}
\usepackage[colorlinks=true,urlcolor=blue,anchorcolor=blue,citecolor=blue,filecolor=blue,linkcolor=blue,menucolor=blue,pagecolor=blue,linktocpage=true,pdfproducer=medialab,pdfa=true]{hyperref}	 
\usetikzlibrary{positioning}

\usepackage{mathtools}

\renewcommand{\c}{\gamma}

\newcommand{\dsl}{\pa \kern-0.5em /}

\newcommand{\pa}{\partial}

\newcommand{\bit}{\begin{itemize}}
\newcommand{\eit}{\end{itemize}}

\def\bZ{\mathbb Z}

\newcommand{\bea}{\begin{eqnarray}}
\newcommand{\eea}{\end{eqnarray}}
\newcommand{\be}{\begin{equation}}
\newcommand{\ee}{\end{equation}}

\renewcommand{\c}{\gamma}

\newcommand{\ba}{\begin{array}}
\newcommand{\ea}{\end{array}}
\def\bZ{\mathbb Z}

\def\mI{\mathcal I}

\makeatletter \@addtoreset{equation}{section} \makeatother
\renewcommand{\theequation}{\thesection.\arabic{equation}}
\newcommand{\lcm}{\operatorname{lcm}} 
\begin{document}

\vskip 0.25in

\newcommand{\todo}[1]{{\bf\color{blue} !! #1 !!}\marginpar{\color{blue}$\Longleftarrow$}}
\newcommand{\comment}[1]{}
\newcommand\T{\rule{0pt}{2.6ex}}
\newcommand\B{\rule[-1.2ex]{0pt}{0pt}}

\newcommand{\CO}{{\cal O}}
\newcommand{\cI}{{\cal I}}
\newcommand{\cM}{{\cal M}}
\newcommand{\cW}{{\cal W}}
\newcommand{\cN}{{\cal N}}
\newcommand{\cR}{{\cal R}}
\newcommand{\cH}{{\cal H}}
\newcommand{\cK}{{\cal K}}
\newcommand{\cT}{{\cal T}}
\newcommand{\cZ}{{\cal Z}}
\newcommand{\cO}{{\cal O}}
\newcommand{\cQ}{{\cal Q}}
\newcommand{\cB}{{\cal B}}
\newcommand{\cC}{{\cal C}}
\newcommand{\cD}{{\cal D}}
\newcommand{\cE}{{\cal E}}
\newcommand{\cF}{{\cal F}}
\newcommand{\cA}{{\cal A}}
\newcommand{\cX}{{\cal X}}
\newcommand{\IA}{\mathbb{A}}
\newcommand{\IP}{\mathbb{CP}}
\newcommand{\IQ}{\mathbb{Q}}
\newcommand{\IH}{\mathbb{H}}
\newcommand{\IR}{\mathbb{R}}
\newcommand{\IC}{\mathbb{C}}
\newcommand{\IF}{\mathbb{F}}
\newcommand{\IS}{\mathbb{S}}
\newcommand{\IV}{\mathbb{V}}
\newcommand{\II}{\mathbb{I}}
\newcommand{\IZ}{\mathbb{Z}}
\newcommand{\re}{{\rm Re}}
\newcommand{\im}{{\rm Im}}
\newcommand{\tr}{\mathop{\rm Tr}}
\newcommand{\ch}{{\rm ch}}
\newcommand{\rk}{{\rm rk}}
\newcommand{\ext}{{\rm Ext}}
\newcommand{\bi}{\begin{itemize}}
\newcommand{\ei}{\end{itemize}}
\newcommand{\beq}{\begin{equation}}
\newcommand{\eeq}{\end{equation}}

\newcommand{\CN}{{\cal N}}
\newcommand{\y}{{\mathbf y}}
\newcommand{\z}{{\mathbf z}}
\newcommand{\C}{\mathbb C}\newcommand{\R}{\mathbb R}
\newcommand{\CA}{\mathbb A}
\newcommand{\CP}{\mathbb P}
\newcommand{\cP}{\mathcal P}
\newcommand{\tmat}[1]{{\tiny \left(\begin{matrix} #1 \end{matrix}\right)}}
\newcommand{\mat}[1]{\left(\begin{matrix} #1 \end{matrix}\right)}
\newcommand{\diff}[2]{\frac{\partial #1}{\partial #2}}
\newcommand{\gen}[1]{\langle #1 \rangle}

\newtheorem{theorem}{\bf THEOREM}
\newtheorem{proposition}{\bf PROPOSITION}
\newtheorem{observation}{\bf OBSERVATION}
\newtheorem{statement}{\bf STATEMENT}

\def\theequation{\thesection.\arabic{equation}}
\newcommand{\setall}{
	\setcounter{equation}{0}
}
\renewcommand{\thefootnote}{\fnsymbol{footnote}}

\begin{titlepage}
\vfill
\begin{flushright}
{\tt\normalsize KIAS-P24045}\\

\end{flushright}
\vfill
\begin{center}
{\Large\bf Web of 4D Dualities, Supersymmetric \\Partition functions and SymTFT}

\vskip 1cm

Zhihao Duan$^\dagger$, Qiang Jia$^{ \natural *}$ and Sungjay Lee$^\natural$

\vskip 5mm
$^\dagger${\it School of Physics and Astronomy, Queen Mary University of London,\\
$~~ $ Mile End Road, London, E1 4NS, UK
}
\vskip 3mm
$^\natural${\it School of Physics,
Korea Institute for Advanced Study,\\
$~~ $85 Hoegiro, Dongdaemun-Gu, Seoul 02455, Korea}
\vskip 3mm
$^*${\it Department of Physics,
Korea Advanced Institute of Science and Technology,\\
$~~ $Daejeon 34141, Korea}

\vskip 5mm
\href{mailto:z.duan@qmul.ac.uk}{z.duan@qmul.ac.uk},
\href{mailto:qjia@kias.re.kr}{qjia@kias.re.kr},
\href{mailto:sjlee@kias.re.kr}{sjlee@kias.re.kr}

\end{center}
\vfill

\begin{abstract}
\noindent
We study $\mathbb{Z}_N$ one-form center symmetries in four-dimensional gauge theories using the symmetry topological field theory (SymTFT). In this context, the associated TFT in the five-dimensional bulk is the BF model. We revisit its canonical quantization and construct topological boundary states on several important classes of four manifolds that are spin, non-spin and torsional. We highlight a web of four-dimensional dualities, which can be naturally interpreted within the SymTFT framework. We also point out an intriguing class of four-dimensional gauge theories that exhibit mixed 't Hooft anomaly between one-form symmetries. In the second part of this work, we extend the SymTFT to account for various quantities protected by supersymmetry (SUSY) in SUSY gauge theories. We proposed that their behaviour under various symmetry operations are entirely captured by the topological boundary of the SymTFT, resulting in strong constraints. Concrete examples are considered, including the Witten index, the lens space index and the Donaldson-Witten and Vafa-Witten partition functions.
\end{abstract}

\vfill
\end{titlepage}

\tableofcontents%\newpage
%\renewcommand{\thefootnote}{\#\arabic{footnote}}
%\setcounter{footnote}{0}
%\vskip 2cm

%%%%%%%%%%%%%%%%%%%%%%%%%%%%%%%%%%%%%%%%%%%%%%%%%%%%%%%%%%%%%%%%%%%%%%%%%%%%%%%%%%%%%%%%%%%
%%%%%%%%%%%%%%%%%%%%%%%%%%%%%%%%%%%%%%%%%%%%%%%%%%%%%%%%%%%%%%%%%%%%%%%%%%%%%%%%%%%%%%%%%%%
%%%%%%%%%%%%%%%%%%%%%%%%%%%%%%%%%%%%%%%%%%%%%%%%%%%%%%%%%%%%%%%%%%%%%%%%%%%%%%%%%%%%%%%%%%%
\section{Introduction and Conclusion}

Symmetries have long played a key role in understanding quantum field theories (QFTs). 
They serve as powerful organizing principles, constrain the dynamics, and often reveal 
profound connections between two seemingly unrelated physical systems. 
In recent years, our understanding of symmetries has significantly evolved. 
It starts by advocating that global ($0$-form) symmetries in QFTs, either continuous or discrete, can be best described by certain topological defects of co-dimension one, known as symmetry operators. 
This geometric perspective on symmetries provides a flexible framework that 
accommodates various generalizations, often referred to as generalized global symmetries. 
They encompass the higher-form or higher-group symmetries \cite{Gaiotto:2014kfa,DelZotto:2015isa,Cordova:2018cvg,Albertini:2020mdx,Morrison:2020ool,Cordova:2020tij,Bhardwaj:2020phs,Apruzzi:2021vcu,Tian:2021cif,DelZotto:2022joo,Cvetic:2022imb,DelZotto:2022fnw,Wang:2023iqt,Bhardwaj:2023zix,Ambrosino:2024ggh,Closset:2024sle}, non-invertible symmetries \cite{Bhardwaj:2017xup,Chang:2018iay,Thorngren:2019iar,Komargodski:2020mxz,Thorngren:2021yso,Choi:2021kmx,Kaidi:2021xfk,Choi:2022zal,Cordova:2022ieu,Choi:2022jqy,Chang:2022hud,Bashmakov:2022uek,Seiberg:2023cdc,Seiberg:2024gek}, subsystem symmetries \cite{Seiberg:2019vrp,Seiberg:2020bhn,Seiberg:2020wsg,Seiberg:2020cxy,Yamaguchi:2021xeq,Stahl:2021sgi,Gorantla:2022eem,Katsura:2022xkg,Gorantla:2022ssr,Yamaguchi:2022apr,Cao:2022lig,Cao:2023doz,Cao:2023rrb}, etc.
A comprehensive summary of the vast literature on this
subject can be found in \cite{Cordova:2022ruw}. See also
\cite{Schafer-Nameki:2023jdn,Brennan:2023mmt,Bhardwaj:2023kri,Shao:2023gho,Luo:2023ive}
for accessible reviews on these topics.

Given a (generalized) global symmetry, one can explore  
various symmetry operations, including gauging and stacking symmetry-protected topological (SPT) phases onto a given system. 
Notable examples include  
the Jordan-Wigner transformation and the Kramers-Wannier duality 
for the two-dimensional theories with $\mathbb{Z}_2$ symmetry. 
Those operations generate a duality web 
of theories that share the same total Hilbert space. Here the total Hilbert space is defined to include 
the quantum states in the untwisted sector and the twisted sectors. Its precise definition 
will be presented later. 
The shared total Hilbert space implies that 
any physical observables of theories in the duality web, such as the thermodynamic 
partition functions, are closely related. 

The symmetry operations are often blind to the specific dynamics but 
strongly tied to the symmetry itself, and hence exhibit universality. 
Recently, the Symmetry Topological Field Theory (SymTFT), also known as the sandwich construction, 
has been emphasized as a powerful framework to fully explore various features of the symmetry operations.

\begin{figure}[!t]
    \centering
    \includegraphics[scale=0.7]{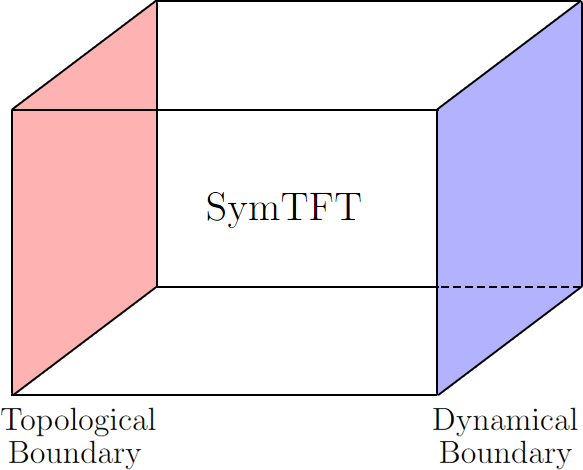}
    \caption{Sandwich construction of SymTFT.}\label{SymTFTsandwitch}
\end{figure}

The idea of the SymTFT can be illustrated as follows. 
We start with a $d$-dimensional quantum field theory with a global symmetry ${\cal S}$. 
As depicted in Figure \ref{SymTFTsandwitch}, 
SymTFT has been proposed to describe the given QFT as a 
$(d+1)$-dimensional topological field theory, specified by the symmetry ${\cal S}$, on a slab with certain boundary conditions imposed. The topological nature of the SymTFT allows us to 
collapse the $(d+1)$-dimensional theory to recover the original QFT in $d$-dimensions.  
The information of symmetry $\mathcal{S}$ and the dynamics are encoded separately on the two boundaries. 
Here we impose a topological boundary condition on the left boundary while 
a dynamical boundary condition on the right boundary. In this description, all symmetry operations
in the QFT are then implemented on the left boundary while the right boundary depends on details of the QFT. 
The SymTFT has played a key role in recent developments \cite{Gaiotto:2020iye,Apruzzi:2021nmk,Lin:2022dhv,Kaidi:2022cpf,Kaidi:2023maf,Zhang:2023wlu,Bhardwaj:2023ayw,Bartsch:2023wvv,Cvetic:2023pgm,Antinucci:2023ezl,Cordova:2023bja,Duan:2023ykn,Cao:2023rrb,Baume:2023kkf,Bhardwaj:2023idu,Bhardwaj:2023bbf,Brennan:2024fgj,Antinucci:2024zjp,Bonetti:2024cjk,GarciaEtxebarria:2024jfv,Apruzzi:2024htg,Antinucci:2024bcm,Heckman:2024zdo,Bhardwaj:2024xcx,Bhardwaj:2024kvy,Choi:2024tri,Choi:2024wfm,Antinucci:2024ltv,Bhardwaj:2024igy}, and arises naturally from geometric engineering or holography \cite{Apruzzi:2022rei,Heckman:2022xgu,Antinucci:2022vyk,vanBeest:2022fss,Bashmakov:2023kwo,Chen:2023qnv,Yu:2023nyn,Argurio:2024oym,Braeger:2024jcj,DelZotto:2024tae,GarciaEtxebarria:2024fuk,Franco:2024mxa,Bergman:2024aly}. 
Furthermore, a closely related and rigorous framework is put forward in \cite{Freed:2022qnc}.

The SymTFT provides a powerful framework for studying various aspects of 
symmetry operations in the QFT. To see this, 
let us now place the given QFT on a compact $d$-dimensional manifold $M_d$, and regard 
the interval of the slab as the time direction. In this setup,
the path-integral of the SymTFT can be interpreted as the inner product between 
two quantum states in the finite-dimensional Hilbert space of the SymTFT on $M_d$. 

One of these states is a topological boundary state 
characterized by the topological boundary condition imposed on the left boundary. In general, the classification of topological boundaries is a challenging problem. For instance, in the case of (2+1)d SymTFT, these boundaries correspond one-to-one with Lagrangian algebras in the category of bulk line operators. For abelian symmetries that are specifically relevant to our study, there are various types of topological boundary states, each 
corresponding to a different set of maximally commuting observables in the $(d+1)$-dimensional SymTFT. 
Here commuting observables refer 
to topological defects that are mutually local. As will be explained in detail later, 
switching the type of topological boundary state leads to 
a change from one QFT to another within the aforementioned duality web generated by 
the symmetry operations. 

How can we characterize the other quantum state on the right boundary? 
This state, known as the dynamical boundary state, is associated with the dynamical boundary condition 
and is defined such that the inner product between the two boundary states gives rise to 
a Euclidean partition function of the QFT on $M_d$. 
It therefore suggests that the partition function of our interest determines the corresponding state 
on the right boundary.

As a demonstration, let us briefly revisit the duality web of two-dimensional theories 
with $\mathbb{Z}_2$ symmetry. The relevant SymTFT is the three-dimensional 
BF model with level two. 
One can show that there are three distinct sets of maximally commuting observables 
in the BF model, consisting of electric, magnetic, and dyonic loop operators. 
Each set corresponds to a canonical basis of the Hilbert space of 
the BF model. When $M_2$ is given by two-torus $T^2$, the dynamical boundary state 
of our interest is associated with the torus partition function. 
In the language of the SymTFT,  
the Jordan-Wigner transformation and the Kramers-Wannier 
duality in two dimensions can then be easily understood as switching between different 
types of topological boundary states. We present more details in Section \ref{sec:4.1}. 

In the present work, we focus on four-dimensional gauge 
theories and the rich physics of one-form symmetries and their 
operations. The one-form symmetry has become central in understanding the global structure of gauge theories, following 
the seminal work of Aharony, Seiberg and Tachikawa \cite{Aharony:2013hda}
where the constraints of one-form symmetries on the spectrum 
of line operators were systematically studied. Their work 
eventually led to the notion of generalized global symmetry  \cite{Gaiotto:2014kfa}.
One of the main goals of the present work is to utilize the powerful framework of SymTFT to revisit 
various aspects of one-form symmetries with emphasis on the
intricate duality web between four-dimensional gauge theories. 

Since we only consider the $\mathbb{Z}_N$ center symmetry, the five-dimensional bulk TQFT can be described by the $\mathbb{Z}_N$ BF model. We carefully work out the canonical quantization and topological boundary states on a selected list of four manifolds: $T^4$, $\mathbb{CP}^2$ and $L(r,s) \times S^1$. They represent three important classes of four-manifolds that are spin, non-spin and torsional respectively, and are crucial for later sections. By selecting distinct topological boundary states, we construct a four-dimensional duality web. In parallel, we clarify the notion of the twisted Hilbert spaces for one-form symmetries that arise in the 4d duality web and demonstrate the equivalence between gauge fields with twisted boundary conditions, non-commutative holonomies, and one-form symmetry backgrounds. We also point out that 
a partial gauging of the one-form center symmetry of $SU(N)$ gives rise to 
a gauge theory that has two one-form global symmetries with a mixed 't Hooft anomaly.
This implies that, under partial gauging, there exists a class of pure Yang-Mills theories with `non-trivial' confining vacua accounting for the anomaly. 
It is therefore an intriguing open problem to identify the infrared (IR) effective topological field theory that saturates this anomaly.

In the past decades, a lot of progress has been made in the study of four-dimensional supersymmetric (SUSY) theories. Although it is often hopeless to compute the partition function directly, people have invented a plethora of quantities protected by supersymmetry that still capture the dynamics of the system. One notable class is the supersymmetric indices, starting from the famous Witten index to more refined ones including the superconformal index, lens space index, etc. Another important class is the partition function of topologically twisted theories which can be defined on curved backgrounds and gives rise to important topological invariants such as Donaldson invariants and Vafa-Witten invariants. All these quantities are robust against small deformations, and teach us valuable lessons in mathematics and physics. Some effects of 1-form symmetries in $\mathcal{N} = 2$ Seiberg-Witten theories are already studied in \cite{Closset:2023pmc,Aspman:2023ate,Furrer:2024zzu}.

Inspired by the above, we would like to generalize the picture of the SymTFT to include those protected quantities. In other words, we propose that we can still use the same five-dimensional slab with the same left topological boundary state $\langle b |$, but we define a SUSY dynamical boundary state $|\chi_{\text{SUSY}}\rangle$ such that
\begin{equation}
    Z_{\text{SUSY}}\big[ b \big] \equiv \big \langle b \big | \chi_{\text{SUSY}} \big\rangle \ .
\end{equation}

Intuitively, $Z_{\text{SUSY}}$ only differs from the thermodynamic partition function 
by modifying the holonomy along non-contractible cycles or adding extra terms in the Lagrangian to change the spin-statistics, which does not change the global symmetry of the theory. 
Therefore, all we need is to modify the dynamical boundaries to suitable SUSY ones. On the other hand, this simple proposal gives us powerful constraints or predictions on the behavior of $Z_{\text{SUSY}}$ under various symmetry operations. In this paper, we will talk about the following cases in turn:

1. Witten index on $T^4$. The study of Witten index of SUSY theories on $T^4$ dates back to one of the first papers by Witten himself \cite{Witten:1982df}, and was later revisited in e.g., \cite{Witten:2000nv,Tachikawa:2014mna}. In this paper, we 
reexamine and extend their results within
the framework of the SymTFT where some of the subtle computations involved 
become clearer. Specifically, we begin with 
simply-connected gauge groups such as $SU(n)$, $Sp(n)$, $Spin(n)$, $E_6$ and $E_7$. They all have non-trivial center groups as shown in the second column of Table \ref{tab:gaugegroup}. Gauging the one-form center symmetry then generates the other theories with non-simply-connected 
gauge groups in the duality web. 
Our proposal based on the SymTFT picture then automatically gives the Witten indices 
of those theories, even when background gauge fields for the one-form symmetry are turned on. 

2. Lens space index on $L(r,1) \times S^1$. The Witten index has been generalized in several directions. Among them is the lens space index \cite{Benini:2011nc}, one interesting 
supersymmetric observable that encodes certain BPS spectrum on $L(r,1)$ which is the lens space. Unlike the superconformal index, the lens space index is known to capture the global properties of four-dimensional gauge theories  \cite{Razamat:2013opa}.  In our work, we revisit the lens space index from the SymTFT perspective. According to our proposal, the lens space indices of theories in the duality web are all related in a uniform manner. However, since the lens space $L(r,1)$ has non-trivial torsional one-cycles, their relationship depends crucially on the value of $r$, as was noticed in \cite{Razamat:2013opa}. To see this, we carefully analyze the topological boundary states and clarify some subtle issues when torsional one-cycles are present. Altogether, we not only find the perfect agreement with previous results in \cite{Razamat:2013opa}, but also obtain new results 
for $USp(2N)$. 

3. Donaldson-Witten (DW) and Vafa-Witten (VW) partition functions. 
In the study of four-dimensional $\mathcal{N} = 2$ SYM, Witten \cite{Witten:1988ze} found that after twisting spacetime symmetry with R-symmetry, the theory becomes topological and gives rise to the famous Donaldson invariants. The global symmetry remains unchanged in the twisting procedure, so the Donaldson invariants are supposed to inherit the information of one-form symmetries. In particular, our proposal ensures the existence of a corresponding dynamical boundary state, and constraints the Donaldson invariants when gauging the one-form symmetries. As a concrete example, we consider the topologically twisted theory of $SU(2)$ gauge group with one adjoint hypermultiplet on the complex projective space $\mathbb{CP}^2$. This theory has a $\bZ_2$ one-form symmetry, which can be gauged to an $SO(3)$ gauge theory. On the other hand, if one takes the mass of the hypermultiplet to zero, one obtains the topologically twisted theory of $\mathcal{N} = 4$ SYM with the same gauge group, which was first studied in \cite{Vafa:1994tf}. Recently, this rich example was thoroughly analyzed in \cite{Manschot:2021qqe}, and we are able to confirm the validity of our proposal explicitly 
with all subtle phase factors matching perfectly. Some recent works that use the SymTFT to study VW partition functions include \cite{Chen:2023qnv,Chen:2024fno}.

This paper is organized as follows. In Section \ref{sec:2}, we introduce the five-dimensional BF model with level $N$ as the bulk SymTFT for one-form $\mathbb{Z}_N$ symmetry in four dimensions. In particular, we start from the canonical quantization and discuss some general properties of topological boundary states. Partial gauging of a subgroup of $\mathbb{Z}_N$ is also considered and examples are provided. Section \ref{sec:3} is devoted to a detailed study of topological boundary states on $T^4$, $\mathbb{CP}^2$ and $L(r,s) \times S^1$, which represent three classes of four-manifolds that are spin, non-spin and torsional respectively. In Section \ref{sec:4}, we generalize the web of 2d dualities to four dimensions, which can be naturally understood as choosing different topological boundary states in the SymTFT. We also point out a class of mixed 't Hooft anomalies in partially gauged $SU(N)$ Yang-Mills theory in the last subsection.

The second part of this paper is about a novel application of the SymTFT to four-dimensional supersymmetric theories. Based on the proposal that there exist supersymmetric boundary conditions corresponding to SUSY protected quantities, by choosing suitable topological boundaries we obtain non-trivial and often new formulas for non-simply-laced gauge groups. For cases that are already known in the literature, we also find a perfect agreement. We analyze a selected representative of the SUSY protected quantities: the Witten index on $T^4$ in Section \ref{sec:5}, the lens space index on $L(r,s) \times S^1$ in Section \ref{sec:6} and the DW and VW partition functions on $\mathbb{CP}^2$ in Section \ref{sec:7}. 

There are many future directions worth exploring. First, in the present work, we only consider the one-form center symmetry, which is always a cyclic group. A natural extension would be theories with more complex symmetry groups, and a good starting point is two-dimensional theories with permutation group symmetries. They appear naturally, for instance, in the symmetric orbifold construction. The corresponding SymTFT becomes the Dijkgraaf-Witten theory, and it would be interesting to explore its implications and gain new insights. Second, we can apply the SymTFT to other SUSY protected quantities such as the topological twisted index on $\Sigma_g \times T^2$ \cite{Benini:2016hjo}. Third, theories with boundaries or interfaces also play an essential role in various contexts. Certain SUSY protected quantities, such as the Witten index, can then be naturally generalized to take into account the effect of boundaries. This generalization is often referred to in the literature as the half-index \cite{Gaiotto_2019}. More recently, the SymTFT for theories with boundaries or interfaces was studied in \cite{Baume:2023kkf,Bhardwaj:2023bbf,Braeger:2024jcj,Heckman:2024zdo,Cvetic:2024dzu,Choi:2024tri,Bhardwaj:2024igy}. It would be valuable to fully explore the boundary SymTFT to learn new lessons about those half-indices. Finally, we highlight an intriguing class of mixed 't Hooft anomalies after partially gauging $SU(N)$ Yang-Mills theory. Determining the low-energy effective theory possibly in the form of a TQFT to saturate the anomaly is a fascinating challenge.

%%%%%%%%%%%%%%%%%%%%%%%%%%%%%%%%%%%%%%%%%%%%%%%%%%%%%%%%%%%%%%%%%%%%%%%%%%%%%%%%%%%%%%%%%%%
%%%%%%%%%%%%%%%%%%%%%%%%%%%%%%%%%%%%%%%%%%%%%%%%%%%%%%%%%%%%%%%%%%%%%%%%%%%%%%%%%%%%%%%%%%%
%%%%%%%%%%%%%%%%%%%%%%%%%%%%%%%%%%%%%%%%%%%%%%%%%%%%%%%%%%%%%%%%%%%%%%%%%%%%%%%%%%%%%%%%%%%
\section{Five-Dimensional BF Model}\label{sec:2}

Let us consider a four-dimensional gauge theory on a compact manifold $M$
without any boundaries. When the gauge theory is coupled to a fermion, the manifold $M$
is restricted to be spin. The global properties of the gauge theory 
depend on the topology of the manifold $M$. The goal of the present 
work is to understand such global structures of the gauge theory 
in the language of the SymTFT. 

The SymTFT of our interest can be described as 
the five-dimensional BF model with level $N$, 
\begin{equation}\label{BF_action}
    S_{BF} = \frac{N}{2\pi} \int \widetilde{B} \wedge d B\ ,
\end{equation}
where $B$ and $\widetilde B$ are two-form gauge fields. The BF model is defined on 
$M \times \mathbb{R}$ where $\mathbb{R}$ describes the time direction.

This model was proposed in \cite{Witten:1998wy} to study the $SL(2,\mathbb{Z})$ duality symmetry of the 
$\CN=4$ super Yang-Mills theories in a holographic manner. To see this, 
we start with the type IIB string theory in the near-horizon geometry of $N$ D3-branes. 
Given the $N$ units of five-form flux, the compactification on $S^5$ leads to a low-energy 
effective action on $AdS_5$ that contains the topological term \eqref{BF_action} 
where two-form gauge fields $B$ and $\tilde B$ are the NS and the RR two-form fields. 
Note also that the level $N$ can be identified as the number of D3-branes. 

We will argue later that the partition function of the gauge theory on $M$
can be specified by a choice of vector in the Hilbert space of the BF model on $M$,
${\cal H}(M)$. Since the BF model is topological, ${\cal H}(M)$ only depends on the topology of $M$. 

Moreover, the action \eqref{BF_action} is invariant 
under $SL(2,\mathbb{Z})$ acting on the two-form gauge fields as follows, 
\begin{equation}\label{SL(2,Z)action}
    \left(\begin{array}{c}
    \widetilde{B}\\B \end{array}\right) \rightarrow  \left(\begin{array}{cc}
        a & b \\
        c & d
    \end{array} \right)\left(\begin{array}{c}
    \widetilde{B}\\B \end{array}\right) 
\end{equation}
with $ad-bc=1$. This symmetry group is generated by the 
$S$- and $T$-transformations 
\begin{align}
    S: \quad  & B \rightarrow \widetilde{B}\ ,
    \quad \widetilde{B} \rightarrow -B\ , 
    \nonumber \\ 
    T: \quad & B \rightarrow B\ ,\quad \widetilde{B} \rightarrow \widetilde{B} + B\ .
\end{align}
They satisfy $S^2 = (ST)^3 = C$ with the charge conjugation $C$-transformation,
\begin{equation}
    C:\quad B \rightarrow -B \ ,\quad \widetilde{B}\rightarrow -\widetilde{B} \ .
\end{equation}

In this section, we first briefly review how to quantize 
the BF model \eqref{BF_action} to obtain the Hilbert space ${\cal H}(M)$.
Then, we discuss how the $SL(2,\mathbb{Z})$ symmetry acts on ${\cal H}(M)$.
We also explain a subtlety in verifying the $SL(2,\mathbb{Z})$ invariance of 
the BF model when $M$ is not a spin manifold.

%%%%%%%%%%%%%%%%%%%%%%%%%%%%%%%%%%%%%%%%%%%%%%%%%%%%%%%%%%%%%%%%%%%%%%%%%%%%%%%%%%%%%%%%%%%

\subsection{Canonical quantization}

We begin by considering the case that the second homology group of $M$ is torsion-free to make the discussion simple. 
Although it turns out that the direct quantization of the BF model on $M$ rather involves 
many subtle details, let us dive into the analysis. 

To quantize the topological field theory \eqref{BF_action}, we first choose a gauge 
where $B$ and $\widetilde B$ have vanishing temporal components. In this gauge, 
the Gauss law constrains that the spatial components of the field strengths 
vanish. The classical field configurations, modulo gauge transformations, 
are thus flat connections on $M_4$, 
\begin{align}\label{zero_energy_config}
    \frac{N}{2\pi} B & = \sum_{i=1}^{h_2} b_i \gamma^i\ , 
    \nonumber \\ 
    \frac{N}{2\pi} \widetilde B & = \sum_{i=1}^{h_2} \tilde b_i \gamma^i\ , 
\end{align}
where $\{ \gamma^i \}$ is a basis of $H^2(M_4)$ whose dimension is $h_2$. 
Here $b_i$ and $\tilde b_i$ are the periods of the flat two-form gauge fields  
\begin{align}\label{def_01}
    b_i = \frac{N}{2\pi} \int_{\Gamma_i} B\ , \qquad
    \tilde b_i = \frac{N}{2\pi} \int_{\Gamma_i} \widetilde B\ ,
\end{align}
where $\{\Gamma_i\}$ are closed two-cycles satisfying 
$ \int_{\Gamma_i} \gamma^j = \delta^j_i$. They are defined modulo $N$ due to 
the large gauge transformations, 
\begin{align}\label{periodicity}
    b_i \simeq b_i + N \ , \qquad \tilde b_i \simeq \tilde b_i + N\ . 
\end{align}
It implies that the flat connections can be characterized by $H^2(M_4,U(1))$.     

Let $b_i(t)$ and $\tilde b_i(t)$ vary slowly over time. Plugging \eqref{zero_energy_config} back
into \eqref{BF_action}, one can obtain the quantum mechanical action   
\begin{equation}\label{BFlowenergy}
    S^{\textrm{eff}}_{BF} = \frac{2\pi}{N} \int d t ~
    \Big(\sum_{i,j} K^{ij} \tilde{b}_i \frac{d}{d t} b_j \Big)\ ,
\end{equation}
where $K^{ij}$ denotes the integer-valued, symmetric and unimodular intersection matrix 
\begin{align}
    K^{ij} = \int_{M_4} \gamma^i \wedge \gamma^j\ . 
\end{align}
The canonical commutation relation then reads 
\begin{equation}\label{BF-Basic-Commutation-Relation}
    \frac{2\pi }{N} \Big[ \hat{b}_i ,\hat{\tilde{b}}_j \Big] = i  K_{ij}\ ,
\end{equation}
where $K_{ij}$ is the inverse of the intersection matrix $K^{ij}$. Since
$K^{ij}$ is unimodular, $K_{ij}$ is also integer-valued. Note that 
the closed two-cycle $\Gamma_i$ is the Poincaré dual of $K_{ij} \gamma^j$ in $M_4$,
and that $K_{ij}$ computes the intersection number between $\Gamma_i$ and $\Gamma_j$.

Upon the canonical quantization \eqref{BF-Basic-Commutation-Relation}, 
the Hilbert space of the BF model on $M_4$ has two canonical bases. 
One of them is a set of ``position'' eigenstates $\{ |b \rangle \}$,
\begin{align}\label{5DBF-Basic-Positioin-States}
   \begin{split}       
        \exp\left[\frac{2\pi i}{N} \hat{b}_i \right] | b \rangle &= \omega^{b_i} |b\rangle\ ,\\
        \exp\left[\frac{2\pi i}{N} \hat{\tilde{b}}_j \right] |b\rangle &= |b - K_j\rangle\ ,
    \end{split}
\end{align}
and the other is a set of ``momentum'' eigenstates $\{ |\tilde b\rangle \} $
\begin{align}\label{5DBF-Basic-Momentum-States}
    \begin{split}
        \exp\left[\frac{2\pi i}{N} \hat{b}_i \right] | \tilde{b} \rangle &=  |\tilde{b} + K_i\rangle\ , \\
        \exp\left[\frac{2\pi i}{N} \hat{\tilde{b}}_j \right] |\tilde{b}\rangle &= \omega^{\tilde{b}_j} |\tilde{b}\rangle\ ,
    \end{split}
\end{align}
where $\omega = e^{2\pi i /N } $ is the $N$-th root of unity and $K_j$
denote the $h_2$-dimensional vectors 
\begin{align}\label{def_of_K}
    K_i = ( K_{i 1}, K_{i2},\cdots,K_{ih_2} )\ .    
\end{align}
Since both positions $\hat b_i$ and momenta $\hat{\tilde{b}}_j$ are periodic, 
their eigenvalues $b_i$ and $\tilde b_j$ are quantized  
\begin{align}        
    \begin{split}
        b_i = 0,1,\cdots,N-1 \ ,\\
        \tilde{b}_i = 0,1,\cdots,N-1 \ .
    \end{split}
\end{align}
Therefore, the Hilbert space is finite-dimensional, $\text{dim}\, {\cal H}(M_4)=N^{h_2}$.  
The two kinds of eigenstates are related by a discrete Fourier transformation,
\begin{equation}\label{transitionmatrix}
    |\tilde{b}\rangle = \frac{1}{\sqrt{N^{h_2}}}\sum_b \omega^{ K(\tilde{b},b) }  |b\rangle\ ,
\end{equation}
where $K(\tilde{b},b) = K^{ij}\tilde b_i b_j $.

In terms of the field variables, one can express the operators in \eqref{5DBF-Basic-Positioin-States} and \eqref{5DBF-Basic-Momentum-States} as
\begin{align}\label{5DBF-Basic-Wilson-Surface-field-variables}
\begin{split}
    U[\Gamma_i] & \equiv \exp\left[i \oint_{\Gamma_i} B \right]  = \exp\left[\frac{2\pi i}{N} \hat{b}_i \right], \\ 
    \widetilde{U}[\Gamma_i] &  \equiv\exp\left[i \oint_{\Gamma_j} \widetilde{B} \right] = \exp\left[\frac{2\pi i}{N} \hat{\tilde{b}}_j \right].
\end{split}
\end{align}
It implies that $\{ U[\Gamma] = e^{i \oint_{\Gamma} B} \} $ ($\{ \widetilde{U}[\Gamma] = e^{i \oint_{\Gamma} \widetilde{B}}\}$), diagonalized by the position basis $\{|b\rangle\}$ (the momentum 
basis $\{|\tilde b\rangle\}$), becomes a complete set of commuting observables in 
the quantum theory of the BF model. Since the $N$-copies of each surface operator 
becomes trivial
\begin{align}
    U[\Gamma]^N= \widetilde{U}[\Gamma]^N = {\bf 1} \ ,
\end{align}
the surface operators $U[N \Gamma]$ and $\widetilde{U}[N\Gamma]$ 
are trivial as well. In other words, they are essentially classified by $H_2(M_4,\mathbb{Z}_N)$. 
From the commutation relation \eqref{BF-Basic-Commutation-Relation}, 
one can also show the surface operators satisfy the following algebra,
\begin{equation}\label{5DBF-Basic-U-UTilde-Algebra}
    U[\Gamma]  \widetilde{U}[\Gamma'] = \omega^{-K(\Gamma,\Gamma')} \widetilde{U}[\Gamma']~ U[\Gamma]\ ,
\end{equation}
where $K(\Gamma , \Gamma')$ is the intersection number between $\Gamma$ and $\Gamma'$. 
This says that $U[\Gamma]$ generates a two-form $\mathbb{Z}_N$ symmetry rotating 
$\widetilde U[\Gamma]$ and vice versa. 

In modern literature, e.g. \cite{Kaidi:2022cpf}, the two canonical bases are also referred to as the Dirichlet and Neumann boundary conditions if we consider the time direction to be inside $M_4$.

%%%%%%%%%%%%%%%%%%%%%%%%%%%%%%%%%%%%%%%%%%%%%%%%%%%%%%%%%%%%%%%%%%%%%%%%%%%%%%%%%%%%%%%%%%%
\subsection{$SL(2,\mathbb{Z}_N)$ and the Pontryagin square}\label{sec:Pontryagin}

As argued above, the action \eqref{BF_action} is invariant under the $SL(2,\mathbb{Z})$ transformation \eqref{SL(2,Z)action}. (We will discuss some subtleties concerning the invariance later.) 
Thus, the transformation \eqref{SL(2,Z)action} can be naturally represented 
on the Hilbert space ${\cal H}(M)$ by a unitary operator $V_\Lambda$. Let $S_{(e,m)}[\Gamma]$
be a generic surface operator given by
\begin{align}
    S_{(e,m)}[\Gamma] =  \exp \left[i \oint_{\Gamma} e B + m \widetilde{B} \right]\ .
\end{align}
For instance, the surface operators in \eqref{5DBF-Basic-Wilson-Surface-field-variables} can be labeled by 
$S_{(1,0)}[\Gamma] = U[\Gamma]$ and $S_{(0,1)}[\Gamma]=\widetilde{U}[\Gamma]$. The 
unitary operator $V_\Lambda$ for a given $\Lambda \in SL(2,\mathbb{Z})$, 
\begin{align}
    \Lambda = 
    \left(
    \begin{array}{cc}
        a & b \\
        c & d
    \end{array} 
    \right)    
\end{align}
with $(ad-bc)=1$, transforms $S_{(e,m)}[\Gamma]$ as follows, 
\begin{align}\label{SL(2,Z)action_on_surfaceop}
\begin{split}
    V_\Lambda S_{(1,0)}[\Gamma] V^\dagger_\Lambda & = S_{(d,c)}[\Gamma]\ ,
    \\ 
    V_\Lambda S_{(0,1)}[\Gamma]  V^\dagger_\Lambda & = S_{(b,a)}[\Gamma]\ . 
\end{split}
\end{align}

Based on \eqref{SL(2,Z)action_on_surfaceop}, one can say that 
the set $\{S_{d,c}[\Gamma]\}$ also plays a role as 
a complete set of commuting observables and the corresponding 
eigenstates can be obtained by acting $V_\Lambda$ on the position
eigenstates $|b\rangle$, 
\begin{align}
    S_{(d,c)}[\Gamma_i] \Big( V_\Lambda |b\rangle \Big) 
    = \omega^{b_i} \Big( V_\Lambda |b\rangle \Big) \ .
\end{align}
The operator $V_\Lambda$ can be constructed as a condensation of line operators\cite{Roumpedakis:2022aik} and we will present an explicit expression of $V_\Lambda$ 
in Appendix \ref{app:A}. Here we consider two special cases $\Lambda=S,T$ 
to illustrate the construction. 

For $\Lambda=S$, the transformation rules become
\begin{align}
    V_S S_{(1,0)}[\Gamma] V^\dagger_S = S_{(0,1)}[\Gamma]\ .
\end{align}
It implies that $V_S$ maps a position eigenstate to a momentum eigenstate
\begin{align}
    \widetilde{U}[\Gamma_i] \Big( V_S |b \rangle \Big) = \omega^{b_i}  
    \Big( V_S |b \rangle \Big)\ . 
\end{align}
Fixing the $U(1)$ phase of the unitary operator $V_S$ such that 
\begin{align}\label{S_on_b0}
    V_S |b=0\rangle \equiv | \tilde b=0 \rangle\ 
\end{align}
without loss of generality, and using the relation 
\begin{align}
    |b\rangle = \widetilde{U}[\Gamma_b] | b=0\rangle\ ,
\end{align}
where $\Gamma_b$ is the Poincaré dual to $(-b_i \gamma^i)$ in $M_4$, 
one can show that 
\begin{align}\label{V_S}
    V_S |b\rangle = \frac{1}{\sqrt{N^{h_2}}}\sum_{b'} \omega^{K(b,b')} |b'\rangle \ . 
\end{align}

On the other hand, we will see that the $T$-transformation involves more subtle details. 
According to \eqref{SL(2,Z)action_on_surfaceop} for $\Lambda=T$, 
\begin{align}
\begin{split}
    V_T U[\Gamma] V^\dagger_T &= U[\Gamma]\ , 
    \\
    V_T \widetilde{U}[\Gamma] V^\dagger_T & = S_{(1,1)}[\Gamma]\ ,
\end{split}
\end{align}
$V_T$ maps a position eigenstate to itself,
\begin{align}
    U[\Gamma_i] \Big(V_T |b\rangle \Big) = \omega^{b_i} \Big(V_T |b\rangle \Big)\ .
\end{align}
In other words, the unitary operator $V_T$ only stacks a phase on $|b\rangle$, 
\begin{align}\label{T_on_b_01}
    V_T |b \rangle = \omega^{\varphi(b)} |b \rangle\ ,  
\end{align}
where the phase of $V_T$ is chosen so that $\varphi(0)=0$. When the manifold $M_4$
is spin, we can determine the phase $\varphi(b)$,
\begin{align}\label{T_phase_01}
    \varphi(b) = - \frac12 K(b,b)\ ,      
\end{align}
where $K(b,b)=K^{ij}b_ib_j$ is even for spin manifold. To see this, let us first rewrite 
the LHS of \eqref{T_on_b_01} as follows 
\begin{align}\label{T_on_b_02}
\begin{split}
    V_T \Big( \widetilde{U}[\Gamma_b] | b=0\rangle \Big) 
    & = S_{(1,1)}[\Gamma_b]  V_T |b=0\rangle \ .  
\end{split}    
\end{align}
Since the Baker-Campbell-Hausdorff (BCH) formula with the canonical commutation relation (naively) implies 
\begin{align}
    S_{(1,1)}[\Gamma_b] = \omega^{-\frac12 K(b,b)} \widetilde{U}[\Gamma_b] U[\Gamma_b]\ ,
\end{align}
one can further massage \eqref{T_on_b_01} into
\begin{align}
    V_T |b \rangle = \omega^{-\frac12 K(b,b)} \widetilde{U}[\Gamma_b] |b=0\rangle\ ,
\end{align}
which asserts \eqref{T_phase_01}. 

We have argued that the holonomies $b_i$ are $\mathbb{Z}_N$-valued 
due to the large gauge symmetry of the two-form field $B$. Let us 
check if the phase $\varphi(b)$ also respects the large gauge symmetry. 
It is sufficient to see how the phase changes under the shift 
\begin{align}\label{largegauge01}
     b \to  b + N e_\ell\ ,
\end{align}
where $e_\ell$ is an $h_2$-dimensional unit vector whose $i$-th component 
is given by $\delta_{i\ell}$. A small computation results in 
\begin{align}\label{xaxa}
    \varphi(b+Ne_\ell) = \varphi(b ) - \frac{N^2}{2} K^{\ell\ell} \text{ modulo } N\ .
\end{align}
Hence, since the self-intersection number $K^{\ell \ell}$ is 
always even for spin manifolds, the $\mathbb{Z}_N$ gauge symmetry  holds for the phase $\omega^{\varphi(b)}$.

As a consistency condition, we should examine if \eqref{V_S} and 
\eqref{T_on_b_01} with $\varphi(b)=-K(b,b)/2$ obey the relations $V_S^2 = V_{ST}^3=V_{C}$. It is trivial to demonstrate that 
\begin{align}
    V^2_S |b\rangle =  | - b\rangle\ , 
\end{align}
which agrees with $V_S^2=V_C$. 
However, confirming the other relation 
requires further elaboration: 
Action of the $(ST)^3$-transformation on the 
topological boundary state $|b\rangle$ results in 
\begin{align}\label{VST3}
\begin{split}
    V_{ST}^3 |b\rangle & = \left( \frac{1}{\sqrt{N^{h_2}}} \sum_{b'} \omega^{-\frac12 K(b',b')} \right) |-b\rangle \ .
\end{split}
\end{align}
Using Proposition $5.42$ of \cite{Hopkins:2002rd}, we then find  
\begin{align}\label{ST3}
    V_{ST}^3 |b \rangle =  
    e^{-2\pi i \sigma(M_4)/8 } | -b \rangle \ ,
\end{align}
where $\sigma(M_4)$ denotes the signature of the manifold $M_4$. 
As the spin manifold has $\sigma(M_4)=0$ modulo $16$, 
it becomes clear that $V_{ST}^3 = V_{C}$. On a non-spin manifold, however, 
a more carefully analysis is required to evaluate the sum in 
\eqref{VST3}. We will address it shortly.  
 
It is obvious that a set of the position eigenstates $\{ |b \rangle \}$ 
is invariant under $\Gamma_0(N)\subset SL(2,\mathbb{Z})$ generated by 
$T$ and $ST^NS$. Thus, one can describe 
a typical topological boundary state in the orbit of $SL(2,\mathbb{Z}_N)$ from $\{|b\rangle\}$
as 
\begin{align}\label{SL_orbit_01}
    V_{ST^k} |b \rangle  = 
    \frac{1}{\sqrt{N^{h_2}}}\sum_{b'} \omega^{K(b,b') - \frac k2 K(b',b')} | b'\rangle\ ,
\end{align}
where $k$ runs from $0$ to $(N-1)$. Note that $\{ V_{ST^k}| b\rangle\}$ for each $k$ becomes 
a basis of the Hilbert space ${\cal H}(M_4)$, and the corresponding 
complete set of commuting observables $\{ S_{(k,1)}[\Gamma]\}$ for any two cycles $\Gamma$:  
for each $\Gamma_i$ of \eqref{def_01}, 
\begin{align}
    S_{(k,1)}[\Gamma_i] \Big( V_{ST^k} |b \rangle \Big)  = \omega^{b_i } \Big( V_{ST^k} |b \rangle \Big)\ .
\end{align}

So far we have assumed that $M_4$ is both spin and torsion-free. For generic manifolds, the self-intersection number $K(b,b)$ is not necessarily even. When $N$ is even \eqref{T_on_b_01} should then be generalized to 
\begin{align}\label{T_on_b_03}
    V_T | b\rangle = \omega^{-\frac12 \mathfrak{P}(b)} | b\rangle\ ,
\end{align}
while the $S$-transformation \eqref{V_S} remains unchanged. 
Here $\mathfrak{P}(b)$ refers to the Pontryagin square which maps an element in $H_2(M_4,\mathbb{Z}_N)$ to $\mathbb{Z}_{2N}$
such that
\begin{align}\label{Pontryaginsqure_def}
    \mathfrak{P}(b) = K(\tilde{b},\tilde{b}) \text{ mod } 
    2 N\ ,
\end{align}
where $\tilde{b}$ is the lift of $b$ from 
$\mathbb{Z}_N$ to $\mathbb{Z}_{2N}$. Note 
that $\mathfrak{P}(b)$ reduces to the conventional  
intersection pairing $K(b,b)$ for spin manifolds. 
The Pontryagin square has the property that 
\begin{align}\label{Pontryagin-Square-Quadratic-Refinement}
    \mathfrak{P}(b+b') =    
    \mathfrak{P}(b) + \mathfrak{P}(b')
    + 2 K(b,b') \text{ mod } 2N\ ,
\end{align}
which is essential to ensure that the invariance of 
\eqref{T_on_b_03} under the large $\mathbb{Z}_N$ 
gauge transformation $b \to b + N e_\ell$. 
We present in Section 3,3 an explicit expression of $\mathfrak{P}(b)$ for $M_4= L(r,1) \times S^1$. 
When $N$ is odd, $\mathfrak{P}(b)/2$ in \eqref{T_on_b_03} reduces to 
the self-intersection number $K(b,b)/2$. The factor of $1/2$ 
should be understood as inverse of $2$, i.e., $(N+1)/2$ in $\mathbb{Z}_N$. Further explanation will be provided below. 

What if $M_4$ is not a spin manifold? We separate the discussion into two cases. If $N$ is even, from \eqref{xaxa} the $\mathbb{Z}_N$ gauge symmetry still holds and equation \eqref{ST3} is still valid. However, the signature $\sigma(M_4)$ is no longer divisible by $16$ but only integer-valued, and thus \eqref{ST3} does not agree with $V_S^2$ in general\footnote{For a given $M_4$ with a specific value of $N$, 
one may adjust overall phases of $V_S$ \eqref{S_on_b0} and $V_T$ \eqref{T_on_b_01} such that the relations
$V_S^2=V_{ST}^3 = V_C$ are obeyed. We will discuss a concrete example in Section \ref{sec:3.2} 
where $M_4=\mathbb{CP}^2$ and $N=2$. }. If $N$ is odd, first 
according to \eqref{xaxa} it seems that the $\mathbb{Z}_N$ gauge symmetry is broken for odd $N$ as the self-intersection numbers are not necessarily even for non-spin manifolds. 
However, this conclusion is rather misleading. For odd $N$, 
$\omega$ and $\omega^{N+1}$ can be interchangeable in all the formulas discussed so far. The substitution then shows
that the phase $(\omega^{N+1})^{\varphi(b)}$ in \eqref{T_on_b_01} 
becomes neutral under the $\mathbb{Z}_N$ gauge transformation.
Namely, $1/2$ should be  understood as the inverse of $2$ 
in $\mathbb{Z}_N$ for odd $N$.
Although the $\mathbb{Z}_N$ gauge symmetry remains consistent with the $T$-transformation on the non-spin manifold $M_4$, 
$V_S$ and $V_T$ fail to satisfy the relations $V_S^2=V_{ST}^3 = V_C$.
Moreover, the overall phase in the right-hand side of \eqref{ST3}
becomes more complicated and even depends on $N$. 
In Section \ref{sec:3.2}, we will explicitly compute 
the overall phase of $V_{ST}^3$ on $M_4=\mathbb{CP}^2$ for odd $N$. 
In short, the above discussion implies that 
the Hilbert space ${\cal H}(M_4)$ does not admit
the genuine unitary representation of $SL(2,\mathbb{Z}_N)$ beyond the spin manifolds.

The reader may wonder why the spin structure of $M_4$ does matter to have 
a unitary representation of  $SL(2,\mathbb{Z}_N)$, given that 
the BF model is topological. The answer is quite subtle and 
requires the rigorous definition of the action \eqref{BF_action}. 
As argued in \cite{Witten:1998wy}, the manifestly gauge-invariant formulation of the BF model
can be described in terms of the Cheeger-Simons cohomology \cite{10.1007/BFb0075216}. Then, the action is not necessarily invariant under $SL(2,\mathbb{Z})$ in general, but is instead invariant under its level-two congruence subgroup $\Gamma_\theta$. %{\color{red}{ZH: $\Gamma_\theta$ notation seems more standard}, all fixed}

One thus expects that when $M_4$ is non-spin, 
the Hilbert space of the BF model 
admits a unitary representation of $\Gamma_\theta$ rather than $SL(2,\mathbb{Z})$.  
The group $\Gamma_\theta$ can be generated by 
$T^2$ and $S$, satisfying two relations below 
\begin{align}
    S^2 T^2 = T^2 S^2 \ , \quad S^4 = {\bf 1}\ .    
\end{align}
It is straightforward to show that $V_T$ and $V_S$ indeed
obey these relations. Since $V_S^4={\bf 1}$ is trivial, 
we focus on the former relation. 
While $V_{S^2 T^2}$ acts on $|b \rangle$ 
as follows, 
\begin{align}
    V_{S^2 T^2} | b \rangle = \omega^{-\mathfrak{P}(-b)} |- b\rangle \ , 
\end{align}
the action of $V_{T^2 S^2}$ becomes
\begin{align}
    V_{T^2 S^2} | b \rangle = \omega^{-\mathfrak{P}(b)} |- b\rangle \ .
\end{align}
By the definition of the Pontryagin square \eqref{Pontryaginsqure_def}, one can see that 
\begin{align}
    \mathfrak{P}(b) = \mathfrak{P}(-b) \text{ mod } 2N\ ,   
\end{align}
which leads to 
\begin{align}
    V_{S^2 T^2} = V_{T^2 S^2}\ . 
\end{align}

We close the subsection with one remark. The relation $T^{2N}=1$ rather than $T^N =1$ of four-dimensional gauge theories on a non-spin manifold has been carefully studied recently in \cite{Ang:2019txy} where either bosonic or fermionic nature of loop operators plays a crucial role.  The $\Gamma_\theta$ symmetry for the BF model on a non-spin manifold combined with $\mathbb{Z}_N$ gauge symmetry can provide an alternative explanation 
for the relation $T^{2N}=1$.

%%%%%%%%%%%%%%%%%%%%%%%%%%%%%%%%%%%%%%%%%%%%%%%%%%%%%%%%%%%%%%%%%%%%%%%%%%%%%%%%%%%%%%%%%%%
\subsection{Partial gauging}\label{sec:partial}

When $N$ is not a prime number, there exist other kinds of topological boundary states. 
As we will discuss later, they correspond to gauging a subgroup 
of the center symmetry of a given four-dimensional gauge theory. 
To describe such states, let $N=PQ$ where $P$ and $Q$ need not be coprime in general.

First, we decompose the $\mathbb{Z}_N$-valued holonomy $b$
as follows, 
\begin{align}\label{decomposition01}
    | c,d \rangle \equiv | b = c + Q d \rangle  \ . 
\end{align}
In this decomposition, one can see that $d$ is $Z_P$-valued, i.e., 
\begin{align}
    |c,d+ Pe_\ell \rangle = | c,d \rangle\ ,
\end{align}
where $e_\ell$ denotes an $h_2$-dimensional unit vector 
with the $i$-th component being $\delta_{\ell i}$. 
On the other hand, $c$ of \eqref{decomposition01} is not
$\mathbb{Z}_Q$-valued, 
\begin{align}\label{shiftofc01}
    | c + Q e_\ell,d  \rangle = |c, d+e_\ell \rangle\ .    
\end{align}

Given that $d$ becomes $\mathbb{Z}_p$-valued, one can 
create a new topological boundary state in a manner analogous to 
the previous construction \eqref{SL_orbit_01}. 
To be more explicit, they can be expressed as, for each $k=0,1,..,(P-1)$,
\begin{align}\label{defofpartial}
    | c , \tilde d   \rangle_{k} \equiv \frac{1}{\sqrt{P^{h_2}}} 
    \sum_{d}  \big( \omega^Q \big)^{K({\tilde d},d) -\frac k2 \mathfrak{P}(d)} | c,d\rangle\ ,
\end{align}
where $\omega^Q$ becomes the $P$-th root of unity. When $\tilde d_P$ is shifted by $P$ units, the state \eqref{defofpartial} remains invariant
\begin{align}\label{tilded}
    | c , \tilde d + P e_\ell \rangle_k  & =    |    c , \tilde d   \rangle_k\ ,
\end{align}
i.e., $\tilde d$ is $\mathbb{Z}_P$-valued. Moreover, 
using \eqref{shiftofc01} and \eqref{Pontryagin-Square-Quadratic-Refinement}, 
one can show that the set $\{ | c , \tilde d   \rangle_{k} \}$ is mapped to 
itself under the shift $c \to c + Q e_\ell$,
\begin{align}\label{cshift}    
\begin{split}
    | c + Q e_\ell , \tilde d  \rangle_k  & =   \frac{1}{\sqrt{P^{h_2}}} 
    \sum_d \big( \omega^Q \big)^{K(\tilde d, d) - \frac k2 \mathfrak{P}(d)} |    c ,  d + e_\ell   \rangle_k 
    \\  & = 
    \big( \omega^Q \big)^{- K(\tilde d, e_\ell) - \frac k2  \mathfrak{P}(e_\ell)} | c, \tilde d + k e_\ell \rangle_k \ .
\end{split}
\end{align}
Since $\{ | c , \tilde d   \rangle_{k} \}$ consists of $(PQ)^{h_2}=N^{h_2}$ 
orthogonal vectors, one can regard it
as a basis of ${\cal H}(M_4)$ for each $k$. 

We can argue that the complete set of 
commuting observables corresponding to the basis 
$\{ |c,\tilde d\rangle_k \}$ can be generated by 
$\{  S_{(k,Q)}[\Gamma_i],  U^P[\Gamma_i]\}$ 
where $\Gamma_i$ are generators of $H_2(M_4)$ defined in \eqref{def_01}. In other words, the basis 
simultaneously diagonalizes $S_{(k,Q)}[\Gamma]$ for any $\Gamma$, 

\begin{align}
\begin{split}
    S_{(k,Q)}[\Gamma_i] | c,{\tilde d} \rangle_{k} & = 
    \frac{\omega^{k c_i}}{\sqrt{P^{h_2}}} 
    \sum_{d}  \big( \omega^Q \big)^{K({\tilde d},d) -\frac k2 \mathfrak{P}(d - K_i) } | c, d-K_i \rangle \ ,
    \\ & =
    \omega^{k c_i+ Q {\tilde d}_i} |c, {\tilde d}\rangle_{k}\ ,
\end{split}
\end{align}
where $K_i$ is the $h_2$-dimensional vector given in \eqref{def_of_K}. It also diagonalizes the other set of surface operators $\{U^P[\Gamma]\}$ as well,
\begin{align}
    U^P[\Gamma_i]  |c,\tilde d\rangle_k = \big( \omega^P \big)^{c_i}  |c,\tilde d_P\rangle_k\ .
\end{align}
\begin{figure}
    \centering
    \begin{subfigure}[b]{0.45\textwidth}
        \centering
        \includegraphics[width=\textwidth]{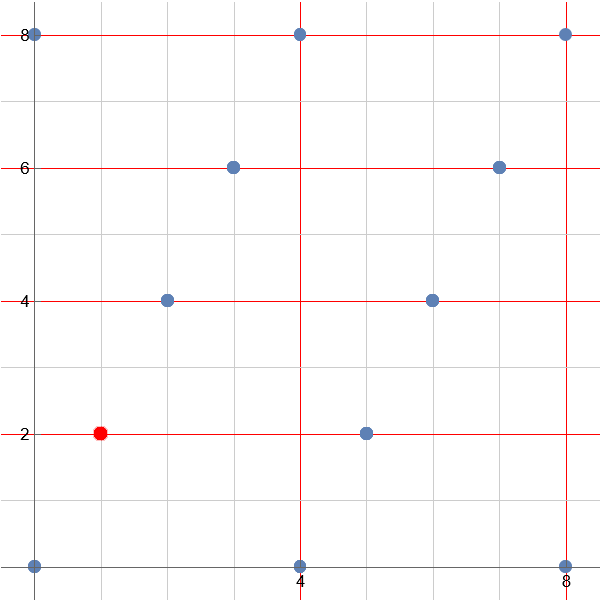}
        \caption{$(k,P,Q) = (1,4,2)$ with symmetry group $\mathbb{Z}_8$ }
    \end{subfigure}
    \hfill
    \begin{subfigure}[b]{0.45\textwidth}
        \centering
        \includegraphics[width=\textwidth]{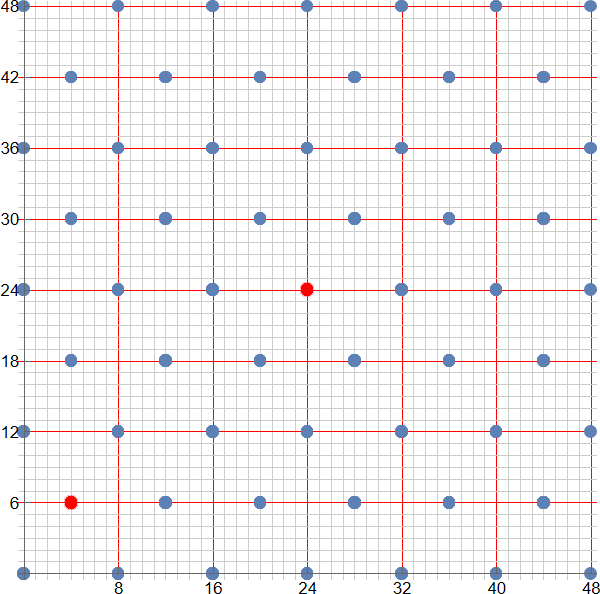}
        \caption{$(k,P,Q) = (4,8,6)$ with symmetry group $\mathbb{Z}_2 \times \mathbb{Z}_{24}$}
    \end{subfigure}
     \caption{Set of commuting observables for $(k,P,Q) = (1,4,2)$ or $(k,P,Q) = (4,8,6)$ and the symmetry group. A point with coordinate $(a,b)$ corresponds to $S_{(a,b)}[\Gamma]$ and the red dot indicates a possible choice of symmetry generators.}
\label{fig:examples}
\end{figure}

The partial gauging does not necessarily lead to 
the topological boundary states associated with 
the $\mathbb{Z}_N$ symmetry. In Figure \ref{fig:examples}, we list some explicit $k$, $P$ and $Q$ and determine their symmetric groups. In fact, for a given $(k, P, Q)$, 
we will show that the symmetry group becomes  
\begin{align}\label{partial_symmetry}
%\begin{cases}
    \mathbb{Z}_{\text{gcd}(k,P,Q)} \times 
    \mathbb{Z}_{N/\text{gcd}(k,P,Q)} \ .
%    & \text{ for } k \neq 0 \\ 
%    \mathbb{Z}_{P} \times \mathbb{Z}_Q & \text{ for } k = 0
%\end{cases}\ . 
\end{align}
Hence, the symmetry group can only be enhanced to 
$\mathbb{Z}_N$ when $\gcd{(k,P,Q)}=1$, e.g., $k=1$. 
We also understand from \eqref{partial_symmetry} that, unless $P$ and $Q$
are relatively prime, the symmetry group always factorizes 
for $k=0$, and becomes 
\begin{align}
    \mathbb{Z}_{\text{gcd}(P,Q)} \times \mathbb{Z}_{N/\text{gcd}(P,Q)}  
    \simeq \mathbb{Z}_{P} \times \mathbb{Z}_{Q}\ .    
\end{align}
One can use the Chinese remainder theorem to understand the above isomorphism. 
The $\mathbb{Z}_P$-valued holonomies $\tilde d$ \eqref{tilded} are account for $\mathbb{Z}_P$ while 
\eqref{cshift} with $k=0$ implies that
\begin{align}\label{ttt}
    | c + Q e_\ell, \tilde d  \rangle_0 = \big( \omega^Q \big)^{- K(\tilde d,e_\ell)}
    | c,\tilde d \rangle_0 \ .
\end{align}
Quantum mechanically, $c$ essentially becomes $\mathbb{Z}_Q$-valued, and thus explains the $\mathbb{Z}_Q$ symmetry.
As we will see later in Section \ref{sec:4}, the $U(1)$ phase in \eqref{ttt} reflects the mixed 't Hooft anomaly of 
four-dimensional gauge theory with $G=SU(PQ)$.
On the other hand, when $P$ and $Q$ are co-prime, 
one can use an alternative decomposition below to remove the $U(1)$ phase,
\begin{align}
    |c',d'\rangle \equiv |b = P c' + Q d' \rangle\ .
\end{align}
Since $c'$ and $d'$ are manifestly $\mathbb{Z}_Q$- and $\mathbb{Z}_P$-valued, so do 
$|c', {\tilde d}'\rangle_0$. The symmetry group is now isomorphic to 
$\mathbb{Z}_N$.

To sketch a proof of \eqref{partial_symmetry}, it is useful to note that 
a surface operator $S_{(e,m)}[\Gamma]$ is a $\mathbb{Z}_{N/\text{gcd}(e,m)}$ symmetry operator 
provided that $\gcd{(e,m)}$ divides $N$.  
This is because the smallest integer $M$ that gives rise to 
\begin{align}
    S_{(e,m)}^M[\Gamma] = S_{M(e,m)}[\Gamma] = 1
\end{align}
has to satisfy the relation below
\begin{align}
    \gcd{(Me,Mm)}=M\gcd{(e,m)} = 0 \text{ mod } N\ , 
\end{align}
i.e., $M=N/\gcd{(e,m)}$. A key idea of the proof      is that 
the complete set of commuting observables contains 
two surface operators $S_{(e,m)}[\Gamma]$ and 
$S_{(e',m')}[\Gamma]$ with
\begin{align}
    \gcd{(e,m)} = \gcd{(k,P,Q)}\ , 
    \quad 
    \gcd{(e',m')} = \frac{N}{\gcd{(k,P,Q)}}\ .
\end{align}

Since the complete set of commuting observables of our interest can be 
generated by $\{  S_{(k,Q)}[\Gamma_i],  U^P[\Gamma_i]\}$, 
any observables can be described as 
\begin{align}\label{observables}
    S_{(e,m)}[\Gamma] \text{ with } (e,m) = (\alpha k + \beta P, \alpha Q)  \ ,
\end{align}
where $\alpha$ and $\beta$ are some integers. Of particular interest is 
the surface operator $S_{(k + n A P, Q)}[\Gamma]$ 
where $A$ is chosen as an integer satisfying the B\'{e}zout identity
\begin{align}
    A P + B Q = \gcd{(P,Q)}\ , 
\end{align}
while the other integer $n$ as 
\begin{align}
    \frac{k}{\gcd{(k,P,Q)}} + n \frac{\gcd{(P,Q)}}{\gcd{(k,P,Q)}} = \mathfrak{p}
\end{align}
with $\mathfrak{p}$ being a certain prime number and assumed to be large enough. The Dirichlet prime number theorem
guarantees the existence of such $n$, since $k/\gcd{(k,P,Q)}$ and $\gcd{(P,Q)}/\gcd{(k,P,Q)}$
are co-prime. One can then show that 
\begin{align}
    \gcd{(k + n A P, Q)} = \gcd{(k + n \gcd{(P,Q)}, Q)} = \gcd{(k,P,Q)}\ . 
\end{align}
That is to say, $S_{(k + n A P, Q)}[\Gamma]$ is a $\mathbb{Z}_{N/\text{gcd}(k,P,Q)}$ 
symmetry operator. 

When $\gcd{(k,P,Q)}\neq 1$, an obvious candidate for the $\mathbb{Z}_{\gcd{(k,P,Q)}}$ symmetry operator  is 
the observable \eqref{observables} with $\alpha = P/\text{gcd}(k,P,Q)$ and $\beta = (Q-k)/\text{gcd}(k,P,Q)$, i.e, 
\begin{align}\label{Zgcdgenerator}
    S_{(\frac{N}{\gcd{(k,P,Q)}},\frac{N}{\gcd{(k,P,Q)}})}[\Gamma]\ .  
 \end{align}
Indeed, one can readily argue that \eqref{Zgcdgenerator} cannot be expressed as a multiple of $S_{(k + n A P, Q)}[\Gamma]$, 
and is therefore independent of $\mathbb{Z}_{N/\text{gcd}(k,P,Q)}$. It completes the proof. 

As a final remark, we propose that any two topological boundary states that share the same 
symmetry group are related by the $SL(2,\mathbb{Z})$ transformation. It sounds plausible, but 
we lack proof for now. 

\subsection{Examples}

To demonstrate the analysis, let us explicitly construct the topological 
boundary states for some examples. 

\begin{figure}
    \centering
    \includegraphics[width=0.95\textwidth]{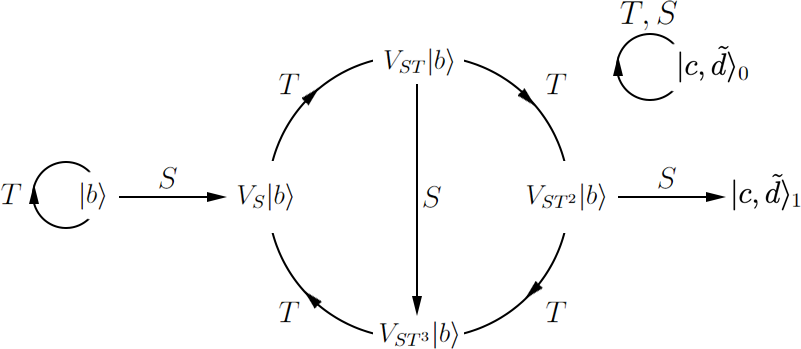}
    \caption{All topological boundary states of the BF model with level 4 and their $SL(2,\mathbb{Z})$ transformations.}
    \label{fig:enter4-label}
\end{figure}

The simplest yet non-trivial example is the BF model with level $N=4$. 
Starting with the position eigenstates $\{|b\rangle\}$, one 
can construct other topological boundary states via 
the $SL(2,\mathbb{Z})$ transformations. As in \eqref{SL_orbit_01},
some of them can be described as  
\begin{align}
    V_{S T^\ell} | b \rangle  =  \frac{1}{2^{h_2}} \sum_{b'}
    \omega^{K(b,b')- \frac \ell2 \mathfrak{P}(b')} |b' \rangle\ , 
\end{align}
where $\omega=e^{i \pi/2}$ is the fourth root of unity, and 
$\ell=0,1,..,3$. Additional topological boundary states 
can be obtained by the partial gauging. Now $P=Q=2$, 
and no other factorization of $N$ is possible. According to 
\eqref{defofpartial}, the method of partial gauging leads to 
two additional topological boundary states 
\begin{align}
    | c, \tilde d \rangle_{k} 
    = \frac{1}{\sqrt{2^{h_2}}} \sum_d (-1)^{K(\tilde d, d) - \frac k2 \mathfrak{P}(d)} 
    | b=c + 2d \rangle 
\end{align}
with $k=0,1$. They are associated with the symmetry groups  
$\mathbb{Z}_2\times \mathbb{Z}_2$ and $\mathbb{Z}_4$, respectively. Altogether, there are seven topological boundary states and their relations under $S$ and $T$ transformations are given in Figure \ref{fig:enter4-label}.

We explain the above figure by one example. Since the basis $\{|c,\tilde d\rangle_{k=1}\}$ corresponds to the $\mathbb{Z}_4$ symmetry, we expect that it can be related to $\{|b\rangle\}$ via
a certain $SL(2,\mathbb{Z})$ transformation. Indeed we can verify that
\begin{align}\label{jjj}
    \{ V_{ST^2S} |b \rangle \} =  \{ |c,\tilde d\rangle_{k=1} \} \ .
\end{align}
Notice that both bases of \eqref{jjj} diagonalize the same complete set of the commuting observables $\{ S_{(1,2)}[\Gamma] \}$. Since $V_{ST^2S} S_{(-1,0)}[\Gamma] V^\dagger_{ST^2S}= S_{(1,2)}[\Gamma]$, 
\begin{align}
    S_{(1,2)}[\Gamma_i] \Big( V_{ST^2S} |b \rangle \Big) = 
    \omega^{-b_i} \Big( V_{ST^2S} |b \rangle \Big)\ . 
\end{align}
On the other hand, by definition,
\begin{align}
    S_{(1,2)}[\Gamma_i]  |c,\tilde d\rangle_{k=1}
    = \omega^{c_i + 2\tilde d_i} |c,\tilde d\rangle_{k=1}\ , 
\end{align}
which implies that the holonomies $b_i$ are translated into 
$(c_i, \tilde d_i)$ as follows 
\begin{align}
    b_i = - c_i + 2 \tilde d_i \text{ mod } 4 \ .     
\end{align}

The explicit matching between those holonomies is
\begin{align}\label{holmatch}
        V_{ST^2S} |b = - c + 2 \tilde d \rangle = 
        e^{-2\pi i \sigma(M_4)/8} \omega^{K(\tilde d,\tilde d)} |  c, \tilde d \rangle_{k=1}\ . 
\end{align}
To show this, let us begin by 
\begin{align}
    V_{ST^2S} |b\rangle = \frac{1}{4^{h_2}} \sum_{b', b''} \omega^{K(b,b') - \mathfrak{P}(b',b') + K (b', b'') } | b'' \rangle
    \nonumber\ .
\end{align}
Decomposing $b=c+ 2d$, $b'=c'+2 d'$, $b''=c''+2d''$ where we restrict 
that $c=0,1$, $c'=0,1$ and $c''=0,-1$, it can be rewritten as
\begin{align}\label{kd}
    V_{ST^2S} |b\rangle =     \frac{1}{4^{h_2}} \sum_{\substack{c',d'\\ c'',d''}} (-1)^{ K (d',c + c'')} (-1)^{-\frac{1}{2}\mathfrak{P}(c',c')}  \omega^{ K (c', c+c'' + 2d + 2d'')}| c'' + 2d'' \rangle \ .
\end{align}
Here we used the fact that $\omega^{\mathfrak{P}(c'+2d')}= \omega^{\mathfrak{P}(c')}$. Summation over $d'$ then imposes a condition $c+c''= 0$, and \eqref{kd} can be further simplified into 
\begin{align}
    V_{ST^2S} |b\rangle = \frac{1}{2^{h_2}} \sum_{d''} 
    \left[ \sum_{c'} (-1)^{-\frac{1}{2}\mathfrak{P}(c',c')} (-1)^{ K (c',d + d'')}
    \right] | -c + 2d'' \rangle\ .
\end{align}
Using Proposition $5.42$ in \cite{Hopkins:2002rd}, the term in the square bracket becomes 
\begin{align}
    \frac{1}{\sqrt{2^{h_2}}} \sum_{c'}  (-1)^{-\frac{1}{2}\mathfrak{P}(c',c')} (-1)^{ K (c',d + d'')}
    = e^{-2\pi i \sigma(M_4)/8} (-1)^{K(d+d'', d+d'')/2}\ ,
\end{align}
where $\sigma(M_4)$ again denotes the signature of $M_4$. Thus, we prove that 
\begin{align}
    V_{ST^2S} |b\rangle =    e^{-2\pi i \sigma(M_4)/8} \omega^{K(d,d)} \left( \frac{1}{\sqrt{2^{h_2}}} \sum_{d''} (-1)^{K(d,d'') + K(d'',d'')/2  } 
    |-c + 2 d'' \rangle \right)\ ,
    \nonumber
\end{align}
which agrees with \eqref{holmatch}.
%\fi

As a second example, we discuss the BF model with level $N=8$, which as far as we know hasn't been worked out explicitly in the literature.  
Starting with the position eigenstates $\{|b\rangle\}$, one 
can construct other topological boundary states via 
the $SL(2,\mathbb{Z})$ transformations. Some of them are $V_{S T^\ell} | b \rangle$ given in \eqref{SL_orbit_01}. Additional topological boundary states 
can be obtained by partial gauging. It turns out that we have either $P = 2Q = 4$ or $Q = 2P = 4$, 
and no further factorization is possible. 

\begin{figure}
    \centering
    \includegraphics[width=1.0\textwidth]{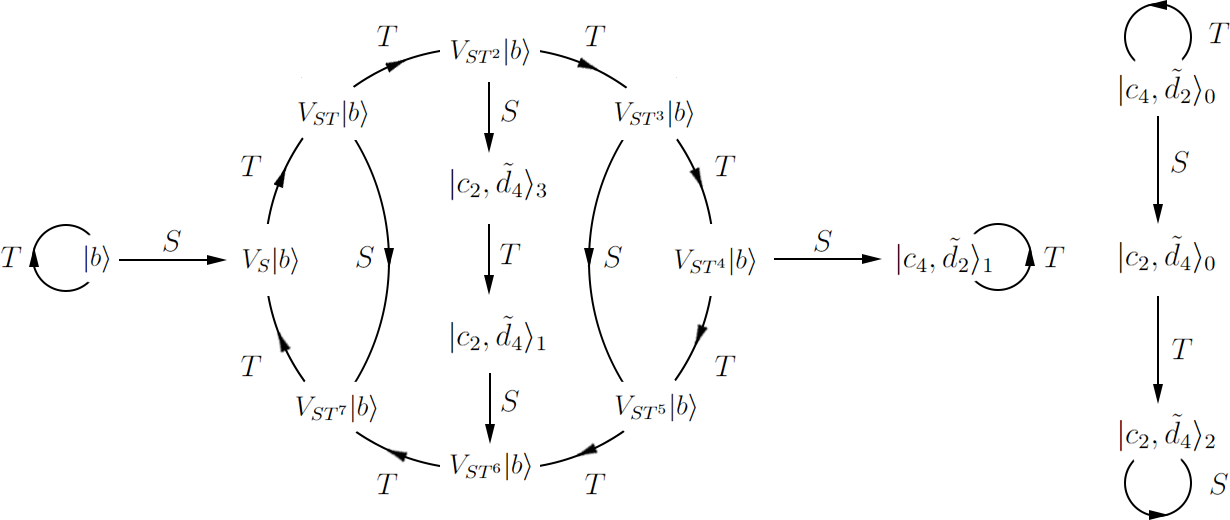}
    \caption{All topological boundary states of the BF model with level 8 and their $SL(2,\mathbb{Z})$ transformations.}
    \label{fig:enter-label4}
\end{figure}

According to \eqref{defofpartial}, the method of partial gauging leads to several additional topological boundary states. For example when $P = 2Q = 4$ we have $| c_2, \tilde d_4 \rangle_{k}$ with $k=0,1,2,3$. The symmetry group is either
$\mathbb{Z}_4\times \mathbb{Z}_2$ or $\mathbb{Z}_8$ from the general formula \eqref{partial_symmetry}. Similarly, for $Q = 2P = 4$ we have two new topological boundary states. Altogether, there are 1+8+4+2 = 15 topological boundary states and their relations under $S$ and $T$ transformations are given in Figure \ref{fig:enter-label4}.

We illustrate how to obtain it by one particular arrow. Since the basis $\{|c_2,\tilde d_4\rangle_{k=1}\}$ has the $\mathbb{Z}_8$ symmetry, we expect that it can be related to $\{|b\rangle\}$ via
a certain $SL(2,\mathbb{Z})$ transformation. Indeed we can verify that
\begin{align}\label{jjj2}
    \{ V_{ST^6S} |b \rangle \} =  \{ |c_2,\tilde d_4\rangle_{k=1} \}\ .
\end{align}
Notice that the both bases of \eqref{jjj2} diagonalize the same complete set of the commuting observables $\{ S_{(1,2)}[\Gamma], S_{(4,0)}[\Gamma] \}$. Since $V_{ST^6S} S_{(-1,0)}[\Gamma] V^\dagger_{ST^6S}= S_{(1,2)}[\Gamma]$, 
\begin{align}
    S_{(1,2)}[\Gamma_i] \Big( V_{ST^6S} |b \rangle \Big) = 
    \omega^{-b_i} \Big( V_{ST^6S} |b \rangle \Big)\ . 
\end{align}
On the other hand, by definition,
\begin{align}
    S_{(1,2)}[\Gamma_i]  |c_2,\tilde d_4\rangle_{k=1}
    = \omega^{(c_2)_i + 2(\tilde d_4)_i} |c_2,\tilde d_4\rangle_{k=1}\ , 
\end{align}
which implies that the holonomies $b_i$ are translated into 
$( (c_2)_i, (\tilde{d}_4)_i)$ as follows 
\begin{align}
    b_i = - (c_2)_i - 2 (\tilde d_4)_i \text{ mod } 8 \ .     
\end{align}

Similarly, we can use the relation
\begin{equation}
    V_{ST^6S} S_{(-4,0)}[\Gamma] V^\dagger_{ST^6S} = (V_{ST^6S} S_{(-1,0)}[\Gamma] V^\dagger_{ST^6S} )^4 = S_{1,2}[\Gamma]^4 \propto S_{4,0}[\Gamma]
\end{equation}
to show that both $V_{ST^6S} |b \rangle$ and $|c_2,\tilde d_4\rangle_{k=1}$ are eigenstates of $S_{4,0}[\Gamma]$. This concludes the proof of \eqref{jjj2}. 

\section{Topological Boundary States}\label{sec:3}
As concrete examples, let us work out the topological boundary states on a selected list of 4-manifolds: $T^4$, $\mathbb{CP}^2$ and $L(r,s) \times S^1$. 
\subsection{Topological boundary states on $T^4$}
Let $x^\mu$ $(\mu=0,1,2,3)$ be the coordinate of $T^4$, where $x^0$ is the time direction
while $x^i$ $(i=1,2,3)$ are the spatial directions. 
The cohomology group $H^2(T^4,\mathbb{Z})$ is  generated by $\gamma^{\mu\nu} = d x^{\mu} \wedge d x^{\nu}$, and is of dimension six, i.e., $h_2=6$.  Accordingly, one can describe the flat connections \eqref{zero_energy_config} as  
\begin{equation}
    \begin{split}
     \frac{N}{2\pi} B  = \sum_i t_i dx^0 \wedge dx^i + \frac{1}{2} \sum_{i,j,k} s_i \epsilon_{ijk} d x^j \wedge d x^k\ , \\
     \frac{N}{2\pi} \widetilde{B}  = \sum_i \tilde{t}_i dx^0 \wedge dx^i + \frac{1}{2} \sum_{i,j,k} \tilde{s}_i \epsilon_{ijk} d x^j \wedge d x^k \  ,        
    \end{split}
\end{equation}
where the totally anti-symmetric tensor $\epsilon_{ijk}$ 
has the convention $\epsilon_{123}=1$. The intersection matrix becomes
\begin{align}
    K\big(\gamma^{\mu\nu}, \gamma^{\rho\sigma} \big) \equiv K^{\mu\nu\hspace*{0.05cm}\rho\sigma}= \epsilon^{\mu\nu\rho\sigma}
\end{align}
where $\epsilon^{\mu\nu\rho\sigma}$ is the totally anti-symmetric tensor with 
$\epsilon^{0ijk}=\epsilon_{ijk}$. The canonical quantization gives the commutation relations below,
\begin{equation}
    \left[\hat{t}_i , \hat{\tilde{s}}_j \right] = \left[\hat{s}_i , \hat{\tilde{t}}_j \right] = i \frac{N}{2\pi} \delta_{ij}\ .
\end{equation}

We describe a generic $2$-cycle $\Gamma$ using a pair of three-vectors $(p_i,q_j)$ such that
\begin{equation}
        \Gamma = \sum_{i} p_i \Gamma_{0i} + \frac{1}{2}\sum_{ijk}  q_i \epsilon_{ijk} \Gamma_{jk} \equiv \Gamma_{p,q}\ .
\end{equation}
Here the $p$-components represent the $2$-cycles having a leg on the time direction $x^0$, 
while the $q$-components label the spatial $2$-cycles. 
The intersection pairing between two 2-cycles $\Gamma$ and $\Gamma'$ is
    \begin{equation}
        K(\Gamma,\Gamma') = p\cdot q' + q \cdot p'\ ,
    \end{equation}
where the inner product is defined as $p \cdot q = \sum_{i} p_i q_i$. The surface operators then satisfy the following algebra
    \begin{equation}
        U[\Gamma] \widetilde{U}[\Gamma'] = \omega^{-p\cdot q' - q \cdot p'} \widetilde{U}[\Gamma'] U[\Gamma]\ , 
    \end{equation}
where $\omega = e^{2\pi i /N}$ is the $N$-th root of unity.

Upon quantization, the Hilbert space of the BF model with level $N$ on $T^4$ 
has two canonical bases, one of which is the position eigenstates and 
the other is the momentum eigenstates. For later convenience, they are denoted by 
$|b\rangle = | (t,s) \rangle$ and $|\widetilde b\rangle= | (\tilde t, \tilde s)\rangle $
in what follows. Here both $t=(t_1,t_2,t_3)$ and $s=(s_1,s_2,s_3)$ are $\mathbb{Z}_N$-valued $3$-vectors 
and the same for $\tilde{t}$ and $\tilde{s}$. 
The position eigenstates are often referred to as the `electric' boundary states
satisfying the relations below, 
\begin{align}\label{5DBF-torus-electric-boundary-state}
\begin{split}
    U[\Gamma_{p,q}] \big|(t,s) \big \rangle & = \omega^{p\cdot t + q \cdot s} \big|(t,s) \big \rangle\ , 
    \\ 
    \widetilde{U}[\Gamma_{p,q}] \big|(t,s) \big\rangle & = \big|(t-q,s-p) \big \rangle\ .
\end{split}
\end{align}
On the other hand, the momentum eigenstates are also known as the  `magnetic' boundary states 
that obeys
\begin{align}\label{5DBF-torus-magnetic-boundary-state}
\begin{split}
        U[\Gamma_{p,q}] \big |(\tilde{t},\tilde{s}) \big \rangle & = 
        \big|(\tilde{t}+q,\tilde{s}+p) \big \rangle\ , \\ 
        \widetilde{U}[\Gamma_{p,q}] \big|(\tilde{t},\tilde{s})\big \rangle & = \omega^{p\cdot \tilde{t} + q \cdot \tilde{s}}
        \big |(\tilde{t},\tilde{s})\big \rangle\ .
\end{split}
\end{align}
The transition matrix between the two bases \eqref{transitionmatrix} can be expressed as  
\begin{equation}
    \big |(\tilde{t},\tilde{s}) \big \rangle = \frac{1}{N^3} \sum_{t,s} \omega^{\tilde{t} \cdot s + \tilde{s} \cdot t}  \big |(t,s) \big \rangle\ .
\end{equation}

The $T$ and $S$ transformation on the topological boundary states are 
\begin{align}
\begin{split}
    V_T \big |(t,s) \big \rangle &= \omega^{- t\cdot s} \big|(t,s) \big \rangle\ ,
    \\
    V_S \big |(t,s) \big \rangle &= \frac{1}{N^3} \sum_{t',s'} \omega^{t\cdot s' + s\cdot t'} \big |(t',s') \big \rangle\ . 
\end{split}
\end{align}
Finally, one can express the typical topological boundary states in the orbit of $SL(2,\mathbb{Z}_N)$ 
as follows, 
\begin{align}\label{SL_orbit_01_torus}
    V_{ST^k} \big |(t,s) \big \rangle  = 
    \frac{1}{N^3}\sum_{t',s'} \omega^{t\cdot s' + s\cdot t' - k t'\cdot s'} \big|(t',s') \big\rangle\ .
\end{align}

\subsection{Topological boundary states on $\mathbb{CP}^2$}\label{sec:3.2}
Consider the $\mathbb{CP}^2$ as a typical non-spin manifold. The 2-cycles are generated by $\Gamma \in H^2(\mathbb{CP}^2,\mathbb{Z}) = \mathbb{Z}$ and the intersection number between $\Gamma$ with itself is simply one. Denote the Poincaré dual of $\Gamma$ as $\gamma$, we can expand
    \begin{equation}
        \frac{N}{2\pi} B = b \gamma\ ,\quad \frac{N}{2\pi} \widetilde{B} = \tilde{b} \gamma\ ,
    \end{equation}
and the canonical quantization gives
    \begin{equation}
        \left[\hat{b},\hat{\tilde{b}} \right] = \frac{2\pi}{N}i\ .
    \end{equation}
The surface operators satisfy
    \begin{equation}
        U[\Gamma] \widetilde{U}[\Gamma] = \omega^{-1} \widetilde{U}[\Gamma] U[\Gamma]\ .
    \end{equation}

Upon quantization, we can denote the position/momentum eigenstates separately as $|b\rangle$ and $|\tilde{b}\rangle$ with $b,\tilde{b}\in \mathbb{Z}_N$. They satisfy
\begin{equation}
    U[\Gamma]|b\rangle = \omega |b\rangle\ , \quad \widetilde{U}[\Gamma]|b\rangle = |b-1\rangle\ ,
\end{equation}
and
\begin{equation}
    U[\Gamma]|\tilde{b}\rangle = |\tilde{b}+1\rangle,\quad \widetilde{U}[\Gamma]|\tilde{b}\rangle = \omega |\tilde{b}\rangle\ .
\end{equation}
with $b,\tilde{b} = 0,\cdots,(N-1)$ and the two kinds of states are related by
    \begin{equation}
    |\tilde{b}\rangle = \frac{1}{\sqrt{N}}\sum_b \omega^{\tilde{b} b}  |b\rangle\ .
\end{equation}
The $T$ and $S$ transformation act on the topological boundary state as
\begin{align}\label{P2VTVS01}
\begin{split}
        V_T |b\rangle & = \omega^{\frac{b^2}{2}} |b\rangle\ ,
        \\ 
        V_S |b\rangle & = \frac{1}{\sqrt{N}}\sum_{b'} \omega^{b b'}  |b'\rangle\ , 
\end{split}
\end{align}
where the Pontryagin square is given by $\mathfrak{P}(b) = b^2$. 
When $N$ is even, it is easy to see $\frac{b^2}{2}$ is invariant under $b\rightarrow b + N$. And when $N$ is odd, since $\gcd(2,N)=1$ in this case the $2^{-1}$ can be understood as inverse of $2$ in $\mathbb{Z}_N$ and $\frac{b^2}{2}$ is still invariant under $b\rightarrow b + N$.  As pointed out in the previous section, $V_T^{2N} =1$ for $\mathbb{CP}^2$.

Since $\mathbb{CP}^2$ is not a spin manifold, the relations 
$V_S^2 = V_{ST}^3= V_C$ are not obeyed in general. See the equation \eqref{ST3} 
and note that the signature is $\sigma(\mathbb{CP}^2)= 1$.
However, for $N=2$, there is a choice of different overall 
phases for $V_S$ and $V_T$ such that the Hilbert space 
of the BF model on $\mathbb{CP}^2$ has no $SL(2,\mathbb{Z})$
anomaly. To see this, let us choose different overall phases as follows, 
\begin{align}\label{CP2VTVS}
\begin{split}
    V_S |b=0\rangle &= (-1) |\tilde b=0\rangle \ , \\ 
    V_T |b=0 \rangle &= e^{-\frac{\pi i}{4}}|b=0 \rangle \ . 
\end{split}
\end{align}
In other words, the $T$ and $S$ transformation now act on the topological boundary state with $N=2$ as
\begin{align}\label{P2VTVS02}
\begin{split}
        V_T |b\rangle & = e^{-\frac{\pi i}{4}} e^{ -\frac{\pi i}{2} b^2} |b\rangle\ ,
        \\ 
        V_S |b\rangle & = - \frac{1}{\sqrt{2}}\sum_{b'=0,1} (-1)^{b b'}  |b'\rangle\ , 
\end{split}
\end{align}
With \eqref{P2VTVS02}, one can verify that 
\begin{align}
    V_S^2 = V_{ST}^3 = {\bf 1} \ . 
\end{align}
$V_T^{4}$ now becomes $(-{\bf 1})\propto {\bf 1}$.

We can also consider the cases with $N$ being odd. In the notation of \cite{Hopkins:2002rd} we can choose the vector space $V$ to be the real numbers, such that the bilinear form, the characteristic element $\lambda$ and the quadratic form are given by
\begin{equation}
    B(x,y) = N x y\ ,\quad \lambda = 1 \in \mathbb{R}\ , \quad q(x) = \frac{B(x,x) - B(x,\lambda)}{2} = N\frac{x(x-1)}{2}\ . 
\end{equation}
The lattice $L$ is nothing but $\mathbb{Z} \in \mathbb{R}$. Then the dual lattice $L^\ast$ is generated by $1/N$, such that we have
\begin{equation}
    \frac{1}{\sqrt{N}}\sum_{x \in L^\ast/L} e^{(-2\pi i q(x))} = e^{2\pi i (B(\lambda,\lambda) - \sigma )/8} = e^{\pi i(N - 1)/4}\ .
\end{equation}
On the other hand, we can rewrite the left-hand side,
\begin{equation}
\begin{aligned}
       &\frac{1}{\sqrt{N}}\sum_{x \in L^\ast/L} e^{(-2\pi i q(x))}\\
       &= \frac{1}{\sqrt{N}} \sum_{k \in \mathbb{Z}_N} e^{(-2\pi i q(k/N))} = \frac{1}{\sqrt{N}} \sum_{k \in \mathbb{Z}_N} e^{(-2\pi i \frac{k(k-N)}{2N})} = \frac{1}{\sqrt{N}} \sum_{k \in \mathbb{Z}_N} e^{(-2\pi i \frac{(N+1)k^2}{2N} )}\ ,
\end{aligned}
\end{equation}
where in the last equality we use the fact that $k(k-N) = (N+1) k^2 $ mod $ 2N$. Combining the above two equations, we can show that 
\begin{equation}\label{xdr}
    V_{ST}^3 |b \rangle = \Big(\frac{1}{\sqrt{N}}\sum_{k \in \mathbb{Z}_N} (\omega^{N+1})^{-k^2/2}\Big) |-b \rangle =  e^{\pi i(N - 1)/4} |-b \rangle\ .
\end{equation}
We comment that the overall phase depends on $N$ when $N$ is odd. It shows that
a universal redefinition of $(V_T, V_S)$, independent of $N$,  that removes the  $SL(2,\mathbb{Z})$ anomaly \eqref{xdr}
is absent.

\subsection{Topological boundary states on $L(r,s)\times S^1$}\label{sec:topboundaryLens}

Finally, we discuss the topological boundary states on $M_4=L(r,s)\times S^1$ where 
$L(r,s)$ is the lens space. Here $r$ and $s$ are relatively prime. 
One can describe the lens spaces as quotients of 
the three-sphere $S^3$ by $\mathbb{Z}_r$ action. To be concrete, 
let us begin with an isomorphism between $SU(2)$ and $S^3$, 
\begin{align}\label{su2s3}
    g(z_1,z_2)= \begin{pmatrix} z_1 & -\bar z_2 \\ z_2 & \bar z_1 \end{pmatrix}
\end{align}
where $z_1$ and $z_2$ are two complex variables subject to the condition $|z_1|^2+|z_2|^2=1$. 
The map \eqref{su2s3} allows us to describe the $SO(4)$ rotation of $S^3$ as 
%e^{-\frac{\pi i}{4}}
\begin{align}
    g(z_1',z_2') = g_L g(z_1, z_2) g_R 
\end{align}
where $g_L$ and $g_R$ denote the left $SU(2)_L$ action and the right $SU(2)_R$ action. 
To define the lens space $L(r,s)$, let us consider a specific 
$SO(4)$ rotation 
\begin{align}\label{quotientmap01}
\begin{split}
    g_L & = e^{2\pi i \frac{s+1}{r} \cdot \frac{\tau^3}{2}}\ , 
    \\
    g_R &= e^{2\pi i \frac{s-1}{r} \cdot \frac{\tau^3}{2}}\ . 
\end{split}
\end{align}
where $\tau^3$ is a Pauli matrix. \eqref{quotientmap01} is 
indeed a $\mathbb{Z}_r$ action on the three-sphere,  
\begin{align}
    (z_1, z_2) \longrightarrow (z_1',z_2') = (e^{2\pi i s/r} z_1, e^{-2\pi i /r} z_2 )\ , 
\end{align}
and the lens space $L(r,s)$ is defined as the quotient of $S^3$ by 
the $\mathbb{Z}_r$ action \eqref{quotientmap01}. 
The lens space $L(r,s)$ has a torsion one-cycle $\mathcal{C}_\tau \in H_1(L(r,s),\mathbb{Z}) = \mathbb{Z}_r$ of order $r$. 
The self torsion linking number of ${\cal C}_\tau$ inside the lens space 
is $s/r$. 

The classical zero-energy configurations are the flat connections of 
$B$ and $\tilde B$ on $L(r,s)\times S^1$. One can characterize 
them by their holonomies along two-surfaces. 
The four-dimensional manifold $L(r,s)\times S^1$ has
two relevant two-surfaces, one of which is 
$\Gamma_1= {\cal C}_\tau \times S^1$. Inside the lens spaces, there exists 
the other two-surface $\Gamma_2$ which has a boundary
\begin{align}
    \partial \Gamma_2 = r {\cal C}_\tau\ . 
\end{align}
Therefore, $\Gamma_2$ is not an element of $H_2(M_4,\mathbb{Z})$ but of $H_2(M_4, \mathbb{Z}_r)$.  As an example, when $r=2,s=1$ the lens space $L(2,1)$ is the 3-dimensional real projective plane $\mathbb{RP}^3$ and $\Gamma_2$ is the 2-dimensional real projective plane $\mathbb{RP}^2$ embedded in $\mathbb{RP}^3$.

The low-energy dynamics \eqref{BFlowenergy} of the BF model on $L(r,s)\times S^1$ 
can be described in terms of the holonomies along $\Gamma_1$ and $\Gamma_2$,  
\begin{align}
    b_i  = \frac{N}{2\pi} \int_{\Gamma_i} B \ , 
    \quad
    \tilde b_i = \frac{N}{2\pi} \int_{\Gamma_i} \widetilde{B}\  .
\end{align}
Here $b_i$ and $\tilde b_i$ are periodic variables \eqref{periodicity}.
Upon the quantization, we have two canonical bases of the Hilbert of 
the BF model on $L(r,s)\times S^1$. Let us focus on the position basis
$\{ |b = (b_1,b_2) \rangle\}$: 
\begin{align}\label{basiclens}
\begin{split}
    U[\Gamma_i] | (b_1,b_2) \rangle  & = \omega^{b_i} | (b_1,b_2)  \rangle\ , 
    \\
    \widetilde{U}[\Gamma_j]  | (b_1,b_2)  \rangle & = | (b_1- K_{j1} ,b_2 - K_{j2})  \rangle\ ,
\end{split}
\end{align}
where $U[\Gamma_i]$ and $\widetilde{U}[\Gamma_i]$ are defined in \eqref{5DBF-Basic-Wilson-Surface-field-variables}   
and the eigenvalues $b_i$ are quantized, $b_i=0,1,..,(N-1)$. 
Here $K_{ij}$ denotes the intersection number between 
$\Gamma_i$ and $\Gamma_j$, 
\begin{align}\label{intersectionlens}
    K_{ij} = \begin{pmatrix} 0 & s \\ s & 0 \end{pmatrix}\ .     
\end{align}

The torsional nature of $\Gamma_1$ and $\Gamma_2$ then further 
restricts the possible values allowed for $\{b_i\}$. Since $\Gamma_1$ 
is the torsion two-cycle,  
\begin{align}\label{operatorconstraint01}
    U[\Gamma_1]^r = \widetilde{U}[\Gamma_1]^r = {\bf 1}\ .
\end{align}
This implies that the eigenvalue $b_1$ has to satisfy the 
relation below
\begin{align}\label{constraints01}
    r b_1 = 0 \text{ mod } N\ .    
\end{align}
On the other hand, the torsional boundary of $\Gamma_2$ 
imposes a constraint on the eigenvalue $b_2$. 
To understand this, we first recall that the quantum theory of the BF model 
preserves the $\mathbb{Z}_N\times\mathbb{Z}_N $ gauge symmetry.
The gauge transformation rules are 
\begin{align}\label{ZNgaugeB}
\begin{split}
    B & \longrightarrow B + \frac1N d\lambda_1 \ , 
    \\
    \widetilde{B} & \longrightarrow \widetilde{B}  +  \frac1N d\lambda_2 \ .
\end{split}
\end{align}
Here two gauge transformation parameters $\lambda_1$ and $\lambda_2$ 
have integer-valued periods: for any one-cycle ${\cal C}$, 
\begin{align}
    \frac{1}{2\pi} \oint_{\cal C} \lambda_1 \in \mathbb{Z}\ , 
    \quad 
    \frac{1}{2\pi} \oint_{\cal C} \lambda_2 \in \mathbb{Z}\ .
\end{align}
Indeed, $U[\Gamma_1]$ and $\widetilde{U}[\Gamma_1]$ 
are invariant under \eqref{ZNgaugeB}. However, since $\Gamma_2$ 
has the boundary, $U[\Gamma_2]$ and $\widetilde{U}[\Gamma_2]$ are no longer 
gauge invariant. Instead, they transform as 
\begin{align}\label{operatorconstraint02}
\begin{split}
    U[\Gamma_2] & \longrightarrow \exp\Big[{\frac{ir}{N}  \oint_{{\cal C}_\tau}  \lambda_1} \Big] U[\Gamma_2] = \omega^{r} U[\Gamma_2] \ , 
    \\
    \widetilde{U}[\Gamma_2] & \longrightarrow \exp\Big[{\frac{ir}{N}  \oint_{{\cal C}_\tau}  \lambda_2} \Big] \widetilde{U}[\Gamma_2] = \omega^{r} \widetilde{U}[\Gamma_2] \ , 
\end{split}
\end{align}
when $\lambda_1$ and $\lambda_2$ have unit periods. 
We denote by $\Lambda_1$ the former $\mathbb{Z}_N$ gauge transformation 
operator that rotates the `electric' surface operator,   
\begin{align}
    \Lambda_1 U[\Gamma_1] \Lambda_1^{-1} = U[\Gamma_1]\ , \quad 
    \Lambda_1 U[\Gamma_2] \Lambda_1^{-1} = \omega^r U[\Gamma_2] \ , 
\end{align}
while leaving the `magnetic' surface operator invariant, 
\begin{align}
    \Lambda_1 \widetilde{U}[\Gamma_1] \Lambda_1^{-1} = \widetilde{U}[\Gamma_1]\ , \quad 
    \Lambda_1 \widetilde{U}[\Gamma_2] \Lambda_1^{-1} = \widetilde{U}[\Gamma_2]\ . 
\end{align}
How does the above $\mathbb{Z}_N$ gauge transformation 
then act on the Hilbert space? We can argue that 
\begin{align}
    \Lambda_1 | (b_1,b_2)  \rangle \propto \big| (b_1, b_2-r ) \big \rangle\ .
\end{align}
This is because the states $\Lambda_1 | (b_1,b_2)  \rangle$ are eigenstates for $U[\Gamma_i]$: specifically
\begin{align}
\begin{split}
    \big(\Lambda_1 U[\Gamma_2] \Lambda_1^{-1} \big) \Lambda_1 |b_1,b_2\rangle 
    &=  \omega^r U[\Gamma_2] \big( \Lambda_1 |b_1,b_2\rangle \big) \ ,
    \\
    & = \omega^{b_2} \Lambda_1 |b_1,b_2\rangle \ .
\end{split}
\end{align}
The gauge equivalence then demands that the eigenvalue $b_2$ obeys,
\begin{align}\label{constraints02}
    | (b_1, b_2) \rangle \simeq | (b_1, b_2 - r ) \rangle\ . 
\end{align}
In fact, the two constraints on the eigenvalues $\{ b_i \}$
\eqref{constraints01} and \eqref{constraints02} reflect the fact that 
\begin{align}
    H^2\big(L(r,s)\times S^1,\mathbb{Z}_N\big) = 
    \textrm{Ker}\ r \oplus \frac{\mathbb{Z}_N}{r \mathbb{Z}_N}\ ,
\end{align}
where $r$ can be understood as mapping $n$ to $rn$ for any $n\in \mathbb{Z}_N$.

Solving the two constraints \eqref{constraints01} and  \eqref{constraints02},
one can show that $b_1$ should be a multiple of $N/\gcd{(r,N)}$, 
and $b_2$ effectively becomes $\mathbb{Z}_{\gcd{(r,N)}}$-valued,
\begin{align}\label{lensposition}
\begin{split}
    b_1 & = \frac{N}{\gcd{(r,N)}} k_1\ , 
    \\
    b_2 & = k_2 \ , 
\end{split}
\end{align}
where $k_1, k_2 = 0,1,.., (\gcd{(r,N)}-1)$. Similarly, the momentum 
eigenvalues are also constrained as
\begin{align}\label{lensmomentum}
\begin{split}
    \tilde b_1 & = \frac{N}{\gcd{(r,N)}} \tilde k_1\ , 
    \\
    \tilde b_2 & = \tilde k_2 \ , 
\end{split}    
\end{align}
where $\tilde k_1, \tilde k_2 = 0,1,.., (\gcd{(r,N)}-1)$. The 
transition matrix between the position and momentum bases now becomes 
\begin{align}\label{transition}
    \big | (\tilde b_1, \tilde b_2) \big \rangle =  \frac{1}{\gcd{(r,N)}} \hspace*{-0.2cm}
    \sum_{k_1, k_2=0}^{\gcd{(r,N)}-1} \hspace*{-0.2cm} \omega ^{\frac{1}{s} ( \tilde b_1 b_2 + \tilde b_2 b_1 )} 
    \big | (b_1, b_2) \big \rangle\ ,  
\end{align}
where $b_i$ and $\tilde b_j$ are parameterized by \eqref{lensposition} and \eqref{lensmomentum} and $\omega$ is the 
$N$-th root of unity. The phase factor in \eqref{transition} can be rewritten as 
\begin{equation}
        \exp\Big[\frac{2\pi i}{\gcd(r,N)} \frac{\tilde{k}_1 k_2 + \tilde{k}_2 k_1}{s} \Big]\ .
\end{equation}
In this expression, we observe that the $N$-th root of unity $\omega$ turns to the $\gcd(r,N)$-th root of unity, and both $k_1,k_2$ and $\tilde{k}_1,\tilde{k}_2$ take values in $\mathbb{Z}_{\gcd(r,N)}$. 
Since $s$ is coprime with $\gcd(r,N)$, we can interpret $1/s$ as the inverse of $s$ under $\mathbb{Z}_{\gcd(r,N)}$. 
Hence, \eqref{transition} can be understood as the standard discrete Fourier transformation. 

These restrictions on the topological boundary states $\{ | b \rangle \}$ are consistent with 
the action of operators $\widetilde{U}$ and $U$. 
Given that $U[\Gamma_i]^N=\widetilde{U}[\Gamma_i]^N={\bf 1}$,
the relation \eqref{operatorconstraint01} leads to   
\begin{align}
    U[\Gamma_1]^{\gcd(r,N)} = \widetilde{U}[\Gamma_1]^{\gcd(r,N)} = {\bf 1}\ . 
\end{align}
while \eqref{operatorconstraint02} implies that the gauge invariant operators should be 
of the form 
\begin{align}\label{gaugeinvariantop}
    U[\Gamma_2]^{\frac{k N}{\gcd(r,N)}}\ , \quad \widetilde{U}[\Gamma_2]^{\frac{\tilde{k}N}{\gcd(r,N)}}\ ,
\end{align}
where $k$ and $\tilde k$ are arbitrary integers. 
We learned that the magnetic operators $\widetilde{U}[\Gamma_i]$
generate the finite translations \eqref{basiclens}.
In particular, the successive actions of $\widetilde{U}[\Gamma_1]$ 
can provide all $\gcd{(r,N)}$ different values of $b_2$,  
\begin{align}
    \widetilde{U}[\Gamma_1]^n \big | (b_1, b_2) \big \rangle = 
    \big | (b_1, b_2- n s) \big \rangle \ , 
\end{align}
given that $s$ is co-prime  with $\gcd{(r,N)}$. 
On the other hand, the translation operator 
$\widetilde{U}[\Gamma_2]$ naively conflicts with 
the fact that $b_1$ has to be a multiple of 
$N/\gcd{(r,N)}$ \eqref{constraints01}. 
However, $\widetilde{U}[\Gamma_2]$ is not a gauge invariant 
operator and thus cannot be regarded as a physical observable. 
The gauge invariant magnetic operator indeed induces 
the finite translation compatible with \eqref{constraints01}
\begin{align}
    \widetilde{U}[\Gamma_2]^{\frac{N}{\gcd(r,N)}} 
    \big | (b_1, b_2) \big \rangle = \big | (b_1 - \frac{N}{\gcd{(r,N)}} s , b_2 ) \big \rangle\ .   
\end{align}
Moreover, it is easy to see that  
\eqref{constraints01} is also consistent with the relation $U[\Gamma_1]^{\gcd{(r,N)}}={\bf 1}$ and the 
gauge invariant electric operator $U[\Gamma_2]^{N/\gcd{(r,N)}}$. 

In order to discuss the other topological boundary states on $L(r,s)\times S^1$, e.g., 
$\{ V_{ST^k} | (b_1,b_2) \rangle \}$, it is crucial to carefully define 
the Pontryagin square $\mathfrak{P}(b)$ for $L(r,s)\times S^1$. 
For later convenience, let us set $s=1$ in what follows. 

We begin by expanding the two-form gauge field $B$ as follows, 
\begin{equation}
    \frac{N}{2\pi} B = b_1 \gamma_2 + b_2 \gamma_1\ ,
\end{equation}
where $\gamma_1$ and $\gamma_2$ are the Poincaré dual to 
$\Gamma_1$ and $\Gamma_2$,
\begin{equation}
    \int_{\Gamma_i} \omega = \int \gamma_i \cup \omega\ , \quad \forall \omega \in C^1(L(r,1)\times S^1)
\end{equation}
with $\Gamma_1 \equiv {\cal C}_\tau \times S^1$ and $\partial \Gamma_2 = r {\cal C}_\tau$. 
Since the manifold of our interest is torsion, the cup product between two-form is not necessarily symmetric. In general, the cup-$1$ product $\cup_1$ 
can measure how non-commutative the cup product is \cite{steenrod1947products}, 
\begin{equation}\label{defcup1}
    f \cup g - (-1)^{pq} g \cup f = (-1)^{p+q+1} \Big( \delta (f\cup_1 g) - \delta f \cup_1 g - (-1)^p f \cup_1 \delta g \Big)\ , 
\end{equation}
where $f \in C^p(M)$ and $g \in C^q(M)$ for a given manifold $M$. 
It is convenient to adopt the Poincaré dual picture and 
think of the cup product and the cup-$1$ product as certain topological intersection numbers within the geometry.  

\begin{figure}
    \centering
    \includegraphics[scale=0.6]{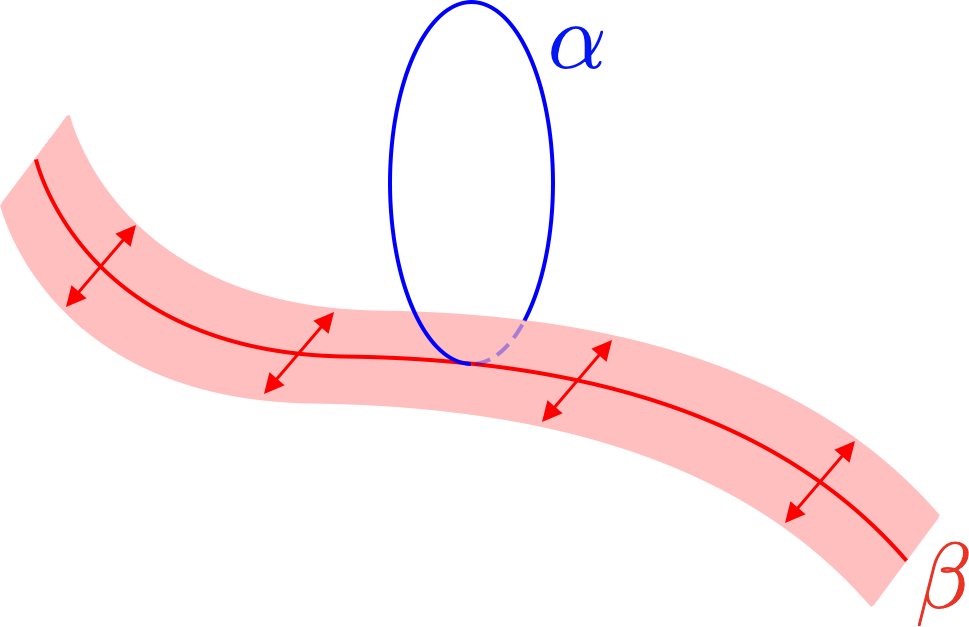}
    \caption{An illustration of the cup-1 product $\int [\alpha] \cup_1 [\beta]$ where $[\cdots]$ denote the Poincaré dual.  The thickening of $\beta$ is given in both the positive and negative directions of the Morse flow (both directions point away from the central red curve). $\int [\alpha] \cup_1 [\beta]$ measures the intersection between $\alpha$ and the thickening of $\beta$. This figure is taken from \cite{Tata:2020qca}. }
    \label{fig:enter-label}
\end{figure}

Since $\partial \Gamma_1 = 0$ and $\partial \Gamma_2 = r {\cal C}_\tau$, 
the relevant cup-$1$ products are 
\begin{equation}
    \gamma_1 \cup_1 \delta \gamma_2\ ,\quad \gamma_2 \cup_1 \delta \gamma_2\ ,
\end{equation}
where $\delta \gamma_2 = r u$ is dual to $\partial \Gamma_2 = r {\cal C}_\tau$ where $u$ is a $3$-cocycle 
dual to ${\cal C}_\tau$. Geometrically the cup-$1$ product between two cycles $\alpha$ and $\beta$ 
can be identified as the intersection number between $\alpha$ and the ``thickening'' of $\beta$, 
as depicted in Figure \ref{fig:enter-label}. 
The thickening of $\beta$ is given in both the positive and negative directions of the Morse flow. 
For more details, the readers are referred to \cite{Tata:2020qca,Thorngren:2018ziu}.

In this picture, the cup-$1$ product between $\gamma_1$ and $\gamma_2$ with $\delta \gamma_2$ 
is defined as  the intersection number between $\Gamma_1$ and $\Gamma_2$ with $\partial \Gamma_2 = r {\cal C}_\tau$ thickened. Since $\Gamma_2$ and $\partial \Gamma_2$ are located at the same point along $S^1$ and the thickening of $\partial \Gamma_2$ generically has a component along $S^1$, we have
\begin{equation}
    \int \gamma_2 \cup_1 \delta \gamma_2 = r\ ,\quad \int \gamma_1 \cup_1 \delta \gamma_2 = 0\ . 
\end{equation}
They make an integral of cup product between $\gamma_1$ and $\gamma_2$ symmetric, 
\begin{equation}
    \int \gamma_1 \cup \gamma_2 - \int \gamma_2 \cup \gamma_1 = \int \delta \gamma_1 \cup_1 \gamma_2 + \int \gamma_1 \cup_1 \delta \gamma_2 = 0\ .
\end{equation}
Since the intersection number between $\Gamma_1$ and $\Gamma_2$ is one \eqref{intersectionlens},
we finally have
\begin{align}
    b_1 = \frac{N}{2\pi} \int_{\Gamma_1} B\ , 
    \quad 
    b_2  = \frac{N}{2\pi} \int_{\Gamma_2} B\ ,  
\end{align}
as required by the definition of the $\mathbb{Z}_N$-valued holonomies $b_1$ and $b_2$. 
    
The Pontryagin square for $H^2(L(r,1)\times S^1,\mathbb{Z}_N)$ is given by \cite{Kapustin:2013qsa}
\begin{align}
    \mathfrak{p}(b) =  
    \begin{cases}
         b \cup b \mod 2N  & \text{ for odd } N \\
         b \cup b + b \cup_1 \delta b \mod 2N \ & \text{ for even } N \ .
    \end{cases}
\end{align}
We focus on the case where $N$ is even, for which $\mathfrak{p}(b)$ has 
an additional term involving the cup-$1$ product $b \cup_1 \delta b$. 
One can argue that this refinement is sufficient (though not necessary) for all four-manifold $M_4$
with torsion. Using the property of the cup-$1$ product, one has 
\begin{align}\label{cup-1-product}
    \mathfrak{p}(b+b') &= (b+b') \cup (b+b') + (b+b') \cup_1 (\delta b + \delta b') \nonumber\\
    &= \mathfrak{p}(b) + \mathfrak{p}(b') + b \cup b' + b' \cup b + b \cup_1 \delta b' + b' \cup_1 \delta b\ .
\end{align}
When $M_4$ is torsion, we emphasize that the cup-product is not symmetric in general and 
$\delta b$ does not necessarily vanish but $\delta b = N u$ for certain $u \in C^3(M_4)$. 
We can further massage \eqref{cup-1-product} into 
\begin{equation}\label{cup-1-product02}
    \mathfrak{p}(b+b') = \mathfrak{p}(b) + \mathfrak{p}(b') + 2b \cup b' + b \cup_1 \delta b' + \delta b' \cup_1 b +2 b' \cup_1 \delta b\ , 
\end{equation}
since, by definition \eqref{defcup1}, one has 
\begin{equation}
    b' \cup b = b \cup b' + \delta b' \cup_1 b + b' \cup_1 \delta b  \ ,
\end{equation}
up to total derivative. The non-commutativity of the cup-$1$ product is characterized by the cup-$2$ product,
\begin{equation}
        f \cup_{i-1} g + (-1)^{pq +i} g \cup_{i-1} f = (-1)^{p+q+i}\left(\delta (f \cup_i g) - \delta f \cup_i g - (-1)^{p} f \cup_i \delta g \right) \ . 
\end{equation}
Thus, \eqref{cup-1-product02} can be rewritten as  
\begin{equation}
        \mathfrak{p}(b+b') = \mathfrak{p}(b) + \mathfrak{p}(b') + 2b \cup b' +2 b' \cup_1 \delta b  + \delta b \cup_2 \delta b'\ .
\end{equation}
up to total derivatives. 
Notice that the fourth term on the RHS satisfies $2 b' \cup_1 \delta b = 0 \textrm{ mod } 2N$. Moreover, 
the last term $\delta b \cup_2 \delta b'$ also vanishes modulo $2N$ for even $N$. 
Therefore, one can verify that 
\begin{equation}
    \mathfrak{p}(b+b') = \mathfrak{p}(b) + \mathfrak{p}(b') + 2b \cup b' \mod 2N\ .
\end{equation}
As a corollary, one can show that 
\begin{align}
    \mathfrak{p}(b + N \alpha) = \mathfrak{p}(b) \mod 2N
\end{align}
for any two-cycle $\alpha$, when $N$ is even. 
On the other hand, the additional term involving the cup-$1$ product 
fails to satisfy the properties of $\mathfrak{p}(b)$ for odd $N$. 

As a warm-up exercise, let us first consider the $\mathbb{Z}_2$ case where the integral of Pontryagin square gives
\begin{equation}\label{Lens-Pontryagin-square}
    \mathfrak{P}(b_1,b_2) \equiv  \int_{M_4} \mathfrak{p}\big(\frac{N}{2\pi} B\big) = 2b_1 b_2 + r b_1^2 \mod 4\  .
\end{equation}
Here $r$ needs to be even otherwise the topological boundary state is trivial. When $r=4k$, the second term does not contribute and we simply have $\mathfrak{P}(b) = 2b_1 b_2 \mod 4$. 
On the other hand, if $r=4k+2$ we should have $\mathfrak{P}(b) = 2 b_1(b_2+b_1)\mod 4$. 
Hence, we can conclude
\begin{align}
        r &= 0 \ (\textrm{mod} \ 4): \quad \frac{1}{2} \mathfrak{P}(b) = \left\{\begin{array}{l}
            1\quad b_1=b_2=1\\
            0\quad \textrm{else}
        \end{array} \right.\nonumber\\
        r &= 2 \ (\textrm{mod} \ 4): \quad \frac{1}{2} \mathfrak{P}(b) = \left\{\begin{array}{l}
            1\quad b_1=1,b_2=0\\
            0\quad \textrm{else}
        \end{array} \right.  \ .      
\end{align}

We then move on to the case where $N$ is even and greater than $2$. 
Recall that $b_1$ and $b_2$ are restricted by
\begin{equation}
    b_1 = \frac{N}{\gcd(r,N)}k_1\ ,\quad b_2 = k_2\ ,\quad (k_1,k_2=0,\cdots,\gcd(r,N)-1)\ ,
\end{equation}
which indicates
\begin{equation}
    \delta b = b_1 \delta \gamma_2 = N \times \frac{rk_1}{\gcd(r,N)} u\ .
\end{equation}
Therefore, the Pontryagin square becomes
\begin{equation}\label{Pontryaginlens}
        \mathfrak{P}(b) = b_1\left( 2b_2 + r b_1\right) \mod 2N \ .
\end{equation}

\subsection{Comments on torsion}
In this section, we make some general comments for torsional manifolds, motivated by the previous discussions on $L(r,s) \times S^1$. For a generic 4-manifold the second homology $H_2(M_4,\mathbb{Z})$ is split into free part $fH_2(M_4,\mathbb{Z})$ and torsion part $\tau H_2(M_4,\mathbb{Z})$,
    \begin{equation}
        H_2(M_4,\mathbb{Z}) = fH_2(M_4,\mathbb{Z}) \oplus \tau H_2(M_4,\mathbb{Z})\ .
    \end{equation}
And there exist two duality pairings involving $H_2(M_4,\mathbb{Z})$
    \begin{itemize}
        \item Intersection pairing $Q$: $f H_2(M_4,\mathbb{Z}) \times f H_2(M_4,\mathbb{Z}) \rightarrow \mathbb{Z}$
        \item Torsion linking form $L$: $\tau H_2(M_4,\mathbb{Z}) \times \tau H_1(M_4,\mathbb{Z}) \rightarrow \mathbb{Q}/\mathbb{Z}$
    \end{itemize}
where $\mathbb{Q}$ is the field of rational number and $\mathbb{Q}/\mathbb{Z}$ stands for the proper fraction. The torsion linking form is calculated as follows: given an $x \in \tau H_2(M_4,\mathbb{Z})$ and $y \in \tau H_1(M_4,\mathbb{Z})$, one can find an integer $n$ such that $n y$ is the boundary of a surface $z$. Then the torsion linking number $L(x,y)$ is the fraction whose numerator is the transverse intersection number of $z$ with $x$, and whose denominator is $n$. The intersection pairing $Q$ will be zero if one of the 2-cycles is torsion due to the linearity of $Q$: if $\Gamma$ is any 2-cycle and $\Gamma'$ is a torsion 2-cycle with order $r$, then the intersection pairing satisfies $Q(\Gamma,r\Gamma')=0$ since $r\Gamma'$ is shrinkable. By the linearity of $Q$ it implies $r Q(\Gamma,\Gamma') = 0$. Since the intersection number is integer-valued we should have $Q(\Gamma,\Gamma')=0$ for any torsion 2-cycle $\Gamma'$. In the following, we will assume the torsion part does not talk to the free part and focus on the torsion 2-cycles only.

We also need to consider the open 2-surfaces whose boundaries are torsion 1-cycles and  introduce the set $\tau H_2(M_4,\mathbb{Z})^D$,
    \begin{equation}
        \tau H_2(M_4,\mathbb{Z})^D = \left\{\Lambda^D \in C_2(M_4,\mathbb{Z}) | \partial \Lambda^D \in  \tau H_1(M_4,\mathbb{Z})\right\}
    \end{equation}
as a dual of $\tau H_2(M_4,\mathbb{Z})$. They are detected by the homology group $H_2(M_4,\mathbb{Z}_k)$ whose coefficients are certain cyclic groups. The torsion linking pairing $L(*,*)$ implies an integer-valued bilinear intersection form $Q_{\tau}(*,*)$,
    \begin{equation}
        Q_{\tau}:\quad \tau H_2(M_4,\mathbb{Z}) \otimes \tau H_2(M_4,\mathbb{Z})^D \rightarrow \mathbb{Z}\ .
    \end{equation}
However, this intersection form is ill-defined as an integer under the shift of torsion cycles in $\tau H_2(M_4,\mathbb{Z})$. Later we will see there is no issue if we focus on the gauge invariant operators.

If $B$ and $\widetilde{B}$ are $U(1)$ two-form gauge fields, the Wilson surfaces supported on $\tau H_2(M_4,\mathbb{Z})^D$ are not invariant under the gauge transformation
    \begin{equation}
        B \rightarrow B + d \lambda\ ,\quad \widetilde{B} \rightarrow \widetilde{B} + d \widetilde{\lambda}\ ,
    \end{equation}
since the surfaces are not closed. Here $\lambda,\widetilde{\lambda}$ are also $U(1)$-valued one-form which means their integral on closed 1-cycles are $U(1)$-valued. However, if we restrict ourselves on $\mathbb{Z}_N$-valued gauge fields we can still construct a set of gauge invariant Wilson surfaces even on non-closed surfaces. 

To be explicit, we choose a set of generators of torsion 2-cycles $\{\Gamma_i\}$ with $i=1,\cdots,2h_{\tau}$ where $\Gamma_{1},\cdots\Gamma_{h_{\tau}}$ form a basis of the torsion part $\tau H_2(M_4,\mathbb{Z})$ while $\Gamma_{h_{\tau}+1}$, $\cdots,\Gamma_{2h_{\tau}}$ generate $\tau H_2(M_4,\mathbb{Z})^D$. Here the dimension of $\tau H_2(M_4,\mathbb{Z})$ is denoted as $h_{\tau}$. Introduce the Wilson surface operators
    \begin{equation}
        U[\Gamma_i] \equiv \exp\left[i \oint_{\Gamma_i} B \right], \quad \widetilde{U}[\Gamma_i] \equiv \exp\left[i \oint_{\Gamma_i} \tilde{B} \right]\ .
    \end{equation}
The prototype algebras are given by
    \begin{equation}\label{Torsion-algebra-general}
        U[\Gamma] \widetilde{U}[\Gamma'] = \omega^{-Q_{\tau}(\Gamma,\Gamma')} \widetilde{U}[\Gamma'] U[\Gamma]\ ,
    \end{equation}
where $Q_{\tau}$ is the intersection pairing defined before. Given any $\Gamma \in \tau H_2(M_4,\mathbb{Z})$ with order $r$ and $\Gamma' \in \tau H_2(M_4,\mathbb{Z})^D$ with $\partial \Gamma' = r' l$ where $l\in \tau H_1(M_4,\mathbb{Z})$ has order $r'$, one needs to consider the following facts.
\begin{itemize}
    \item Since $\Gamma'$ has a boundary, $U[\Gamma']$ is not gauge invariant under the gauge transformation $B \rightarrow B + d \lambda, \widetilde{B} \rightarrow \widetilde{B} + d \widetilde{\lambda} $ with $\lambda,\widetilde{\lambda}$ one-forms with $\mathbb{Z}_N$ period. It gives the gauge transformation
        \begin{equation}
            U[\Gamma'] \sim e^{2\pi i\frac{r'}{N}} U[\Gamma']\ ,\quad \widetilde{U}[\Gamma'] \sim e^{2\pi i\frac{r'}{N}} \widetilde{U}[\Gamma']\ .
        \end{equation}
    \item Since $r\Gamma$ is trivial, we need to impose $U[\Gamma]^r = \widetilde{U}[\Gamma]^r = 1$. They introduce further restrictions from the algebra
        \begin{equation}
            U[\Gamma']\sim e^{2\pi i \frac{rQ_\tau(\Gamma,\Gamma')}{N}} U[\Gamma']\ ,\quad \widetilde{U}[\Gamma']\sim e^{2\pi i \frac{rQ_\tau(\Gamma,\Gamma')}{N}} \widetilde{U}[\Gamma']\ .
        \end{equation}
\end{itemize}
Actually, the second is automatically satisfied providing the first one, which is argued as follows. Given $x\in \tau H_2(M_4,\mathbb{Z})$ with order $r$ and $y\in \tau H_1(M_4,\mathbb{Z})$ with order $r'$, the linking form is symmetric and satisfy $L_{\tau}(x,y) = L_{\tau}(y,x)$. By definition of the linking form, one should have,
    \begin{equation}
        r L_{\tau}(x,y) \in \mathbb{Z}\ ,\quad r' L_{\tau}(x,y) \in \mathbb{Z}\ ,
    \end{equation}
and it indicates
    \begin{equation}
        L_{\tau}(x,y) = \frac{k}{\gcd(r,r')}\ ,
    \end{equation}
with some integer $|k| < \gcd(r,r')$. According to this, we have,
    \begin{equation}
        r Q_{\tau}(\Gamma,\Gamma') = \frac{r r' k}{\gcd(r,r')} = \frac{r k}{\gcd(r,r')} \times r'\ ,
    \end{equation}
which is some integer multiply $r'$.

Since we have a non-trivial gauge transformation of $U[\Gamma']$ and $\widetilde{U}[\Gamma']$ with $\Gamma' \in \tau H_2(M_4,\mathbb{Z})^D$, we need to consider only the gauge invariant operators,
    \begin{equation}
        U[\Gamma']^{\frac{N}{\gcd(N,r')}}\ , \quad  \widetilde{U}[\Gamma']^{\frac{N}{\gcd(N,r')}}\ ,
    \end{equation}
and there are $\gcd(N,r')$ of them and the total numbers of independent operators are $\prod \gcd(N,r')$ for each type where the product runs over all generators in $\tau H_2(M_4,\mathbb{Z})^D $ or $\tau H_1(M_4,\mathbb{Z})$. On the other hand, for $\Gamma \in \tau H_2(M_4,\mathbb{Z})$ the restriction $U[\Gamma]^r=\widetilde{U}[\Gamma]^r=1$ also implies $U[\Gamma]^{\gcd(r,N)}=\widetilde{U}[\Gamma]^{\gcd(r,N)}=1$ and there should be $\gcd(N,r)$ of them. Then the total numbers of independent operators are $\prod \gcd(N,r)$ for each type where the product runs over all generators in $\tau H_2(M_4,\mathbb{Z})$. Since we have $\tau H_2(M_4,\mathbb{Z}) \cong \tau H_1(M_4,\mathbb{Z})$ by Poincaré duality, it implies $\prod \gcd(N,r') = \prod \gcd(N,r)$ as expected.

The last step is to find a minimal representation of the reduced algebras subject to the above restrictions. This automatically fixes the ambiguity of the intersection pairing $Q_{\tau}$. Recall that if we shift $\Gamma \rightarrow (r+1)\Gamma$ in \eqref{Torsion-algebra-general}
    \begin{equation}
        U[\Gamma] \widetilde{U}[\Gamma'] = \omega^{-Q_{\tau}(\Gamma,\Gamma') - r Q_{\tau}(\Gamma,\Gamma')} \widetilde{U}[\Gamma'] U[\Gamma]\ ,
    \end{equation}
which renders the algebra ambiguous. However, if we only focus on the gauge invariant operators we may find
    \begin{equation}
         U[\Gamma] \widetilde{U}[\Gamma']^{\frac{N}{\gcd(N,r')}} = \omega^{-\frac{N}{\gcd(N,r')}Q_{\tau}(\Gamma,\Gamma') - \frac{N r}{\gcd(N,r')} Q_{\tau}(\Gamma,\Gamma')} \widetilde{U}[\Gamma']^{\frac{N}{\gcd(N,r')}} U[\Gamma]\ .
    \end{equation}
As discussed before we have $r'|r Q_{\tau}(\Gamma,\Gamma')$ so that the additional phase induced by the ambiguity of the torsion cycle vanishes automatically. We can fix the value of $Q_{\tau}$ arbitrarily since we will only consider gauge invariant operators eventually.

\section{Web of Four-Dimensional Dualities}\label{sec:4}
In the last section we construct the topological boundary states on some simple concrete 4d manifolds $M_4$ including $T^4,\mathbb{CP}^2$ and $L(r,s) \times S^1$ where $L(r,s)$ is the lens space. They are the typical examples of spin, non-spin, and torsional 4-manifold. In this section, we will mainly work with $T^4$ and consider the physical consequence of changing topological boundary states. In particular, we will focus on the Yang-Mills theory and see how topological boundary states determine the global structures of the gauge groups and the spectrum of dyonic line operators.

\subsection{Web of dualities}\label{sec:4.1}
We will first review the web of 2d dualities \cite{Karch:2019lnn,Hsieh:2020uwb} including the Kramers-Wannier duality and the Jordan-Wigner transformation, and the interpretation using SymTFT. It will provide a guideline that can be generalized to four dimensions straightforwardly. 

\paragraph{2d dualities}

Consider a 2-dimensional bosonic theory with a non-anomalous $\mathbb{Z}_2$ global symmetry such as Ising CFT, denoted by ${\cal B}$.
We put the theory on a circle, and impose the periodic 
boundary condition. We then obtain the  
`untwisted' Hilbert space, ${\cal H}({\cal B})_U$. 
The Hilbert space can be further decomposed into two superselection 
sectors, of which the one is for the even states  
under $\mathbb{Z}_2$ while the other for the odd states. 
We denote the former by ${\cal H}({\cal B})_U^e$ and the latter 
by ${\cal H}({\cal B})_U^o$,
\begin{align}
    {\cal H}({\cal B})_U   = 
    {\cal H}({\cal B})_U^e\oplus {\cal H}({\cal B})_U^o\ . 
\end{align}

One can utilize the $\mathbb{Z}_2$ symmetry 
to define a new theory where the boundary condition 
along the circle is twisted by the $\mathbb{Z}_2$ charge. 
The corresponding Hilbert space is often referred to as the  
`twisted' Hilbert space of ${\cal B}$. The twisted 
Hilbert space ${\cal H}({\cal B})_T$ can be decomposed into 
two parts, 
\begin{align}
    {\cal H}({\cal B})_T   = 
    {\cal H}({\cal B})_T^e\oplus {\cal H}({\cal B})_T^o \ ,
\end{align}
where the former contains the even states under $\mathbb{Z}_2$
and the latter is for the odd states. 

Strictly speaking the spectrum of the bosonic theory ${\cal B}$
only refers to the states in the untwisted Hilbert space. 
As we will argue shortly, it is however useful to extend the notion of the Hilbert space of ${\cal B}$ to include  
both the untwisted and twisted Hilbert spaces as follows,  
\begin{align}\label{deftotHilbertspace}
    {\cal H}({\cal B})_\text{tot} 
    = {\cal H}({\cal B})_U \oplus {\cal H}({\cal B})_T\ . 
\end{align}

We can make use of the Kramers-Wannier transformation  
to obtain a new bosonic theory $\widetilde{\cal B}= {\cal B}/\mathbb{Z}_2$. 
The spectrum of $\widetilde{\cal B}$ can be identified 
with the states invariant under the $\mathbb{Z}_2$ symmetry
in both the untwisted and twisted Hilbert space of ${\cal B}$, 
\begin{align}
    {\cal H}(\widetilde{\cal B})_U =  {\cal H}({\cal B})_U^e \oplus {\cal H}({\cal B})_T^e\ . 
\end{align}
A notable feature of the bosonic theory $\widetilde{\cal B}$ is the emergence of the quantum $\widetilde{\mathbb{Z}}_2$ 
symmetry. One can demonstrate that ${\cal H}({\cal B})_U^e$ has the even states under the quantum $\widetilde{\mathbb{Z}}_2$ while ${\cal H}({\cal B})_T^e$ has the odd states. 
The quantum $\widetilde{\mathbb{Z}}_2$ also allows us to define the twisted Hilbert space of 
$\widetilde{\cal B}$. It turns out that the twisted Hilbert space of 
$\widetilde{\cal B}$ is composed of the odd states of ${\cal B}$ prior to the $\mathbb{Z}_2$ gauging: 
\begin{align}
    {\cal H}(\widetilde{\cal B})_T =  {\cal H}({\cal B})_U^o \oplus {\cal H}({\cal B})_T^o\ , 
\end{align}
where  the former is invariant under the quantum $\widetilde{\mathbb{Z}}_2$ while the latter is charged. 
We notice that the total Hilbert space of the $\mathbb{Z}_2$ 
orbifold $\widetilde{\cal B}$ then becomes identical to that of ${\cal B}$,
\begin{align}
    {\cal H}(\widetilde{\cal B})_\text{tot} = {\cal H}({\cal B})_\text{tot} \ .
\end{align}

On the other hand, the Jordan-Wigner transformation maps a bosonic 
theory ${\cal B}$ to a fermionic theory ${\cal F}$. Every fermionic 
theory possesses a natural $\mathbb{Z}_2$ symmetry, the fermion parity 
generated by $(-1)^F$ where $F$ is the fermion number operator. 
Again, we can consider the boundary 
condition along the circle without or with the $(-1)^F$ twist. 
Accordingly, one can define the untwisted Hilbert space ${\cal H}({\cal F})_{NS}$ and the twisted Hilbert space ${\cal H}({\cal F})_{R}$, also referred to as the Neveu-Schwarz sector and the Ramond sector. 
Depending on the charge under the fermion parity, each Hilbert space can be decomposed 
into the two parts, one for bosonic states and the other for fermionic states. 
It was argued by Gliozzi, Scherk and Olive (GSO) that the bosonic states of ${\cal F}$ 
are precisely matching with the untwisted states of the corresponding bosonic theory ${\cal B}$, 
\begin{align}
    {\cal H}({\cal F})_{NS}^{b} = {\cal H}({\cal B})_U^e \ , \qquad 
    {\cal H}({\cal F})_{R}^{b} = {\cal H}({\cal B})_U^o\ .
\end{align}
On the other hand, the fermionic states of ${\cal F}$ correspond to the twisted states 
of ${\cal B}$,
\begin{align}
    {\cal H}({\cal F})_{NS}^{f} = {\cal H}({\cal B})_T^o \ , \qquad 
    {\cal H}({\cal F})_{R}^{f} = {\cal H}({\cal B})_T^e\ .
\end{align}
One can see that the total Hilbert space of ${\cal F}$ is the same as that of ${\cal B}$,
\begin{align}
    {\cal H}({\cal F})_\text{tot} = {\cal H}({\cal F})_{NS} 
    \oplus  {\cal H}({\cal F})_{R} =   {\cal H}({\cal B})_\text{tot}\ . 
\end{align}
\begin{figure}[t!]
    \centering
    \includegraphics[scale=0.7]{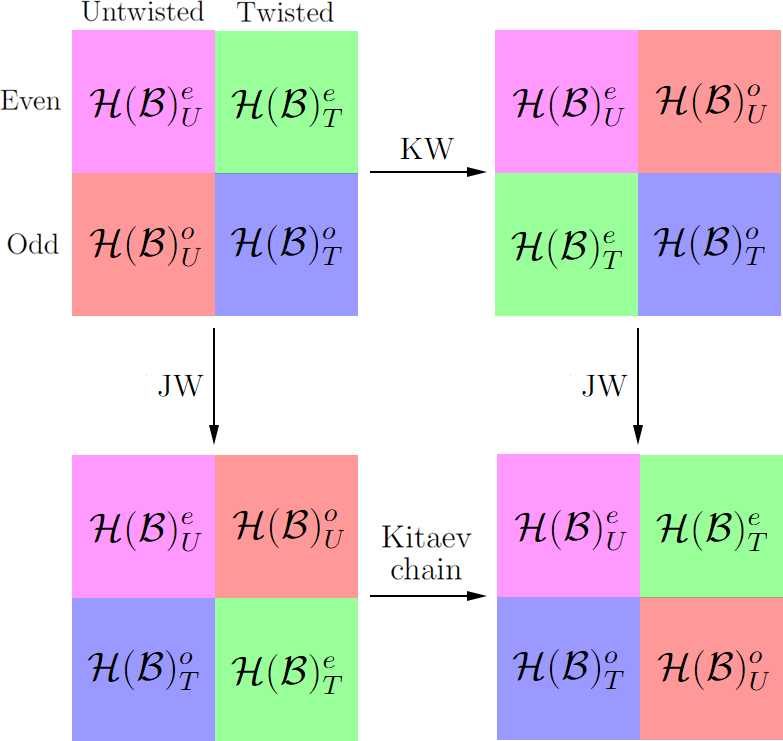}
    \caption{2d duality web and the mapping of Hilbert spaces.}
    \label{fig:2d-duality}
\end{figure}
 
In summary, the Kramers-Wannier duality and the Jordan-Wigner transformation 
generate the so-called `web of dualities' of the two-dimensional theories 
that share the equivalent total Hilbert space modulo some rearrangements, as depicted in 
Figure \ref{fig:2d-duality}.  

The two-dimensional theories within the duality web all exhibit the $\mathbb{Z}_2$ global symmetry.
Let us weakly gauge the $\mathbb{Z}_2$ symmetry. 
One can then define the twisted thermodynamic partition functions 
of a given theory in the presence of the background $\mathbb{Z}_2$ gauge fields,
\begin{align}\label{2dtwistedpartitionfunction}
    Z\big[ (t,s) \big] = \text{Tr}_{{\cal H}_s} \Big[ D_t e^{-\beta H} \Big]\ . 
\end{align}
Here the background $\mathbb{Z}_2$ gauge fields 
along the temporal and spatial circle 
are labeled by the $\mathbb{Z}_2$-valued holonomies 
$t$ and $s$.
Turning on a background gauge field introduces 
a corresponding symmetry defect. For instance, 
the background gauge field with $t=1$ 
introduces the symmetry defect $D_t$ inside the trace. 
On the other hand, the background gauge field with $s=1$ 
essentially defines the twisted Hilbert space.  
This is because we can perform a singular gauge transformation 
so that the background gauge field is gauged away and 
the charged fields are subject to the boundary condition twisted by 
their $\mathbb{Z}_2$ charges. 

Since the total Hilbert spaces of ${\cal B}$, $\widetilde{\cal B}$, ${\cal F}$, 
and $\widetilde{\cal F}$ in the duality web are all the same, 
one can expect that their twisted partition functions \eqref{2dtwistedpartitionfunction} are all interrelated. 
Based on the mappings in Figure \ref{fig:2d-duality}, one can indeed show that  
\begin{align}\label{relation01}
    Z_{\widetilde{\cal B}}\big[(\tilde t, \tilde s)\big] 
    = \frac{1}{2}\sum_{s,t=0,1} (-1)^{\tilde t s + \tilde s t} Z_{\cal B}\big[ (t,s) \big] \ ,
\end{align}
and 
\begin{align}\label{relation02}
    Z_{\cal F}\big[ (a, b) \big] = 
    \frac{1}{2}\sum_{s,t=0,1} (-1)^{ (a+ t) ( b+ s) } Z_{\cal B}\big[ (t,s) \big]\ .
\end{align}
Here $(\tilde t, \tilde s)$ and $(a,b)$ denote the 
background gauge fields for the quantum $\widetilde{\mathbb{Z}}_2$ 
and the fermion parity. The summation over $(t,s)$ reflects 
the integration over all possible $\mathbb{Z}_2$ gauge field configurations 
for the gauging process. 

The SymTFT then provides a novel perspective on 
how to understand the above relations between various partition functions.  
The SymTFT relevant to our discussion is the three-dimensional 
BF model with level two. From this viewpoint, the relative phases in \eqref{relation01} and \eqref{relation02}
can be understood as transformation matrices between two different bases 
of the Hilbert space of the BF model on the two-torus $T^2$. 
The Hilbert space of the BF model on $T^2$ is four-dimensional and has 
three canonical bases. Such bases are also referred to as the topological boundary states in modern literature, e.g., \cite{Duan:2023ykn}:
\begin{itemize}
    \item One basis consists of the Dirichlet boundary states denoted by $\{|t,s\rangle\}$.
    
    \item The other basis consists of the Neumann boundary states denoted by $\{|(\tilde{t},\tilde{s})\rangle\}$. 
    This basis is related to $\{ |(t,s)\rangle\} $ as follows, 
    \begin{equation}
        \big |(\tilde{t},\tilde{s}) \big\rangle = \frac{1}{2} \sum_{t,s=0,1} (-1)^{\tilde{t} s + \tilde{s} t} |(t,s)\rangle\ , 
    \end{equation}
    which agrees perfectly with the prefactors of \eqref{relation01}
    
    \item The last canonical basis is composed of the fermionic boundary states $\{|(a,b)\rangle\}$ where $(a,b)$
    specifies the spin structure of the torus. It is related to $\{ |(t,s)\rangle \} $ as follows, 
    \begin{equation}
       |(a,b)\rangle = \frac{1}{2} \sum_{t,s=0,1} (-1)^{(a+t)(b+s)} |(t,s)\rangle\ ,
    \end{equation}
    which matches with the prefactors of \eqref{relation02}. 
    
\end{itemize}
It implies that every theory in the duality web 
can be associated with one of the above bases, 
and its partition function 
can be interpreted as the inner product between 
two states in the Hilbert space of the SymTFT. 
One state corresponds to the chosen topological boundary state, and 
the other is the so-called dynamical boundary state 
specified by the physical observable of our interest \eqref{2dtwistedpartitionfunction}. 
For instance, 
\begin{align}
    Z_{\widetilde{\cal B}} \big[ (\tilde t, \tilde s) \big] 
    = \big \langle  (\tilde t, \tilde s)  \big | 
    \chi \big \rangle\ , 
\end{align}
where the dynamical boundary state $|\chi \rangle $ corresponding to the 
torus-partition function can be expressed as 
\begin{align}
    \big| \chi \big \rangle \equiv \sum_{t,s} Z_{\cal B}\big[(t,s) \big] \big | (t,s) \big \rangle  \ .
\end{align}

This example illustrates that the SymTFT is a powerful tool for exploring two-dimensional dualities 
and examining how the physical observables therein, such as the torus partition function, are related.
We extend this approach to investigate the duality web among the four-dimensional gauge theories with non-anomalous one-form global symmetry.

\paragraph{4d dualities}

Now let us turn to a four-dimensional gauge theory placed on a compact 
four-manifold $M_4$. The gauge theory preserves a discrete one-form symmetry $H$. 
The charged objects under the one-form symmetry are genuine line operators. 
Since $M_4$ is compact, the spectrum of the theory includes such line operators.

To explore theories sharing the same total Hilbert space in the sense of \eqref{deftotHilbertspace}, 
we employ the SymTFT approach.  
The SymTFT of our interest is the five-dimensional BF model \eqref{BF_action} 
studied in the previous section. We quantize the BF model on $M_4$ 
to obtain the Hilbert space, ${\cal H}(M_4)$. One 
canonical basis for ${\cal H}(M_4)$ is the position basis $\{ |b\rangle \}$ 
\eqref{5DBF-Basic-Positioin-States} where the surface operators 
$U[\Gamma]$ \eqref{5DBF-Basic-Wilson-Surface-field-variables} are simultaneously diagonalized. 
We also learned that other topological boundary states can be obtained 
from $\{ |b\rangle\} $ via the $SL(2,\mathbb{Z})$ action or the partial gauging. 

What can we expect to learn from the SymTFT? 
As in two dimensions, we expect that 
different choices of the topological boundary 
states correspond to distinct gauge theories in the 
duality web. This correspondence thus implies  
that, starting with a given gauge theory, the others in the web 
can be constructed by gauging the discrete 
one-form global symmetry. In other words, the 
only difference between those gauge theories is not local but global properties of the gauge group. 
The local physics of all gauge theories 
in the duality web should be identical. 
Their subtle difference can be probed by analyzing 
the partition function on the compact space or 
the set of allowed line operators. In fact, there is 
an one-to-one correspondence between 
the line operators allowed by the gauge theory 
and maximally commuting observables chosen by 
the corresponding topological boundary state. 

To illustrate this, let us consider a gauge theory with 
$\widetilde{G}= SU(8)$. The gauge theory has the $H=\mathbb{Z}_8$ discrete 
one-form global symmetry. There are numerous ways to 
gauge the one-form symmetry consistently, 
which leads to a duality web. We claim 
that this duality web is identical to that in 
Figure \ref{fig:enter-label4} describing the topological boundary states of the 
corresponding SymTFT with their $SL(2,\mathbb{Z})$ actions. 
In other words, each topological boundary state 
can specify the gauge theory in correspondence, and 
the $SL(2,\mathbb{Z})$ action of the BF model 
is indeed closely related to the $SL(2,\mathbb{Z})$ action of the gauge theory
in four dimensions. 

To give an example, let us consider gauging the full $\mathbb{Z}_8$ center symmetry and obtain $PSU(8)$ gauge theory, which is also referred to as 
$\big( SU(8)/\mathbb{Z}_8 \big)_0$ gauge theory.
Now the theta angle is enlarged to range from 0 to $16\pi$. 
Different choices of gauging $\mathbb{Z}_8$ 
result in distinct gauge theories, labeled by $\big( SU(8)/\mathbb{Z}_8 \big)_k$, in the duality web. 
They are permuted by shifting the theta angle by $2\pi$,
\begin{align}
    \big( SU(8)/\mathbb{Z}_8 \big)^{\theta+2\pi}_k =
    \big( SU(8)/\mathbb{Z}_8 \big)^\theta_{k+1}\ . 
\end{align}
One can argue that $\big( SU(8)/\mathbb{Z}_8 \big)_k$ corresponds to the  topological boundary state $\{ V_{ST^k}|b\rangle\}$. 
It was shown in \cite{Aharony:2013hda} that the $\big( SU(8)/\mathbb{Z}_8 \big)_k$ gauge theory 
can be characterized by the consistent choice of 
line operators carrying electric and magnetic $\mathbb{Z}_8$ charges.
One can also see that those charges exactly match
with the $\mathbb{Z}_8$ charges $(e,m)$ of 
maximally commuting observables $S_{(e,m)}[\Gamma]$  
for the corresponding topological boundary states $\{V_{ST^k}|b\rangle\}$.  
For instance, the line operators allowed for $k=1$
carry the charges identical to those in Figure \ref{fig:examples} (a). Similarly, if we partially gauge $\mathbb{Z}_2$ or $\mathbb{Z}_4$ subgroup, we find the other gauge theories associated with the remaining three topological boundary states in Figure \ref{fig:enter-label4}.

The Euclidean partition function on $M_4$ is a good 
physical observable to examine the difference between the
theories in the duality web. However, their partition functions on $M_4$
are also intimately related, as 
they all have the same total Hilbert space. From the perspective of 
the SymTFT, computing such partition functions and deriving 
the relations between them boils down to a two-step process. 
First, we need to construct the dynamical boundary state 
corresponding to a physical observable of our interest. 
The next step is to 
select the topological boundary state 
associated with the gauge theory under investigation. 

For instance, if a physical observable 
of our interest is the supersymmetric partition function on $M_4$
such as the Witten index for $M_4=T^4$, 
\begin{align}
     \big | \chi \big \rangle =  \sum_{b} Z\big[ b \big] \big | b \big \rangle\ .
\end{align}
Here $Z\big[b\big]$ denotes the supersymmetric partition function 
on $M_4$ for the gauge theory associated with the topological 
boundary state $|b\rangle$ with the background two-form gauge fields $b$ turned on. 
Projecting the above state onto various topological boundary states then generates 
a collection of supersymmetric partition functions of 
the other theories dual to the original one. 

The question is how to compute $Z\big[b\big]$.
When $M_4 = S^1 \times M_3$, the partition function on $M_4$
admits the Hamiltonian interpretation
\begin{align}
    Z\big[ b=(t,s) \big] = \text{Tr}_{{\cal H}_s} \big[ D_t e^{-\beta H} \big]\ , 
\end{align}
where $t,s$ are the $\mathbb{Z}_N$-valued holonomies
labeling the background two-form gauge fields
with and without the temporal component. 
The background gauge field with $t\neq0$  
introduces the co-dimension-two defect $\mathcal{D}_t$
of the $\mathbb{Z}_N$ one-form symmetry acting on the quantum states of the gauge theory. 
By definition, the symmetry defects $\mathcal{D}_t$ form an abelian group and 
satisfy the fusion rule below
\begin{equation}
    \mathcal{D}_t \times \mathcal{D}_{t'} = \mathcal{D}_{t+t'}\ ,\quad \mathcal{D}_t^N = 1\ .
\end{equation}
On the other hand, the background gauge field with 
$s\neq 0$ essentially defines the twisted Hilbert space 
of the given gauge theory. We will discuss 
what the twisted Hilbert space means in the four-dimensional gauge theories
in more detail below.

\subsection{Twisted Hilbert space}\label{sec:twistedbc}

In \cite{tHooft:1981nnx} 't Hooft introduced the twisted boundary condition for the gauge field $A_{\mu}(x)$ in $SU(N)$ gauge theory living on $T^4$, which is now known as discrete 't Hooft flux. There are several equivalent definitions for the discrete 't Hooft flux:
\begin{itemize}
    \item \textbf{Twisted bundle} : In \cite{tHooft:1981nnx}, it is defined by requiring the gauge fields $A_{\mu}(x)$ to be periodic along the $T^4$ up to some gauge transformation. Namely, we impose twisted boundary conditions.
    \item \textbf{Non-commutative holonomies}: When the gauge bundle is flat, the discrete 't Hooft fluxes can be introduced by imposing holonomies $H_{\mu}$ along the 1-cycles such that they commute up to some $\mathbb{Z}_N$ phase\cite{Borel:1999bx, Witten:2000nv}.
    \item \textbf{one-form symmetry backgrounds}: The discrete 't Hooft fluxes can also be described by turning on two-form backgrounds $B_{\mu \nu}$ of one-form symmetry whose charged objects are Wilson lines\cite{Gaiotto:2014kfa}.
\end{itemize}
We will demonstrate their equivalences in the following order: first, we review the $SU(N)$ twisted boundary condition and show that it can also be described using non-commutative holonomies with a flat gauge bundle. Next, we argue that the twisted boundary condition can also be replaced by performing a singular gauge transformation and activating the background for one-form symmetry. In practice, the equivalence between non-commutative holonomies and one-form symmetry backgrounds proves to be the most useful and is used extensively throughout the main text.

Consider the $SU(N)$ theory living on $T^4$ with lengths $a_{\mu}$ for $\mu=1,2,3,4$. The gauge field $A_{\mu}(x)$ is hermitian and traceless. On the four walls located at $x_{\mu} = a_{\mu}$, consider a set of $SU(N)$ matrices $\Omega_{\mu}(x)$ with $x=\{x_{\nu \neq \mu}\}$ defined on the wall. 
We can build a twisted gauge bundle by imposing a twisted boundary condition
\begin{equation}\label{Twisted-boundary-condition-1}
    A_{\nu} (x_{\mu}=a_{\mu}) = \Omega_{\mu}\left(A_{\nu}(x_{\mu}=0) - i \frac{\partial}{\partial x_{\nu}} \right) \Omega^{-1}_{\mu}\equiv \Omega_{\mu} A_{\nu}(x_{\mu}=0)\ ,
\end{equation}
where the gauge field $A_{\mu}(x)$ is periodic only up to some gauge transformation. Here we write $\Omega_{\mu}A_{\nu}$ for short and require $\partial_{\mu} \Omega_{\mu} = 0$ (no summation) for each $\mu$. For any 2-torus $T^2 \subseteq T^4$ parameterized by 
$x_{\mu},x_{\nu}$ with $\mu \neq \nu$, the gauge field $A(x_{\mu} = a_{\mu},x_{\nu} = a_{\nu})$ can be obtained from $A(x_{\mu}=0,x_{\nu}=0)$ in two different ways. Starting at $x_{\mu}=x_{\nu}=0$, we can first move along $x_{\nu}$ direction and then $x_{\mu}$ direction which corresponds to,
\begin{align}
    A(x_{\mu}=a_{\mu},x_{\nu}=a_{\nu}) &= \Omega_{\mu}(x_{\nu}=a_{\nu}) A(x_{\mu}=0,x_{\nu}=a_{\nu})\nonumber\\ &= \Omega_{\mu}(x_{\nu}=a_{\nu}) \Omega_{\nu}(x_{\mu}=0) A(x_{\mu}=0,x_{\nu}=0)\ .
\end{align}
Alternatively, we can first move along $x_{\mu}$ direction and then $x_{\nu}$ direction,
\begin{align}
    A(x_{\mu}=a_{\mu},x_{\nu}=a_{\nu}) &= \Omega_{\nu}(x_{\mu}=a_{\mu}) A(x_{\mu}=a_{\mu},x_{\nu}=0)\nonumber\\ &= \Omega_{\nu}(x_{\mu}=a_{\mu}) \Omega_{\mu}(x_{\nu}=0) A(x_{\mu}=0,x_{\nu}=0)\ .
\end{align}
They must produce the same result and since $\Omega$ acts on the gauge field in the adjoint representation, we should have
\begin{equation} \label{Twisted-boundary-condition-2}
    \Omega_{\mu}(x_{\nu}=a_{\nu}) \Omega_{\nu} (x_{\mu}=0) = \Omega_{\nu} (x_{\mu}=a_{\mu}) \Omega_{\mu}(x_{\nu}=0) Z_{\mu \nu}\ ,
\end{equation}
and $Z_{\mu\nu}$ belongs to the center $\mathbb{Z}_N$ of $SU(N)$,
\begin{equation}
    Z_{\mu \nu} = \exp \left(\frac{2\pi i}{N} n_{\mu \nu} \right)\ ,
\end{equation}
where $n_{\mu \nu}$ is a $\mathbb{Z}_N$-valued anti-symmetric $4\times 4$ matrix known as the discrete 't Hooft flux. We will parameterize $n_{\mu \nu}$ as,
    \begin{equation}
        n_{\mu \nu} = \left(\begin{array}{cccc}
            0&t_1&t_2&t_3\\
            -t_1&0&s_3&-s_2\\
            -t_2&-s_3&0&s_1\\
            -t_3&s_2&-s_1&0
        \end{array} \right)
    \end{equation}
where $t=(t_1,t_2,t_3)$ are the discrete electric fluxes and $s=(s_1,s_2,s_3)$ are the discrete magnetic fluxes. Moreover, if we consider a gauge transformation $A\rightarrow \Lambda A$, the $\Omega$ 
 functions transform according to $\Omega_{\mu} \rightarrow \Lambda(x_{\mu}=a_{\mu}) \Omega_{\mu} \Lambda^{-1}(x_{\mu}=0)$ but one can check $Z_{\mu \nu}$ and thus discrete fluxes $n_{\mu\nu}$ are invariant.

In the following, we will mainly work with $x$-$y$ plane for simplicity and denote for example $\Omega_{x}(y=a_y) \equiv \Omega_{x}(a_y)$. The holonomies along $x$ and $y$ directions are defined as\cite{Tanizaki:2022ngt}
    \begin{equation}\label{Holonomies}
        \left\{ \begin{split}
        H_x(y) = \Omega_x(y) \mathcal{P} \exp\left(i \int_{0}^{a_x} A_x(y) d x \right)\\ H_y(x) = \Omega_y(x) \mathcal{P} \exp\left(i \int_{0}^{a_y} A_y(x) d y \right)
        \end{split} \right.
    \end{equation}
where $\mathcal{P}$ is the path-ordering operator and the insertion of $\Omega_x(y),\Omega_y(x)$ are understood as transition functions connecting two patches. When the gauge bundle is flat, we can set $A_{\mu}=0$ in a single patch and $\Omega_{\mu}(x_{\nu})$ are constant so that \eqref{Twisted-boundary-condition-1} is satisfied. Then we can write $H_x = \Omega_x$ and $H_y = \Omega_y$ and from \eqref{Twisted-boundary-condition-2} we learn that the two holonomies $H_x,H_y$ do not commute and satisfy
    \begin{equation}
        H_x H_y = e^{\frac{2\pi i}{N} n_{xy}} H_y H_x\ .
    \end{equation}
The discrete fluxes $n_{\mu \nu}$ can be understood as the twisted boundary conditions for Wilson loops in $SU(N)$ gauge theory. The Wilson loops $W_x(y),W_y(x)$ are trace of holonomies 
    \begin{equation}
        W_x(y) = \textrm{Tr} H_x(y)\ ,\quad W_y(x) = \textrm{Tr} H_y(x)\ .
    \end{equation}
From \eqref{Holonomies} it is straightforward to check that the Wilson loops are gauge invariant and satisfy
    \begin{equation}
        W_x(a_y) = e^{-\frac{2\pi i}{N} n_{xy} } W_x(0)\ ,\quad W_y(a_x)=e^{\frac{2\pi i}{N} n_{xy} } W_y(0)\ .
    \end{equation}
    
\begin{figure}[!t]
    \centering
    \includegraphics[scale=1]{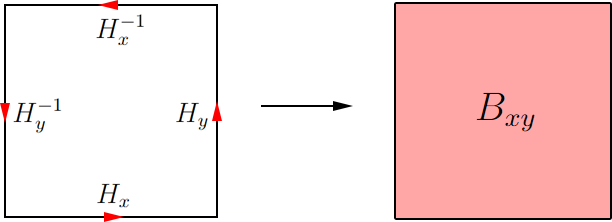}
    \caption{One can perform a singular one-form transformation which trivializes the holonomy $H_y^{-1} H_x^{-1} H_y H_x = e^{-\frac{2\pi i}{N}n_{xy}}$ around the square and turns on a two-form background $B_{xy} = -\frac{2\pi}{N a_x a_y} n_{xy}$. }
    \label{Fig-holonomy-background}
\end{figure}

The nature of $n_{\mu \nu}$ as boundary conditions of Wilson loops implies we can alternatively turn on a background of one-form symmetry $B_{\mu \nu} = -\frac{2\pi}{N a_x a_y} n_{\mu \nu}$ whose charged objects are Wilson loops. Before talking about the details, we illustrate the idea in Figure \ref{Fig-holonomy-background}. Consider the 1-from gauge transformation
    \begin{equation}
        A \rightarrow A - \Lambda \mathbf{1}_{N\times N}\ ,\quad B\rightarrow B + d \Lambda\ , 
    \end{equation}
where $\Lambda$ is a one-form gauge parameter and we will choose $\Lambda = \frac{2\pi }{N a_x a_y} n_{xy} y dx $. The gauge transformation is singular since $\Lambda$ is not single-valued on the torus. One may expect the gauge transformation will convert the non-commutative of holonomies $H_x, H_y$ into the two-form background field $B_{xy}$.

To be concrete, let us consider the $SU(2)$ case as an illustration. Assuming we have an $SU(2)$ gauge configuration with a non-trivial 't Hooft flux on the $x$-$y$ torus defined by the boundary condition,
    \begin{equation}
        A(a_x,y) = -i \sigma_2 A(0,y)\ ,\quad A(x,a_y) = -i \sigma_1 A(x,0)\ ,
    \end{equation}
such that
    \begin{equation}
        \Omega_y(x) = -i \sigma_1\ ,\quad \Omega_x(y) = -i \sigma_2\ ,
    \end{equation}
with $\sigma_i (i=1,2,3)$ the Pauli matrices. One has
    \begin{equation}
        \Omega_x(a_y) \Omega_y(0) = - \Omega_y(a_x) \Omega_x(0)\ ,
    \end{equation}
and the 't Hooft flux is characterized by $Z_{xy}=-1$ or $n_{xy}=1$. 

Next, let us lift the $SU(2)$ gauge field $A$ to a $U(2)$ gauge field $\widetilde{A}$, and couple $\widetilde{A}$ with the two-form gauge field $B$ such that we have the one-form gauge transformation
    \begin{equation}
        \widetilde{A} \rightarrow \widetilde{A} - \Lambda \mathbf{1}_{2\times 2}\ ,\quad B\rightarrow B + d \Lambda\ , 
    \end{equation}
with $\Lambda$ a differential one-form on $T^4$. The $U(2)$ gauge field $\widetilde{A}$ now enjoys a $U(2)$ gauge transformation
    \begin{equation}
        \widetilde{A} \rightarrow g\widetilde{A} g^{-1} - i g d g^{-1}\ ,
    \end{equation}
where now $g$ is $U(2)$-valued.  A singular one-form gauge transformation given by $\Lambda = \frac{\pi}{a_x a_y} y dx$ leads to
    \begin{equation}
        \widetilde{A} = A - (\frac{\pi}{a_x a_y}y dx) \mathbf{1}_{2\times 2}\ ,\quad B = -\frac{\pi}{a_x a_y}dx \wedge dy\ .
    \end{equation}
One can check that the boundary conditions for $\widetilde{A}$ are characterized by $\widetilde{\Omega}$ as
    \begin{equation}
        \widetilde{\Omega}_y(x) = -i \sigma_1 e^{\frac{i \pi x}{a_x}},\quad \widetilde{\Omega}_x(y) = -i \sigma_2\ ,
    \end{equation}
 which satisfy
    \begin{equation}
        \widetilde{\Omega}_x(a_y) \widetilde{\Omega}_y(0) = \widetilde{\Omega}_y(a_x) \widetilde{\Omega}_x(0)\ .
    \end{equation}
In other words, the 't Hooft fluxes become trivial. One can show that we can do a further $U(2)$ gauge transformation to make $\widetilde{A}$ periodic on $T^4$. In summary, we trivialize the twisted boundary condition at the expense of introducing the two-form background field
\begin{equation}
    B = -\frac{\pi}{a_x a_y} dx \wedge dy\ .
\end{equation}

For general gauge groups $G$ with simply connected covering $\widetilde{G}$, the one-form symmetry group for $\widetilde{G}$ is identified with the center $\mathcal{Z}(\widetilde{G})$. Therefore the discrete 't Hooft fluxes also take values in $\mathcal{Z}(\widetilde{G})$ and we have $t_i,s_j \in \mathcal{Z}(\widetilde{G})$.

Finally, we comment on the geometric picture to show the equivalence between the twisted bundle and background for one-form symmetry. Since Wilson loops are constructed using local gauge potential $A$, the twisted boundary condition for Wilson loops reflects a twisted boundary condition for the gauge potential. In the latter case, the gauge bundle becomes twisted $\widetilde{G}$-bundle. Consider the manifold $T^4$ as a union of patches $T^4 = \bigcup X_i$ with the transition functions $t_{ij}\in \widetilde{G}$ defined in $X_i \bigcap X_j$. The coboundary $f_{ijk} \equiv t_{ij} t_{jk} t_{ki}$ should be unity for a $\widetilde{G}$-bundle. However, when the gauge potential satisfies a twisted boundary condition, $[f_{ijk}]$ will define a non-zero element in $H^2(T^4, \mathcal{Z}(\widetilde{G}))$ which equals the two-form background we turned on. 

With this understanding between two-form background and twisted $\widetilde{G}$-bundle, gauging $\mathcal{Z}(\widetilde{G})$ one-form symmetry is equivalent to summing over twisted $\widetilde{G}$-bundle with $[f_{ijk}] \in H^2(T^4, \mathcal{Z}(\widetilde{G}))$ in the path integral. The twisted $\widetilde{G}$ bundles are legal $\widetilde{G}/\mathcal{Z}(\widetilde{G})$ bundles since $[f_{ijk}]$ is trivialized after $\mathcal{Z}(\widetilde{G})$ quotient. Therefore gauging $\mathcal{Z}(\widetilde{G})$ one-form symmetry changes the gauge group from $\widetilde{G}$ to $\widetilde{G}/\mathcal{Z}(\widetilde{G})$.

\subsection{Mixed 't Hooft anomaly}
Recall that in two dimensions, if we gauge a normal subgroup $H$ of the whole symmetry group $G$, there could exist a mixed anomaly between the quantum symmetry $\widetilde{H}$ (which is the dual group of $H$) and the quotient group $G/H$. As a simple example, take $G$ to be $\mathbb{Z}_N$ with $N = PQ$ and $H = \mathbb{Z}_P$. If $P$ and $Q$ are not coprime, then there is an anomalous phase if we turn on both background fields. For more details, we refer the readers to e.g. \cite{Duan:2023ykn}.

Similar story happens in four dimensions if we replace zero-form symmetry by one-form symmetry. First as a concrete example, let us consider the $SU(4)$ pure YM theory. It has a $\mathbb{Z}_{4}$ one-form center symmetry and we can gauge the $\mathbb{Z}_{2}$ subgroup. Since details are already worked out in Section \ref{sec:partial}, we can just quote the answer from there.

After performing the partial gauging, there are two possible theories denoted by $\big(SU(4)/\mathbb{Z}_2\big)_0$ and $\big(SU(4)/\mathbb{Z}_2\big)_1$. 
The former one has one-form symmetry $\mathbb{Z}_2 \times \mathbb{Z}_2$ and is invariant under $SL(2,\mathbb{Z})$ action. In Figure \ref{fig:enter4-label}, it corresponds to the topological boundary states $\{|c,\tilde{d} \rangle_0\}$. The latter one has one-form symmetry $\mathbb{Z}_4$ and corresponds to $\{|c,\tilde{d} \rangle_1\}$.

Now let us focus on $\big(SU(4)/\mathbb{Z}_2\big)_0$. Let us label the background field of two $\mathbb{Z}_2$ symmetries by $c$ and $\tilde{d}$ separately. The former $\mathbb{Z}_2$ is electric and rotates 
the Wilson loop operators while the latter is magnetic and rotates 
the 't Hooft loop operators. 
In the presence of the magnetic 
$\mathbb{Z}_2$ background $d$, we learn from 
\eqref{ttt} that
the partition function is not invariant 
under the large electric $\mathbb{Z}_2$ gauge transformation,  
\begin{equation}
    Z_{(SU(4)/\mathbb{Z}_2)_0}\big[ c + 2 e_l , \tilde{d}\big] = (-1)^{-K(\tilde{d}, e_l)} Z_{(SU(4)/\mathbb{Z}_2)_0}\big[ c  , \tilde{d}\big]\ ,
\end{equation}
where $e_\ell$ being an $h_2$-dimensional unit vector in the $\ell$-th direction. This $U(1)$ phase signals a mixed 't Hooft anomaly between the two $\mathbb{Z}_2$ one-form symmetries. On the other hand,  the magnetic $\mathbb{Z}_2$ is non-anomalous. 

In general, we could consider $SU(PQ)$ pure YM theory where 
$P$ and $Q$ are not relatively prime. Gauging a $\mathbb{Z}_P$ center subgroup
leads to the $\big(SU(PQ)/Z_P\big)_0$ gauge theory. It has 
$\mathbb{Z}_Q \times \mathbb{Z}_P$ one-form symmetry, where 
the former is electric and the latter is magnetic. 
Parallel to the above example, one can easily argue that $\big(SU(PQ)/Z_P\big)_0$ theory has a mixed 't Hooft anomaly between $\mathbb{Z}_Q$ and $\mathbb{Z}_P$ one-form symmetries, which is dictated by an overall phase 
\begin{equation}
    Z_{(SU(PQ)/\mathbb{Z}_P)_0}\big[ c + Q e_l , \tilde{d}\big]  = e^{-2\pi i K(\tilde{d}, e_l)/P} Z_{(SU(PQ)/\mathbb{Z}_P)_0}\big[ c , \tilde{d}\big]\ .
\end{equation}

Now we let the $\big(SU(PQ)/Z_P\big)_0$ theory flow to the deep IR. 
Since the theory is in the confinement phase, 
the theory develops a mass gap and everything above the ground state is integrated out.
On the other hand, due to the 't Hooft anomaly matching condition, 
the vacuum cannot be trivial, but should be described by a certain
topological field theory that accounts for the mixed 't Hooft anomaly. 
By placing the $\big(SU(PQ)/Z_P\big)_0$ gauge theories on a three-dimensional 
space with boundaries and increasing the temperature, one 
may observe the non-trivial nature of their confining vacua. 

We leave it as a future project to identify this TQFT and if possible, write down its effective Lagrangian.

%%%%%%%%%%%%%%%%%%%%%%%%%%%%%%%%%%%%%%%%%%%%%%%%%%%%%%%%%%%%%%%%%%%%%%%%%%%%%%%%%%%%%%%%%%%
%%%%%%%%%%%%%%%%%%%%%%%%%%%%%%%%%%%%%%%%%%%%%%%%%%%%%%%%%%%%%%%%%%%%%%%%%%%%%%%%%%%%%%%%%%%
%%%%%%%%%%%%%%%%%%%%%%%%%%%%%%%%%%%%%%%%%%%%%%%%%%%%%%%%%%%%%%%%%%%%%%%%%%%%%%%%%%%%%%%%%%%
\section{Witten Index}\label{sec:5}

In this section, we will mainly consider the Witten index of four-dimensional $\mathcal{N}=1$ pure Yang-Mills theory on $T^4$ and study how it depends on the global structure of the gauge group $G$. We denote by $\widetilde{G}$ a simply-connected covering of $G$ such that $G = \widetilde{G}/H$ where $H$ is a subgroup of the center $\mathcal{Z}(\widetilde{G})$ of $\widetilde{G}$.

\subsection{Witten index}

In this section, we study microscopic computations of the Witten index in the Hamiltonian approach rather than the path-integral method. Specifically, we focus on constructing the ground states by quantizing the space 
of classical zero-energy gauge configurations, which is the only contribution to the indices in the small four-torus limit.

The Witten index for a theory with gauge group $\widetilde{G}$ is defined as the four-torus partition function 
endowed with the periodic boundary condition. One can define it as 
\begin{equation}
    Z_{\widetilde{G}}[(0,0)] = \textrm{Tr}_{\mathcal{H}(T^3)} \Big[ (-1)^F e^{-\beta H}\Big] \ ,
\end{equation}
where the trace is performed over the Hilbert space of the gauge theory defined on 
the spatial three-torus $T^3$. For ease of notation, we omit the dependence on $T^3$ in what follows.  
Here $\beta$ is the size of the temporal circle and 
the inverse of the temperature. We normalize it by $2\pi$ in the present work. 
It is well-known that the supersymmetric index only receives contributions from the ground states, 
as the excited states always appear in pairs with an equal number of 
bosons and fermions. 

To compute the Witten index of a theory with gauge group $G$, 
one also has to compute the partition function in the presence of 
the background gauge fields for the $\mathcal{Z}(\widetilde{G})$ one-form symmetry.
In the Hamiltonian interpretation, such partition functions can be 
expressed as 
\begin{align}
   Z_{\widetilde{G}}\big[(t,s) \big] = 
    \textrm{Tr}_{\mathcal{H}_s} \Big[ {\cal D}_t (-1)^F e^{-\beta H}\Big]
\end{align}
where ${\cal D}_t$ stands for a $\mathcal{Z}(\widetilde{G})$ one-form symmetry 
defect. We shall present shortly how ${\cal D}_t$ acts on 
the gauge holonomies. The trace is now performed over the Hilbert space 
endowed with the twisted boundary condition characterized by the background $s$, 
as explained in Section \ref{sec:twistedbc}. 

For later convenience, it is useful to decompose the Hilbert space ${\cal H}_s$ 
into its superselection sectors 
\begin{align}
    {\cal H}_s = \mathop{\oplus}_u {\cal H}_{u,s}
\end{align}
where ${\cal H}_{u,s}$ is the subspace of ${\cal H}_s$ consisting of  
states with charge $u=(u_1,u_2,u_3)$ under ${\cal D}_t$ : for any state $|v_s \rangle \in {\cal H}_{u,s}$, we have
\begin{align}
    {\cal D}_t |v_s \rangle = \omega^{t \cdot u} |v_s \rangle\ ,
\end{align}
where $t\cdot u= t_1 u_1 + t_2 u_2 + t_3 u_3$. 
Accordingly, one can express the twisted partition function as
\begin{align}
    Z_{\widetilde{G}}\big[(t,s) \big] = \sum_{u} \omega^{ t\cdot u} I_{\widetilde{G}}\big[(u,s) \big]
\end{align}
where $I_{\widetilde{G}}\big[(u,s) \big]$ denotes an index defined by 
\begin{align}
   I_{\widetilde{G}}\big[(u,s) \big] =   \textrm{Tr}_{\mathcal{H}_{u,s}} \Big[(-1)^F e^{-\beta H}\Big]\ .
\end{align}

The computation of the indices $I_{\widetilde{G}}\big[(u,s) \big]$ 
was originally performed in the seminal work by Witten \cite{Witten:2000nv}. Here, we
revisit the microscopic computation from the perspective 
of SymTFT. For simplicity, we first restrict our attention to 
the gauge theory with $\widetilde{G}=SU(N)$.

\paragraph{$s=0$ background} 

It can be argued that the classical zero-energy field configurations must satisfy 
the flat-connection condition and hence can be described, up to gauge transformation, 
by the holonomies around the three directions in the spatial $T^3$, denoted by 
$g_i$ $(i=1,2,3)$. These three holonomies also commute unless the one-form global 
symmetry is weakly gauged, i.e., $s=(0,0,0)$. More precisely, they can be determined by 
a constant gauge field 
\begin{align}\label{cl_vac_s0}
    A_i = \sum_{a=1}^{\text{rk}(\tilde G)} \vartheta_i^a H^a \ ,
\end{align}
where $\{H^a\}$ denotes a Cartan subalgebra of the gauge group.  
Here $\vartheta_i^a$ are normalized to have $2\pi$ periodicity. 
Since each holonomy can be simultaneously conjugate to the 
Cartan torus ${\cal T}$ of $\widetilde G=SU(N)$, the space of classical zero-energy configurations 
for $s=0$ becomes 
\begin{align}\label{maximaltorus}
    {\cal M}_{s=0} =  {\cal T}\times  {\cal T} \times  {\cal T} /W\ ,
\end{align}
where $W$ is the Weyl group of $\widetilde G$. 

Taking into account the fermionic zero modes around \eqref{cl_vac_s0}, one 
obtains the supersymmetric quantum mechanics on the moduli space ${\cal M}_{s=0}$
that describes the low-lying states of the four-dimensional system. 
As shown in \cite{Witten:2000nv}, the quantum mechanical model possesses degenerate supersymmetric ground 
states whose wavefunctions are constant on the moduli space ${\cal M}_{s=0}$. In particular, 
the space of ground states is generated by $N$ vacua $|\Omega,r\rangle$ ($r=0,1,..,(N-1)$). 
Moreover, it was argued in \cite{Witten:2000nv,Tachikawa:2014mna} that all supersymmetric vacua 
are bosonic for odd $N$ and fermionic for even $N$.

One can identify the one-form symmetry operator ${\cal D}_t$  
as a translation operator in the above supersymmetric quantum mechanics on ${\cal M}_{s=0}$. 
To see this, recall that 
${\cal D}_t$ shifts the three holonomies $g_i$ at a given point $\vartheta_i^a$ 
on ${\cal M}_{s=0}$ by 
\begin{align}
    {\cal D}_t :~ g_i(\vartheta_i^a) \longrightarrow \big( e^{2\pi i/N} \big)^{t_i} g_i(\vartheta_i^a)\ ,
\end{align}
which are nothing but the holonomies at a different point ${\vartheta'}_i^a$,
\begin{align}
    g_i({\vartheta'}_i^a) = \big( e^{2\pi i/N} \big)^{t_i} g_i(\vartheta_i^a)\ .
\end{align}
Since each ground state wavefunction is constant on ${\cal M}_{s=0}$, i.e., 
$\langle \vartheta_i^a | \Omega , r\rangle =\text{const}.$,  
it thus carries zero charge under ${\cal D}_t$ for all $t$. 

Based on the Hilbert space obtained by the Born-Oppenheimer approximation, 
one can conclude that the indices in the $s=0$ sector are given by
\begin{align}
\begin{split}
    I_{\widetilde{G}}[u=0,s=0] & = (-1)^{N-1} N \ , 
    \\
    I_{\widetilde{G}}[u\neq0,s=0] & = 0 \ .
\end{split}
\end{align}

\paragraph{$s\neq0$ background} Let us now move onto the computation of 
the indices in the presence of the background two-form gauge fields along the spatial $T^3$,
$s \neq 0$. 

For a given background $s=(s_1,s_2,s_3)$, one can always find a frame 
in which $s=\big(0,0,\gcd(s_1,s_2,s_3)\big)$ by performing a suitable 
$SL(3,\mathbb{Z})$ transformation. Therefore, it is sufficient 
to consider the case where $s=(0,0,s_3)$ without loss of generality.

We first compute the index $I(u,s)$ when $\gcd(s_3,N)=1$. As explained in Section 4,
one can gauge the two-form background away, and the holonomies no longer commute but 
are subject to the twisted boundary conditions 
\begin{align}\label{twistedbd01}
\begin{split}
    g_1 g_2 &= \omega^{s_3} g_2 g_1    \  ,
    \\
    g_3 g_{1} & = g_{1} g_3\ ,
    \\
    g_3 g_{2} & = g_{2} g_3\ .
\end{split}
\end{align}
Here $\omega=e^{2 \pi i /N}$. One can argue that the classical 
vacuum field configurations obeying the condition \eqref{twistedbd01} can be described as 
\begin{align}\label{cl_vac_sneq0}
    g_1^{(\ell)} = C_N^{s_3}\ ,  ~ g_2^{(\ell)} = S_N \ , ~ g_3^{(\ell)} = \omega^\ell {\bf 1}_N \ ,
\end{align}
where $\ell=0,1,..,(N-1)$ and $C_N, S_N$ are the clock and shift matrices 
\begin{align}\label{clockshift}
\begin{split}
    C_{N}& = \epsilon_N  
    \left(\begin{array}{cccc}
        1&&&\\
        &\omega&&\\
        &&\ddots&\\            
        &&&\omega^{N-1}
    \end{array} \right)\ ,
    \\
    S_{N} & = \epsilon_N 
    \left(\begin{array}{ccccc}
        0&1&0&\cdots&0\\
        0&0&1&\cdots&0\\
        \vdots&\ddots&\ddots&\ddots&\vdots\\            
        1&0&0&\cdots&0
    \end{array} \right)\ 
\end{split}
\end{align}
with $\epsilon_N= 1$ for odd  $N$ and $\epsilon_N = \omega^{1/2}$ for even $N$. 
The clock and shift matrices satisfy the relation below
\begin{align}
    C_N S_N = \omega S_N C_N\ . 
\end{align}
All other classical zero-energy states obeying the condition \eqref{twistedbd01} can be rotated 
back to \eqref{cl_vac_sneq0} by performing a suitable $SU(N)$ gauge transformation. 
Since \eqref{cl_vac_sneq0} are all isolated, the quantization then gives rise to 
$N$ independent degenerate quantum states $|\ell \rangle $ $(\ell=0,1,..(N-1))$. 

Note that $|\ell\rangle$ are invariant under ${\cal D}_t$ when $t_3=0$. 
This is because ${\cal D}_t$ maps each configuration $g_i^{(\ell)}$ to itself modulo $SU(N)$ gauge 
transformation as long as $t_3=0$. On the other hand, we can show that the $N$ vacua  $|\ell\rangle$
can be cyclically permuted by the $\mathbb{Z}_N$ symmetry
operators ${\cal D}_t$ with $t_3\neq 0$, 
\begin{align}
    {\cal D}_t | \ell \rangle = | \ell + t_3 \rangle\ .
\end{align}
Here $|\ell+N\rangle = |\ell \rangle$. It implies that the Bloch state for each $u\in \mathbb{Z}_N$
\begin{align}
    | \Theta_u \rangle = \sum_{\ell=0}^{N-1} \omega^{- u \ell} | \ell \rangle\ 
\end{align}
can simultaneously diagonalize ${\cal D}_t$
\begin{align}
    {\cal D}_t | \Theta_u \rangle = \omega^{ut_3}  | \Theta_u \rangle\ .
\end{align}
In other words, $|\Theta_u\rangle$ is the unique supersymmetric ground state 
of the charge $(u_1,u_2,u_3)=(0,0,u)$. 

As a result, when $\gcd{(s,N)}=\gcd{(s_1,s_2,s_3,N)}= 1$, the indices become 
\begin{align}
\begin{split}
    I_{\widetilde{G}}\big[ u , s \big] & = (-1)^{N-1}  \text{ for } u \parallel s \ ,
    \\
    I_{\widetilde{G}}\big[ u , s \big] & = 0 \text{ otherwise .} 
\end{split}
\end{align}

It is not difficult to extend the above results to the case 
where $s_3$ is not relatively prime to $N$. We first notice that 
the $s_3$ copies of clock matrix $C_N$ become
\begin{align}
    C_N^{s_3} =  \big(C_{\frac{N}{\gcd{(s_3,N)}}}\big)^{\frac{s_3}{\gcd{(s_3,N)}}}
    \otimes {\bf 1}_{\gcd{(s_3,N)}} \ .
\end{align}
It implies that the general solution of the conditions \eqref{twistedbd01} 
can be described as
\begin{align}\label{cl_vac_sneq0_02}
\begin{split}
    g_1^{(\ell)} & = \big(C_{\frac{N}{\gcd{(s_3,N)}}}\big)^{\frac{s_3}{\gcd{(s_3,N)}}} \otimes {g'}_1 \ , 
    \\
    g_2^{(\ell)} & = S_{\frac{N}{\gcd{(s_3,N)}}} \otimes {g'}_2 \ ,
    \\
    g_3^{(\ell)} & = \omega^\ell {\bf 1}_{\frac{N}{\gcd{(s_3,N)}}} \otimes {g'}_3\  ,
\end{split}
\end{align}
where ${g'}_i$ are simultaneously conjugate to the Cartan torus 
of $\widetilde{G}'=SU\big(\gcd{(s_3,N)}\big)$ parameterized by angle variables ${\vartheta}_i^b$ 
$\big(b=1,2,..,(\gcd{(s_3,N)}-1)\big)$ modulo $SU(N)$ gauge transformations. 
To see this, we first note that in the parameterization \eqref{cl_vac_sneq0_02}, 
the determinants of $g_1'$ and $g_2'$ should be $(N/\gcd{(s_3, N)})$-th root of unity. 
Let us then consider an $SU(N)$ gauge transformation below, 
    \begin{equation}\label{gauge-transformation}
        U = \left(\begin{array}{ccccc}
            \mathbf{1}_{\frac{N}{\gcd{(s_3,N)}}}&&&&\\
            &\mathbf{1}_{\frac{N}{\gcd{(s_3,N)}}}&&&\\
            &&\ddots&&\\
            &&&\mathbf{1}_{\frac{N}{\gcd{(s_3,N)}}}&\\
            &&&&S_{\frac{N}{\gcd{(s_3,N)}}}^{\alpha}
        \end{array} \right)
    \end{equation}
where $\alpha$ and $\beta$ are certain integers satisfying
    \begin{equation}
        \alpha s_3 + \beta N = \gcd(s_3,N)\ . 
    \end{equation}
Under the gauge transformation $U$,  ${g'}_2$ remains invariant while  ${g'}_1$ 
transforms into
    \begin{equation}
        \det {g'}_1 \rightarrow 
 e^{-\frac{2\pi i}{N} \gcd(s_3,N)} \det {g'}_1\ ,
    \end{equation}
Thus, one can always set $\det {g'_1}=1$. One can also
set $\det{g'_2}=1$
by performing the gauge transformation similar to \eqref{gauge-transformation}
with $S_{\frac{N}{\gcd{(s_3,N)}}}$ and  $C_{\frac{N}{\gcd{(s_3,N)}}}$ exchanged.

The space of the classical zero-energy configurations is $N/\gcd{(s_3,N)}$ copies of 
the moduli space \eqref{maximaltorus} for $\widetilde{G}'$. This is because, since 
$\omega^{N/\gcd{(s_3,N)}}$ is a center element of $\widetilde{G}'$, the shift 
$\ell \to \ell + N/\gcd{(s_3,N)}$ leads to a translation on the moduli space,
\begin{align}
    \big( {g'}_1(\vartheta_1) ,~ {g'}_2(\vartheta_2), ~ \omega^{N/\gcd{(s_3,N)}}  {g'}_3(\vartheta_3) \big) 
    = 
    \big( {g'}_1(\vartheta_1) ,~ {g'}_2(\vartheta_2), ~ {g'}_3(\vartheta'_3) \big)\ .
\end{align}

Repeating the previous analyses with \eqref{cl_vac_sneq0_02}, 
the quantization of the vacuum moduli space for each $\ell$ then gives us 
the space of supersymmetric ground states spanned by  
\begin{align}
    \Big\{ |\ell,r\rangle \equiv | \ell \rangle \otimes |\Omega , r\rangle \Big\} 
\end{align}
where $r=0,1,..(\gcd{(s_3,N)}-1)$ labels the vacua of $\widetilde{G}'$. Note that $\ell$ now effectively 
runs from $0$ to $(N/\gcd{(s_3,N)}-1)$ rather than $(N-1)$,
\begin{align}
    |\ell + N/\gcd{(s_3,N),r} \rangle = |\ell, r\rangle\ , 
\end{align}
because the ground states $|\Omega,r\rangle$
have constant wavefunctions on the moduli space. One can also show that 
the $\mathbb{Z}_N$ symmetry operator ${\cal D}_t$ acts on 
each quantum ground state as follows, 
\begin{align}
    {\cal D}_t | \ell , r \rangle = |\ell+t_3 , r\rangle\ .
\end{align}
${\cal D}_t$ becomes trivial when $t_3$ is a multiple of 
$N/\gcd{(s_3,N)}$. Therefore, the Bloch states of our interest 
are 
\begin{align}\label{}
    | \Theta_\lambda, r\rangle = \sum_{\ell=0}^{N/\gcd{(s_3,N)}-1} 
    \Big( \omega^{\gcd{(s_3,N)}}\Big)^{- \lambda  \ell} | \ell,r \rangle \ .
\end{align}
They have no $u_1$ and $u_2$ charges and only carries $u_3$  
charge that must be a multiple of $\gcd{(s_3,N)}$, i.e., $u_3=\lambda \gcd{(s_3,N)}$,
\begin{align}
    {\cal D}_t | \Theta_\lambda, r\rangle  = \omega^{(\lambda \gcd{(s_3,N)}) t_3}  
    | \Theta_\lambda, r\rangle  \ .
\end{align}
$\{ | \Theta_\lambda, r\rangle \}$ for each $\lambda$ 
then spans the space ${\cal H}_{u,s}$ where $u=(0,0,\lambda \gcd{(s_3,N)})$
and $s=(0,0,s_3)$. Using $SL(3,\mathbb{Z})$ transformations, one can conclude that 
${\cal H}_{u,s}$ for $s=(s_1,s_2,s_3)$ is $\gcd{(s,N)}$-dimensional only when $u=\lambda s$, and becomes empty otherwise. \

Therefore, the indices are 
\begin{align}
\begin{split}
    I_{\widetilde{G}}\big[ u , s \big] & = (-1)^{N-1} \gcd{(s,N)} \text{ for } u = \lambda s \ ,
    \\
    I_{\widetilde{G}}\big[ u , s \big] & = 0 \text{ otherwise}\ .
\end{split}
\end{align}
Using them, one can obtain the twisted partition functions,
\begin{align}
\begin{split}
    Z_{\widetilde{G}}\big[(t,s)\big] & = \sum_{u} \omega^{t \cdot u} I_{\widetilde{G}}\big[u,s\big]      
    = (-1)^{N-1} \hspace{-0.4cm}
    \sum_{\lambda=0}^{\frac{N}{\gcd{(s,N)}}-1} \hspace{-0.4cm}
    \big( \omega^{\gcd{(s,N)}} \big)^{\lambda \frac{t\cdot s}{\gcd{(s,N)}} } \gcd{(s,N)}\ ,
    \\
    &= 
    \begin{cases}
        (-1)^{N-1} N  & \text{ if }  t\cdot s  = 0 \text{ mod } N \\ 
        0 & \text{ otherwise}  
    \end{cases} \ .
\end{split}
\end{align}
Thus, the SUSY dynamical boundary state corresponding to the Witten index 
for $\widetilde{G}=SU(N)$ is 
\begin{align}\label{dynbdstateSU(N)}
    \big| \chi_\text{W} \big\rangle_{SU(N)} = 
    (-1)^{N-1} N \sum_{t,s} \delta_{t\cdot s,0} \big| (t,s) \big\rangle \ ,
\end{align}
where the Kronecker delta $\delta$ is defined modulo $N$.

Using \eqref{dynbdstateSU(N)}, one can easily show that the Witten index for $G=\big( SU(N)/\mathbb{Z}_N\big)_k$ 
is independent of the value of $k$. The four-dimensional gauge theory with $G$ 
has the $\mathbb{Z}_N$ one-form symmetry, and is related to the theory with $SU(N)$ 
via the $SL(2,\mathbb{Z})$ transformation $ST^k$. We learned that the corresponding 
topological boundary state is given by \eqref{SL_orbit_01} and \eqref{SL_orbit_01_torus} for $T^4$. The twisted partition functions 
for $G$ therefore become 
\begin{align}
\begin{split}
    Z_G\big[ (t,s) \big]  = 
    \big\langle (t,s) \big| V_{ST^k}^\dagger \big| \chi_\text{W} \big \rangle 
    & = \frac{(-1)^{N-1}}{N^2} \sum_{t',s'} \omega^{- t \cdot s' - s \cdot t' + k t'\cdot s' } \delta_{t'\cdot s',0} \ ,
    \\ 
    & = \frac{(-1)^{N-1}}{N^3} \sum_{j=0}^{N-1} \sum_{t',s'} \omega^{-t\cdot s' - s\cdot t' + j t'\cdot s'} \ , 
    \\
    & = (-1)^{N-1} \sum_{j=0}^{N-1} \sum_{s'} \omega^{-t \cdot s'} \delta_{js',s} \ ,
\end{split}
\end{align}
where $\delta_{js',s} = \delta_{j{s'}_1,s_1}\delta_{j{s'}_2,s_2}\delta_{j{s'}_3,s_3}$. 
When the background $\mathbb{Z}_N$ gauge field is turned off, one can further 
simplify the above expression and manage to express the Witten index as  
\begin{align}\label{macroresult}
    Z_{G}\big[ (0,0) \big] = (-1)^{N-1} \sum_{j=0}^{N-1} \big( \gcd{(j,N) }\big)^3. 
\end{align}
The result \eqref{macroresult} matches with the 
macroscopic computation of the Witten index \cite{Witten:2000nv}. 
In the infinite volume limit, 
the ${\cal N}=1$ super Yang-Mills theory with $SU(N)/\mathbb{Z}_N$  
has the $N$ different confining vacua. This is because the local physics 
should not be aware of the difference between $SU(N)$ and 
$SU(N)/\mathbb{Z}_N$. Here the $j$-th vacuum preserves 
the $\mathbb{Z}_{\gcd{(j,N)}}$ magnetic gauge symmetry. It was argued in \cite{Aharony:2013hda}
that, upon the large $T^3$ compactification, the $j$-th vacuum splits into  
$\gcd{(j,N)}^3$ zero-energy states which differ by the vacuum 
expectation value of the 't Hooft operator wrapping $S^1$ inside $T^3$. 
Thus, the total number of vacua is given by \eqref{macroresult}. 

Based on the index computation $I_{\widetilde{G}}[(u,s)]$
for other simply-connected Lie groups \cite{Witten:2000nv}, one can also read off 
the corresponding SUSY dynamical boundary states  $|\chi_\text{W} \rangle_{\widetilde{G}}$. 
For completeness, they are summarized in the Table \ref{tab:gaugegroup}. 
\begin{table}[t!]
\renewcommand{\arraystretch}{1.3} 
\begin{center}
\begin{tabular}{c|c|c|c}
    $\widetilde{G}$ & ${\cal Z}(\widetilde{G})$ & SUSY Dynamical Boundary State $|\chi_\text{W} \rangle_{\widetilde{G}}$ & Fractional Instanton\\
    \hline
    $SU(N)$ & $\mathbb{Z}_N$ & $ (-1)^{N-1} N  \sum_{t,s} \delta_{t\cdot s,0} |(t,s)\rangle$ & $s\cdot t/N$ \\
    $Sp(N)$ & $\mathbb{Z}_2$ & $(-1)^N (N+1) \sum_{t,s} \delta_{N t\cdot s,0} |(t,s)\rangle$ & $N (s\cdot t)/2$ \\
    $Spin(2N+1)$ & $\mathbb{Z}_2$ & $(-1)^N (2N-1) \sum_{t,s} |(t,s)\rangle$ & 0 \\
    $Spin(4N+2)$ & $\mathbb{Z}_4$ & $-4N \sum_{t,s} \delta_{t\cdot s,0} |(t,s)\rangle$ & $\pm s\cdot t/4$ \\
    $Spin(8N+4)$ & $\mathbb{Z}_2\times \mathbb{Z}_2$ & $(8N+2)\sum_{t,s;t',s'}  \delta_{t\cdot s + t'\cdot s',0}|(t,s);(t',s')\rangle$ & $s\cdot t/2 + s' \cdot t'/2$ \\
    $Spin(8N)$ & $\mathbb{Z}_2 \times \mathbb{Z}_2$ & $(8N-2)\sum_{t,s;t',s'}  \delta_{t\cdot s' + t'\cdot s,0}|(t,s);(t',s')\rangle$ & $s\cdot t'/2 + s'\cdot t/2$ \\
    $E_6$ & $\mathbb{Z}_3$ & $12\sum_{t,s} \delta_{2 t\cdot s,0} |(t,s)\rangle $ & $2 (s\cdot t)/3$ \\
    $E_7$ & $\mathbb{Z}_2$ & $-18\sum_{t,s} \delta_{t\cdot s,0} |(t,s)\rangle$ & $s\cdot t/2$
\end{tabular}
\end{center}
\caption{The center group, the SUSY dynamical boundary state $|\chi_\text{W} \rangle_{\widetilde{G}}$, and
the fractional instanton number are presented for each simply-connected Lie group $\widetilde{G}$.}
\label{tab:gaugegroup}
\end{table}

\subsection{Example}

As an illustration, let us compute the Witten indices for $\widetilde{G}=SU(4)$ 
and its cousins for $G=SU(4)/H$ where $H \subset {\cal Z}(\widetilde{G})$.  

The SUSY dynamical boundary state of our interest is 
\begin{align}
    \big| \chi_\text{W} \big\rangle = - 4 \sum_{t,s} \delta_{t\cdot s, 0} 
    \big| (t,s) \big\rangle\ , 
\end{align}
which implies that the Witten index of the ${\cal N}=1$ super Yang-Mills 
theory with the gauge group $SU(4)$ becomes 
\begin{align}
    Z_{SU(4)} \big[ (0,0) \big] = - 4 \ . 
\end{align}
We also argued in \eqref{macroresult} that the Witten indices for $(SU(4)/\mathbb{Z}_4)_k$
($k=0,1,2,3$) are identical 
\begin{align}
     Z_{(SU(4)/\mathbb{Z}_4)_k}\big[ (0,0) \big] = - \sum_{j=0}^{3} \big( \gcd{(j,4)} \big)^3
     = - 74\ .
\end{align}
There are two more topological boundary states associated with partial gauging
\begin{align}
    \big|c=(t_1,s_1), \widetilde{d}=(t_2,s_2) \big\rangle_k 
    = \frac{1}{2^3} \sum_{t'_2,s'_2} (-1)^{t_2\cdot s'_2+s_2\cdot t'_2  -k t'_2\cdot s'_2}
    \big|c=(t_1,s_1), d=(t'_2,s'_2) \big\rangle_k 
    \nonumber 
\end{align}
where $k=0,1$, and the corresponding partition functions become
\begin{align}
    Z_{(SU(4)/\mathbb{Z}_2)_k}\big[c=(t_1,s_1), \widetilde{d}=(t_2,s_2) \big] 
    = - \frac{1}{2} \sum_{t_2',s_2'} (-1)^{t_2\cdot s_2' + s_2 \cdot t_2' - k t_2'\cdot s_2'}\  . 
\end{align}
Hence, the Witten indices for $(SU(4)/\mathbb{Z}_2)_k$ are 
\begin{align}
    Z_{(SU(4)/\mathbb{Z}_2)_k} \big[ (0,0), (0,0) \big]& = 
    - 4 \sum_{s_2'} \delta_{ks'_2,0} = 
    \begin{cases}
        -32 & \text{ for } k=0  \\
        -4 & \text{ for } k=1
    \end{cases}\ . 
\end{align}
%

%%%%%%%%%%%%%%%%%%%%%%%%%%%%%%%%%%%%%%%%%%%%%%%%%%%%%%%%%%%%%%%%%%%%%%%%%%%%%%%%%%%%%%%%%%%
%%%%%%%%%%%%%%%%%%%%%%%%%%%%%%%%%%%%%%%%%%%%%%%%%%%%%%%%%%%%%%%%%%%%%%%%%%%%%%%%%%%%%%%%%%%
%%%%%%%%%%%%%%%%%%%%%%%%%%%%%%%%%%%%%%%%%%%%%%%%%%%%%%%%%%%%%%%%%%%%%%%%%%%%%%%%%%%%%%%%%%%
\section{Lens Space Index}\label{sec:6}

The second application of the SymTFT framework is to compute the lens space index, i.e.,
$L(r,1)\times S^1$ partition function of the ${\cal N}=1$
supersymmetric gauge theories with gauge group $G$. As in the previous Section, 
the gauge group $G$ can be written as $\widetilde{G}/H$ where 
$\widetilde{G}$ is simply-connected and $H$ is a subgroup of the center ${\cal Z}(\widetilde{G})$. 
The computation was carried out for simple cases in previous work \cite{Razamat:2013opa},  
and we revisit it from the SymTFT point of view. 

For simplicity, we consider a system with a vector multiplet for  $\widetilde{G}=SU(N)$ and a chiral multiplet in the $SU(N)$ representation $\rho$. 
The representation $\rho$ is reducible in general. 
To preserve the one-form symmetry associated to $H$, we stress that the matter field must be neutral under $H$. 

When the given gauge theory flows to a superconformal 
theory in the infrared limit, one can define the lens space index as
\begin{align}
     Z_{\widetilde{G}}\big[ (0,0) \big] 
     = \text{Tr}_{{\cal H}(L(r,1))}\Big[ 
     (-1)^F e^{-\beta \{ Q,S \} } q_1^{J^3_L + J^3_R + R/2} q_2^{J^3_R - J^3_L + R/2} \Big]\ ,
\end{align}
where we follow the convention in \cite{Hosomichi:2014hja}. $Q$ and $S$ are supercharges of the  ${\cal N}=1$ superconformal algebra $SU(2,2|1)$
which obey the anti-commutation relation below
\begin{align}
    \{ Q, S \} = D - 2 J_R^3 - \frac32 R\ .     
\end{align}
Here $D$, $R$ denote the dilation and $U(1)_R$ charge
while $J_L^3$ and $J_R^3$ refer to the angular momenta 
of $SO(4)=SU(2)_L \times SU(2)_R$. They are bosonic generators 
of $SU(2,2|1)$. The trace is now performed over the Hilbert space 
of the gauge theory on the lens space $L(r,1)$, denoted by ${\cal H}(L(r,1))$.  
Again, we omit the dependence on $L(r,1)$ below 
unless we need it for clarification. Using the 
supersymmetric localization technique, one can obtain the lens space index \cite{Benini:2011nc} 
which can be expressed schematically as 
\begin{align}\label{lensindex00}
    Z_{\widetilde{G}} \big[ (0,0) \big] = 
    \sum_{  \mathbf{m} } \int 
    d \bm{\vartheta}   
    \hspace*{0.2cm}
    \frac{\Delta_{\mathbf{m}}}{\left| W(\bm{m}) \right|} 
    \prod_{\alpha \in \textrm{roots}} \hspace*{-0.2cm} \mI_V \left( \mathbf{m}(\alpha) ,e^{i \bm{\vartheta}(\alpha)} \right)  \prod_{w \in \rho} \mI_{\chi} \left(\mathbf{m}(w),e^{i \bm{\vartheta}(w)} \right),
\end{align}
where $\mI_V$ and $\mI_\chi$ are the contributions 
to the partition function from the vector 
multiplet and the chiral multiplet.  Their explicit 
expressions are 
\begin{align}\label{lensvector}
    \mI_V \left(m, u \right) =  \left( \frac{1}{1-u^{-1}} \right)^{\delta_{m.0}} 
    \frac{\mI_V^{(0)}}{\Gamma \left[ q_2^{m} u^{-1},q_2^r,q_1q_2 \right]\Gamma \left[ q_1^{r-m} u^{-1},q_1^r,q_1q_2 \right]}   
\end{align}
and
\begin{equation}\label{lenschiral}
    \mI_{\chi}\left( m , u \right) = 
    \mI_{\chi}^{(0)} \left(m,u \right) 
    \Gamma\left[(q_1 q_2)^{\frac{R}{2}} q_2^{r-m} u ,q_2^r,q_1 q_2 \right] \Gamma\left[(q_1 q_2)^{\frac{R}{2}} q_1^m u,q_1^r,q_1 q_2 \right].
\end{equation}
Here $\mathbf{m}$ is a set of integers 
labelling the gauge holonomy $g_1$ along the torsion 
cycle ${\cal C}_\tau$,
\begin{align}\label{holonomyuntwist01}
    g_1 = \exp{\Big[2\pi i \frac{m_a H_a}{r}\Big]} \ ,   
\end{align}
where $\{ H_a\}$ ($a=1,2,..,\text{rk}(\widetilde{G})$) denotes the Cartan subalgebra of $\widetilde{G}$.
The periodic variables $\bm{\vartheta}$ 
parameterize the gauge holonomy $g_2$ along the temporal circle $S^1$,
\begin{align}\label{holonomyuntwist02}
    g_2 = \exp{\Big[i \vartheta_a H_a \Big]}\ . 
\end{align}
Note that two holonomies commute with each other,  
\begin{align}
\begin{split}
    g_1 g_2 & = g_2 g_1 \ , 
    \\
    g_1^r & = 1 \ . 
\end{split}    
\end{align}

To compute the lens space index for the gauge group $G=\widetilde{G}/H$, one has to 
gauge the one-form global symmetry by summing over all possible configurations of 
the two-form gauge fields.  To this end, let us compute the partition function 
of the gauge theory coupled to the background two-form $\mathbb{Z}_N$ gauge field,
\begin{align}
   Z_{\widetilde{G}} \big[ b = ( b_1, b_2) \big]\ . 
\end{align}
where the two-form gauge fields are specified 
by the holonomies $b=(b_1,b_2)$ around the two surfaces $\Gamma_1$ and $\Gamma_2$.

Based on the arguments in Section \ref{sec:twistedbc}, the background gauge field $b=(b_1,b_2)$ can be gauged away, and
the two holonomies $g_1$ and $g_2$ instead satisfy the twisted boundary conditions 
\begin{align}\label{twistedbd02}
\begin{split}
    g_1 g_2 &= \omega^{b_1} g_2 g_1 \ , 
    \\
    g_1^r &  = \omega^{b_2}\ , 
\end{split}
\end{align}
where $\omega=e^{2\pi i /N}$ denotes the $N$-th root of unity. 
On $L(r,1)\times S^1$, we examined in Section \ref{sec:topboundaryLens} that  
$b_1$ has to be a multiple of $N/\gcd{(r,N)}$,
\begin{align}\label{condition01}
    b_1 = \frac{N}{\gcd{(r,N)}} k_1\ , 
\end{align}    
where $k_1=0,1,..,(\gcd{(r,N)}-1)$ while $b_2$ essentially becomes $\mathbb{Z}_{\gcd{(r,N)}}$-valued, 
\begin{align}\label{condition02}
    b_2 \simeq b_2 + \gcd{(r,N)}\ . 
\end{align}

When $b_1=0$ and $b_2 \neq 0$, $g_1,g_2$ commute and they are still solved by \eqref{holonomyuntwist01} and \eqref{holonomyuntwist02} where the integers $m_a$ in \eqref{holonomyuntwist01} will get a fractional shift due to $b_2$. For general case, as shown in the previous Section, one can express 
the general solutions to the first condition in \eqref{twistedbd02} as
\begin{align}\label{soltotwisted02}
\begin{split}
    g_1 & = \Big( C_\frac{N}{\gcd{(b_1,N)}} \Big)^{\frac{b_1}{\gcd{(b_1,N)}}}
    \otimes g_1' 
    \\
    g_2 &= S_{\frac{N}{\gcd{(b_1,N)}}  }\otimes g_2' 
\end{split}\ ,
\end{align}
modulo the $SU(N)$ gauge transformation. 
Here $C_M$ and $S_M$ are the $M$-dimensional clock and shift matrices 
defined in \eqref{clockshift}, and   $g_1', g_2'$ 
are diagonal matrices in $\gcd{(b_1, N)}$ dimensions.  
Since $g_1,g_2$ are elements of $SU(N)$, the determinants of $g_1',g_2'$
should be the $\big(N/\gcd{(b_1,N)}\big)$-th root of unity,  
\begin{align}\label{detcondition01}
   \left( \det{g_{1,2}'} \right)^{\frac{N}{\gcd{(b_1,N)}}} = 1\ . 
\end{align}
Using the $SU(N)$ gauge transformations, 
we can make both $g_1'$ and $g_2'$ have determinant one. 
For now, we choose a gauge such that $g_2'$ only has the determinant one.

The second condition in  \eqref{twistedbd02} constrains 
the possible form of $g_1'$. We  notice that, due to 
\eqref{condition01}, the diagonal matrix $g_1'$ obeys  
the relation below
\begin{align}
    \omega^{b_2} = \Big(\epsilon_{\frac{N}{\gcd{(b_1,N)}}} \Big)^{\frac{rb_1}{\gcd{(b_1,N)}}} {\bf 1}_{\frac{N}{\gcd{(b_1,N)}}}
    \otimes {g_1'}^r \ . 
\end{align}
Hence, one can describe the matrix $g_1'$ as 
\begin{equation}
    g_1' =  \omega^{\tilde b_2/r} 
    \begin{pmatrix}
    e^{\frac{2\pi i}{r} n_1} & 0 &  \cdots & 0 \\
    0 & e^{\frac{2\pi i }{r} n_2} &  \cdots & 0 \\
    \vdots & \vdots  & \ddots & \vdots \\
    0 & 0 &  \cdots & e^{\frac{2\pi i}{r} n_{\gcd{(b_1,N)}} }
    \end{pmatrix}
\end{equation}
with
\begin{align}\label{tildeb2def}
    \tilde b_2   =  
    \begin{cases} 
        b_2 & \text{ when }  \frac{N}{\gcd{(b_1,N)}} \text{ is odd } \ , \\ 
        b_2 - r \frac{b_1}{2}& \text{ otherwise}  \ .
    \end{cases} 
\end{align}
Here $\{ n_{a'} \}$ is a set of integers satisfying \eqref{detcondition01}, i.e.,
\begin{align}\label{conditionforg1'}
    \sum_{a'=1}^{\gcd{(b_1,N)}} \hspace{-0.3cm} n_{a'} + \tilde b_2 \frac{\gcd{(b_1,N)}}{N} = 0 \ \text{ mod } \ \frac{r \gcd{(b_1,N)}}{N}\ .
\end{align}
Note that $\frac{r \gcd{(b_1,N)}}{N}$ is always an integer due to \eqref{condition01}. Since $n_{a'}$ are all integers, the constraint, in particular, implies that a solution to \eqref{twistedbd02} exists only when 
$\tilde b_2$ is a multiple of $N/\gcd{(b_1,N)}$ modulo $r$. 
Clearly, \eqref{conditionforg1'} is compatible with the shift $b_2 \to b_2 + r$ \eqref{condition02}. 
One can also show that each integer $n_{a'}$ 
can be shifted by
\begin{align}
    n_{a'} \longrightarrow n_{a'} + \frac{r \gcd{(b_1,N )}}{N}
\end{align}
under an $SU(N)$ gauge transformation similar to \eqref{gauge-transformation}. Thus, we use the $SU(N)$ gauge transformation to restrict the range of $\{n_{a'}\}$ to 
\begin{align}
    0 \leq n_1 \leq n_2 \leq ... \leq n_{\gcd{(b_1,N )}} < \frac{r \gcd{(b_1,N )}}{N}\ ,
\end{align}
rather than to fix  $\det{g_1'}=1$.  On the other hand, $g_2'$ has determinant one and can be described simply as
\begin{align}\label{g'2}
    g_2' = \exp{\Big[i \theta_{a'} H_{a'} \Big]}\ . 
\end{align}
where $\{ H_{a'}\}$ denotes the Cartan subalgebra of $SU(\gcd{(b_1,N)})$.

Given the solution \eqref{soltotwisted02}, we discuss how to compute the twisted lens space index $Z_{\widetilde{G}}[b]$. 
Let us first recall that the contributions to the untwisted lens space index \eqref{lensindex00}
are given by the action of two commuting holonomies $g_1$ \eqref{holonomyuntwist01} and 
$g_2$ \eqref{holonomyuntwist02} on various fields. 
In particular, one used the weight space decomposition 
of quantum fields where $g_1$ and $g_2$ act on  
each component of weight $w$ as 
$e^{2\pi i \mathbf{m}(w)/r}$ and $e^{i\bm{\vartheta}(w)}$.

In the presence of $b=(b_1,b_2)$, the two holonomies 
\eqref{soltotwisted02} cannot be simultaneously conjugated to the Cartan torus,  
and thus their action is not diagonal in the weight space basis. 
Nonetheless, we can argue that their action becomes diagonal in a certain basis, since the matter fields are in a representation 
$\rho$ where the center $\mathbb{Z}_N$ acts trivially. Let us first restrict to the case $\gcd(b_1,N)=1$ where $g_1 = C_N^{b_1}$ and $g_2 = S_N$. To address a convenient choice of basis, we notice that the clock $C_N$
and the shift $S_N$ generate the same group action on $\rho$ as $\mathbb{Z}_N\times \mathbb{Z}_N$ due to $(C_N)^N=(S_N)^N = 1$. Provided that $C_N$ and $S_N$ act diagonally on the chiral multiplet of the representation $\rho$, 
we can decompose it into irreducible representations of the two cyclic groups,
\begin{align}
    \rho \longrightarrow \bigoplus\limits_{c,s=0}^{N-1} d_\rho(c,s) \times (c,s)
\end{align}
where $(c,s)$ denotes a one-dimensional representation 
of $\mathbb{Z}_N\times\mathbb{Z}_N$ on which 
$C_N$ and $S_N$ become $\omega^c$ and $\omega^s$. Here the coefficient
$d_{\rho}(c,s)$ is the multiplicity of the representation 
$(c,s)$ in the decomposition of $\rho$, and $\sum_{c,s}d_\rho(c,s)=\dim{\rho}$. Notice that for fundamental representation expanded by the basis $\psi_a (a=0,\cdots,N-1)$, the $S_N$ and $C_N$ act on the basis as
    \begin{equation}
        S_N \psi_a = \psi_{a+1},\quad C_N \psi_a = e^{\frac{2\pi i a }{N}} \psi_a,
    \end{equation}
where we assume $\psi_{a+N} \equiv \psi_a$. Other representation can be constructed via tensor products of fundamental representations and from which we can read off the eigenvalues. When $\gcd(b_1,N) \neq 1$ the basis of fundamental representation are labeled by $\psi_{a,i} (a=0,\cdots,\frac{N}{\gcd(b_1,N)}-1, i =1,\cdots ,\gcd(b_1,N))$ such that
    \begin{equation}
        g_1 \psi_{a,i} = \omega^{\tilde{b}_2/r} e^{\frac{2\pi i n_i}{r}} \psi_{a+\frac{b_1}{\gcd(b_1,N)},i},\quad g_2 \psi_{a,i} = e^{\frac{2\pi i \gcd(b_1,N) a}{N} + i \theta(i)} \psi_{a,i}
    \end{equation}
where $\theta(i)$ is the eigenvalue of $g'_2$ in \eqref{g'2}. A general representation $\rho$ one can decomposed according to
\begin{equation}
    \rho \longrightarrow \bigoplus\limits_{c,s=0}^{\frac{N}{\gcd(b_1,N)}-1} \bigoplus\limits_{\lambda(g'_1),\lambda(g'_2)} d_\rho(c,s;\lambda(g'_1),\lambda(g'_2))  \times (c,s;\lambda(g'_1),\lambda(g'_2))
\end{equation}
where $c$ and $s$ are the eigenvalues of $\widetilde{C}_{\frac{N}{\gcd(b_1,N)}}$ and $\widetilde{S}_{\frac{N}{\gcd(b_1,N)}}$, respectively. Similarly, $\lambda(g'_1)$ and $\lambda(g'_2)$ are the eigenvalues of $g'_1$ and $g'_2$. All of them can be read off from the tensor product of fundamental representations.

Based on the above basis, one can decompose 
the chiral multiplet into components labeled by 
the eigenvalues of two holonomies. 
The contribution from each component is then given by $\mI_\chi(m,u)$ \eqref{lenschiral} where $e^{2\pi i m/r}$ and $u$ 
are identified as the eigenvalues of $g_1$ and $g_2$. 

As an illustration, let us consider an ${\cal N}=1$ supersymmetric 
gauge theory with $\widetilde{G}=SU(3)$ placed on $L(3,1)\times S^1$.
The gauge theory is coupled to a chiral multiplet in 
the representation $\rho= \scalebox{0.5}{\yng(3)}$. 
Since $\gcd{(r,N)}=3$, both $b_1$ and $b_2$ are now $\mathbb{Z}_3$-valued. 
When $b_1\neq 0$, the solution to \eqref{twistedbd02} only exists for $b_2=0$ and 
becomes 
\begin{align}\label{holexam01}
    g_1 = (C_3)^{b_1}\ , \quad  g_2 = S_3\ .      
\end{align}
The chiral multiplet can be decomposed into irreducible representations of $\mathbb{Z}_3\times \mathbb{Z}_3$ 
as follows,
\begin{align}
    \rho \longrightarrow \bigoplus\limits_{c,s=0}^{2} d_{\scalebox{0.3}{\yng(3)}}(c,s) \times (c,s)\ ,
\end{align}
where $d_{\scalebox{0.3}{\yng(3)}}(c,s)=1$ for all pairs of $(c,s)$ except $d_{\scalebox{0.3}{\yng(3)}}(0,0)=2$. 
In this basis, the action of $g_1$ and $g_2$ is diagonal, 
\begin{align}
\begin{split}
    g_1 & : (c,s) \to \omega^c (c,s)\ , 
    \\ 
    g_2 & : (c,s) \to \omega^s (c,s)\ , 
\end{split}
\end{align}
where $\omega=e^{2\pi i/3}$ is the third root of unity. 
The contribution from the component $(c,s)$ is therefore $\mI_\chi(m, u)$
with 
\begin{align}
    e^{ 2\pi i \frac{m}{3}} & = \omega^{b_1 c}    \  , 
    \quad     u=\omega^s \ . 
\end{align}
Similarly, one can show that the contribution from the vector 
multiplet is given by 
\begin{align}
    \prod\limits_{c,s=0}^2 \Big[ \mI_V( b_1 c, \omega^s) \Big]^{d_{\scalebox{0.3}{\yng(2,1)}}(c,s)} 
\end{align}
where $d_{\scalebox{0.3}{\yng(2,1)}}(c,s)=1$ for all pairs of $(c,s)$ except $d_{\scalebox{0.3}{\yng(2,1)}}(0,0)=0$. 
Combining all the results, the twisted lens space index for $b=(b_1\neq0,b_2)$ 
takes the following form 
\begin{align}
    Z_{SU(3)}\big[ (b_1,b_2) \big] &= 
        \begin{cases}
        \prod\limits_{c,s=0}^2 \Big[ \mI_V( b_1 c, \omega^s) \Big]^{d_{\scalebox{0.3}{\yng(2,1)}}(c,s)} 
        \prod\limits_{c,s=0}^2 \Big[ \mI_\chi( b_1 c , \omega^s) \Big]^{d_{\scalebox{0.3}{\yng(3)}}(c,s)} & \text{ for } b_2=0  
        \\[15pt]
        ~ 0 & \text{ for } b_2 \neq 0
    \end{cases} \nonumber
\end{align}
Note that there is neither an integral over $\bm{\vartheta}$ nor a sum over $\bm{m}$, 
since the gauge holonomies $g_1$ and $g_2$ are completely fixed by \eqref{holexam01}. 

Once we have all the twisted lens space indices, the SUSY dynamical boundary state 
can be described as 
\begin{align}
     \big| \chi_\text{L} \big\rangle_{SU(N)} = \sum_{b_1,b_2}
     Z_{SU(N)} \big[ (b_1,b_2) \big] \big|  b_1, b_2 \big\rangle  \ , 
\end{align}
where two $\mathbb{Z}_N$ holonomies $b_1$ and $b_2$ are constrained by \eqref{condition01}
and \eqref{condition02}. 

We now proceed to compute the lens space index for $G=\big( SU(N)/\mathbb{Z}_N\big)_k$.  
The supersymmetric gauge theory with $G$ is related to the theory with $SU(N)$ 
via the $SL(2,\mathbb{Z})$ transformation $ST^k$. It leads to  
\begin{align}
   \begin{split}
    Z_G\big[ (b_1,b_2) \big]  & = 
    \big\langle (b_1,b_2) \big| V_{ST^k}^\dagger \big| \chi_\text{L} \big \rangle_{SU(N)}  
    \\
    & = \frac{1}{\gcd{(r,N)}} \sum_{b_1',b_2'} \omega^{-b_1b_2' - b_2 b_1' + \frac{k}{2}\mathfrak{P}(b')}
    Z_{SU(N)}\big[ (b_1', b_2') \big]\ , 
\end{split} 
\end{align}
where $b_i$ and $b_i'$ obey \eqref{condition01} and \eqref{condition02}. 
Here the Pontryagin square $\mathfrak{P}(b)$ 
for $L(r,1)\times S^1$ is given by 
\begin{align}
    \mathfrak{P}(b) = 
    \begin{cases} 
        b_1 (2 b_2 - r b_1) \text{ mod } 2N  & \text{ when } N \text{ is even\ ,} \\
        2 b_1 b_2  \text{ mod } 2N & \text{ when } N \text{ is odd\ .} 
    \end{cases}
\end{align} 
It is noteworthy that $\mathfrak{P}(b)$ remains intact under the shift $b_2 \to b_2+r$:
due to \eqref{condition01}, 
\begin{align}
\begin{split}
    \mathfrak{P}(b') - \mathfrak{P}(b) & = 2 b_1 r 
    \\ & = 
    2 \lcm{(r,N)} k_1 = 0  \text{ mod } 2N
\end{split}
\end{align}
where $b=(b_1,b_2)$ and $b'=(b_1,b_2+r)$.  For the detailed derivation of 
the Pontryagin square on $L(r,1)\times S^1$, we refer readers to Section \ref{sec:3}. 
Interestingly, we will soon demonstrate that the Pontryagin square $\mathfrak{P}(b)$ vanishes 
if and only if a solution to \eqref{twistedbd02} exists. In other words, the twisted 
lens space index $Z_{\widetilde G}[b]$ is identically zero whenever $\mathfrak{P}(b)\neq 0$.  
Thus, we can conclude that the lens space index for $G=\big( SU(N)/\mathbb{Z}_N\big)_k$ 
is independent of the value of $k$. 

Why does $\mathfrak{P}(b)$ vanish if and only if a solution to \eqref{twistedbd02} exists? 
We begin by considering the case where $N$ is even. 
Under this assumption, one can say that the Pontryagin square vanishes if and only if either $b_1=0$ or 
\begin{align}\label{eqn01}
    b_2 - \frac{rb_1}{2} = 0 \text{ mod } \frac{N}{\gcd{(b_1,N)}} \ .     
\end{align}
Since it is evident that \eqref{twistedbd02} has a solution for $b_1=0$, 
let us focus on the condition \eqref{eqn01}. \eqref{eqn01} is actually 
identical to the condition that ensures the existence of a solution,
stated below the equation \eqref{conditionforg1'},  
\begin{align}\label{eqn03}
    \tilde b_2 = 0 \text{ mod } \frac{N}{\gcd{(b_1,N)}}\ ,
\end{align}
where $\tilde b_2$ is defined in \eqref{tildeb2def}. 
They may look different when $N/\gcd{(b_1,N)}$ is odd. However the discrepancy is illusory as we shall 
demonstrate: Given $N$ is even and $N/\gcd{(b_1,N)}$ is odd, $b_1$ has to be even. 
Therefore, $b_2 - \frac{rb_1}{2} \simeq b_2$ due to \eqref{condition02}. 
On the other hand, $\mathfrak{P}(b)$ equals zero modulo $2N$ for odd $N$
if and only if either $b_1=0$ or $b_2 = 0$ modulo $N/\gcd{(b_1,N)}$. 
The former case always guarantees the solution exists. The latter condition
is again the same with \eqref{eqn03}. This is because $N/\gcd{(b_1,N)}$ 
should be odd and $\tilde b_2=b_2$.  
\subsection*{$USp(2N)$ group}
Before ending this section, we will also present the example of unitary symplectic group $USp(2N)$ which is not discussed in \cite{Razamat:2013opa}. It is convenient to choose the symplectic invariant tensor as
    \begin{equation}
        \Omega = \epsilon \oplus \epsilon \oplus \cdots \oplus \epsilon = \epsilon \otimes I_{N\times N}
    \end{equation}
and the group elements $g$ are $2N \times 2N$ matrix satisfying
\begin{equation}
    g^T \Omega g = \Omega\ .
\end{equation}
The center of $USp(2N)$ is $\mathbb{Z}_2$ and $b_1,b_2=0,1$. When $b_1=0$, $g_1,g_2$ take values in Cartan torus and are given by \eqref{holonomyuntwist01} and \eqref{holonomyuntwist02}. When $b_1=1$, we can embed a pair of $USp(2)=SU(2)$ configuration $C_2,S_2$ into $USp(2N)$ group according to
    \begin{equation}
        g_1 = C_2 \otimes g'_1\ ,\quad g_2 = S_2 \otimes g'_2\ ,
    \end{equation}
where $S_2,C_2$ are $SU(2)$ matrices satisfying $S_2 C_2 = - C_2 S_2$, $g'_1,g'_2$ are $O(N)$ matrices commuting with each other. Since $O(N)$ has two disjoint pieces, there are two solution for $g'_2$ (and $g'_1$). For the first solution $g'_2$ is connected to the identity of $O(N)$ and lying in the Cartan torus of $SO(N)$ 
    \begin{equation}
        g'_2 = \left(\begin{array}{ccc}
            \Lambda_{\theta_1}&&\\
            &\Lambda_{\theta_2}&\\
            &&\ddots
        \end{array} \right)\ ,
    \end{equation}
where $\Lambda_{\theta}$ is the $2\times 2$ rotation matrix with angle $\theta$. 
For the second solution, $g'_2$ is connected to the reflection element $\textrm{diag}(-1,1,1,\cdots,1)$ in $O(N)$ and lying in the Cartan torus of $SO(N-1)$
    \begin{equation}
        g'_2 = \left( \begin{array}{cccc}
            -1&&&\\
            &\Lambda_{\theta_1}&&\\
            &&\Lambda_{\theta_2}&\\
            &&&\ddots
        \end{array}\right)\ .
    \end{equation}
However, one can apply a $USp(2N)$ gauge transformation given by
    \begin{equation}
        U = \left(\begin{array}{cc}
            C_2 &  \\
             & \mathbf{1}_{2N-2}
        \end{array} \right)
    \end{equation}
such that $g'_1$ is left invariant and sign of the first element in $g'_2$ is flipped. Therefore the second solution is actually equivalent to the first one up to gauge transformation. 

Now $g'_2$ can still be described by $g_2' = \exp{\Big[i \theta_{a'} H_{a'} \Big]}$ where $\{ H_{a'}\}$ denotes the Cartan subalgebra of $SO(N)$. The second condition in \eqref{twistedbd02} constrains the possible form of $g'_1$. We have
    \begin{equation}
        (-1)^{b_2} = i^r \mathbf{1}_{2} \otimes {g_1'}^r\ .
    \end{equation}
When $N$ is even, we can adjust $g'_1$ such that 
    \begin{equation}
        g'_1 = \left( \begin{array}{cccc}
            \Lambda_{\theta'_1} & & &\\
            & \Lambda_{\theta'_2} & &\\
            & & \ddots &\\
            & & & \Lambda_{\theta'_{N/2}}
        \end{array}\right)
    \end{equation}
and $\theta'_{a'}$ are constrained by
    \begin{equation}
        b_2 \pi = r \theta'_{a'} + \frac{r}{2}\pi \mod 2\pi
    \end{equation}
where $r\theta'_a$ must be an integer multiple of $\pi$ in order the matrix to be diagonal. When $r$ is odd there is no solution and we only need to consider even $r$. We can solve $\theta'_{a'}$ as
    \begin{equation}
        \theta'_{a'} = (\frac{b_2}{r}-\frac{1}{2})\pi \mod \frac{2\pi}{r}
    \end{equation}
which is also compatible with $b_2 \rightarrow b_2 + 2$ \eqref{condition02}. On the other hand, when $N$ is odd we can write $g'_1$ similarly as
    \begin{equation}
        g'_1 = \left( \begin{array}{ccccc}
            1&&&&\\
            &\Lambda_{\theta'_1} & & &\\
            && \Lambda_{\theta'_2} & &\\
            & & & \ddots &\\
            & & & & \Lambda_{\theta'_{(N-1)/2}}
        \end{array}\right)\ .
    \end{equation}
Due to the first element, $r\theta'_{a'}$ must be an integer multiple of $2\pi$ and one finds $b_2$ is restricted to
    \begin{equation}
        b_2 = \frac{r}{2} \mod 2\ ,
    \end{equation}
and $\theta'_{a'}$ are solved by
    \begin{equation}
        \theta'_{a'} = 0 \mod \frac{2\pi}{r}\ .
    \end{equation}
    
In both cases one can also show that each angle $\theta'_{a'}$ can be shifted by
    \begin{equation}
        \theta'_{a'} \rightarrow \theta'_{a'}+\pi
    \end{equation}
under an $Sp(N)$ gauge transformation similar to \eqref{gauge-transformation}. Thus, we can do a gauge transformation to restrict the range of $\theta'_{a'}$ to
    \begin{equation}
        0 \leq \theta'_1 \leq \theta'_2 \leq \cdots \leq \theta'_{\frac{N}{2}/\frac{N-1}{2}} < \pi.
    \end{equation}

When $b_1\neq 0$, for any representations $\rho$ invariant under the center $\mathbb{Z}_2$ of $USp(2N)$, we need to decompose it as
\begin{equation}
    \rho \longrightarrow \bigoplus\limits_{c,s=0,1} \bigoplus\limits_{\lambda(g'_1),\lambda(g'_2)} d_\rho(c,s;\lambda(g'_1),\lambda(g'_2))  \times (c,s;\lambda(g'_1),\lambda(g'_2))
\end{equation}
where $c$ and $s$ are the eigenvalues of $C_{2}$ and $S_{2}$, respectively. Similarly, $\lambda(g'_1)$ and $\lambda(g'_2)$ are the eigenvalues of $g'_1$ and $g'_2$. Then the lens space index is written in the same way as the $SU(N)$ case.

\section{Donaldson-Witten and Vafa-Witten Partition Functions}\label{sec:7}

Donaldson invariants are diffeomorphism invariants of four manifolds and can be nicely incorporated into the framework of SymTFT. Historically, Witten \cite{Witten:1988ze} first discovered a QFT interpretation of Donaldson invariants using topological twisted 4d $\mathcal{N} = 2$ pure $SU(2)$ supersymmetric Yang-Mills theory. More precisely, the Donaldson invariants can be understood as correlation functions of certain cohomologically nontrivial observables. Later, the ground-breaking works of Seiberg and Witten \cite{Seiberg:1994rs,Seiberg:1994aj} teach us how to understand the IR dynamics of those theories. Then it was realized in \cite{Witten:1994cg} that after topological twisting in the presence of matter, the low energy effective field theory can also give rise to four-manifold invariants, which are conjectured to be equivalent to the Donaldson invariants.

\begin{figure}
    \hspace{-0.5cm}
    \includegraphics[scale=0.7]{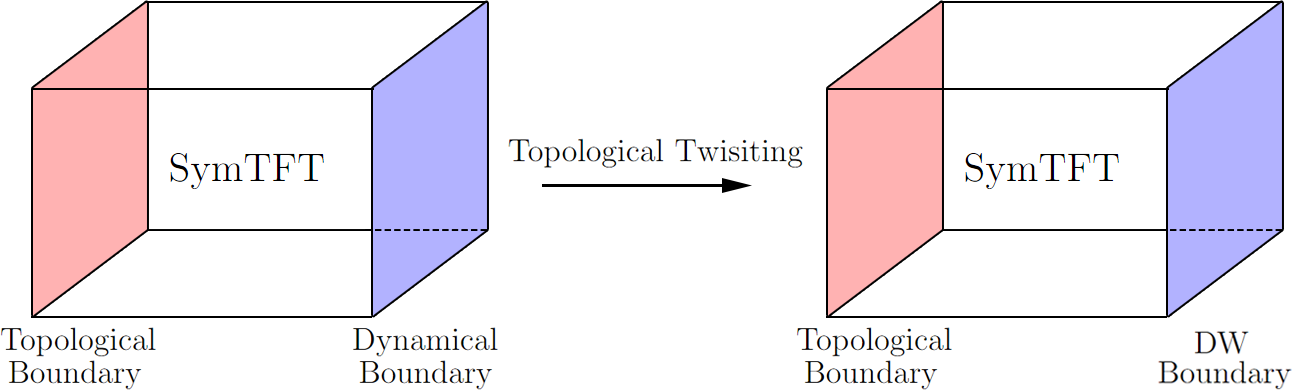}
    \caption{At the level of the SymTFT, topological twisting amounts to changing the dynamical boundary to the DW boundary.}
    \label{figure:twisting}
\end{figure}

From the SymTFT point of view, the crucial point of the topological twisting is that global symmetries of our theory remain intact. Therefore, we can effectively think of it as only changing the dynamical boundary to the DW boundary while keeping the topological boundary fixed. Pictorially, we have Figure \ref{figure:twisting}.

With the picture in mind, let us work out the detail. Consider a smooth, oriented, and compact four-manifold $M_4$. Let $b_2^+(M_4)$ be the rank of the positive definite subspace of $H^2(M_4, \mathbb{Z})$ under the Poincaré pairing. Now we put the topological twisted theory on $M_4$. It turns out that for each class in $H^0(M_4, \mathbb{Z})$ denoted as $[pt]$ and $H^2(M_4, \mathbb{Z})$ denoted as $[S]$, we can associate observables that are fields integrate over them. Moreover, we can also turn on non-trivial background for the one-form symmetry, denoted abstractly as $b$. Then the DW partition is nothing but the path integral in the topological field theory with certain choice of observables and background $Z_{DW}([pt], [S], b)$. Hence the DW boundary state $| \chi_\text{DW} \rangle$ can be written as
\be
    \big | \chi_\text{DW}; [pt], [S] \big \rangle = \sum_{b} Z_\text{DW}([pt], [S], b)\, \big|  b \big\rangle\ .
\ee

On the other hand, if we gauge the one-form symmetry, we just need to change the topological boundary state and obtain for free the DW partition functions for the gauged theory. We emphasis that the relationship between them is universal and purely determined by the symmetry, just in the case of ordinary partition function.

Now let us consider a concrete example. Take gauge group to be $SU(2)$ and matter in the adjoint representation. It is clear that there is a $\bZ_2$ one-form symmetry given by the center of $SU(2)$, and we assume it is non-anomalous. We put the theory on the complex projective space $\mathbb{CP}^2$. Reference \cite{Manschot:2021qqe} meticulously works out the whole procedure, to which readers can refer for all the details. As discussed before, if one wants to gauge the center symmetry and obtain the DW partition function for $SO(3)$ gauge group, there are two different ways. Namely, we can choose to perform a $T$ transformation before gauging, resulting in $SO(3)_+$ or $SO(3)_-$ gauge group. Their difference lies in an additional phase given by one-half of the Pontryagin square $\mathfrak{P}$,  
\be
\begin{aligned}\label{DWSO(3)}
{\widehat Z}_\text{DW}^{SO(3)_+}([pt], [S],b^\prime) &= - \frac{1}{\sqrt 2}\sum_{b} e^{-\pi i\langle b', b \rangle } {\widehat Z}_\text{DW}^{SU(2)} ([pt], [S],b)\ ,\\
{\widehat Z}_\text{DW}^{SO(3)_-}([pt], [S], b') &= - \frac{e^{i\pi/4}}{\sqrt 2}\sum_{b} e^{-\pi i \langle b', b \rangle + \frac{\pi i}{2} \mathfrak{P}(b)}  {\widehat Z}_\text{DW}^{SU(2)} ([pt], [S], b)\ .
\end{aligned}
\ee
In order to simplify the relations  \eqref{DWSO(3)}, we utilize the eta function $\eta(\tau)$ to normalize 
the DW partition functions ${\widehat Z}_\text{DW}$ to be of weight $0$ under $SL(2,\mathbb{Z})$. Note also that, for \eqref{DWSO(3)}, 
we used \eqref{CP2VTVS} in order to have
consistent $SL(2,\mathbb{Z})$ transformations
on the non-spin manifold $\mathbb{CP}^2$.
For $\mathbb{CP}^2$, $H^2(\mathbb{CP}^2, \bZ)$ is clearly 
 torsion free. So the phase involving the Pontryagin square is
\be
\exp\big(-\frac{\pi i}{2} \mathfrak{P}(b)\big) = \exp\Big(-\frac{\pi i}{2} \int_{\mathbb{CP}^2} b^\prime \cup b^\prime\Big)\ ,
\ee
where $b'$ is a lift to integral 2-cocycle and the whole expression is independent of the lifting. The effect of $SL(2,\bZ)$ transformation on the DW partition function was studied in Section 5.2 of \cite{Manschot:2021qqe} (see in particular the last part of that section). The presence of Pontryagin square in $Z_\text{DW}^{SO(3)_-}$ is shown explicitly in Section 6 of \cite{Manschot:2021qqe}. 

Furthermore, if we take the mass of the adjoint matter to zero, supersymmetry enhances from $\mathcal{N} = 2$ to $\mathcal{N} = 4$. The topological twisting then becomes the VW twist, and the resulting VW partition function enjoys $SL(2,\bZ)$ modularity, reminiscent of the $SL(2,\bZ)$ duality in $\mathcal{N} = 4$ super Yang-Mills theory. It is still labelled by the $\bZ_2$ one-form symmetry background $b$, but there is no dependence on $H^0(M_4, \bZ)$ or $H^2(M_4, \bZ)$ any more. For $M_4 = \mathbb{CP}^2$, we have naively,
\begin{align}\label{eq.VW}
{\widehat Z}_\text{VW}^{SU(2)}\big[ b \big](\tau) \   ``=" \  \frac{g_b(\tau)}{\eta^3(\tau)}\ ,
\end{align}
where $\eta(\tau)$ denotes the Dedekind eta function and $g_b(\tau)$ is the generating function of Hurwitz class numbers $H(4n - b)$ for $b = 0, 1$,
\be
g_b(\tau) = 3 \sum_{n \geq 0}  H(4n -b) q^{n-b/4}.
\ee
However this cannot be the final answer. It is known that VW partition suffers a holomorphic anomaly \cite{Vafa:1994tf}, and here indeed $g_b(\tau)$ is known to be a weight $3/2$ mock modular form which has a non-holomorphic but modular completion \cite{ZAGIER1974-1975}. We give the first few terms of $g_b$,
\be
\begin{aligned}
g_{0}(q) &= -\frac{1}{4} + \frac{3q}{2} + 3 q^2 + 4 q^3 + \frac{9q^4}{2} + \cdots\\
g_{1}(q) &= q^{3/4} + 3 q^{7/4} + 3 q^{11/4} + 6q^{15/4} + \cdots\ .\\
\end{aligned}
\ee

The modular completion of $g_b(\tau)$ requires a non-holomorphic shadow,
\begin{equation}
    \widehat{g}_b(\tau, \bar{\tau}) = g_b(\tau) - \frac{3 i}{4\sqrt{2} \pi} \int_{-\bar{\tau}}^{i \infty} \frac{\Theta_b(y)}{(-i(y+\tau)^\frac{3}{2})} dy\ ,
\end{equation}
where $\Theta_b(y)$ is the theta function
\begin{equation}
    \Theta_b(y) = \sum_{k \in \bZ + b/2} q^{k^2}\ .
\end{equation}
For the modular invariant partition function, we should replace $g_b(\tau)$ by $\widehat{g}_b(\tau)$ in equation \eqref{eq.VW},
\begin{align}\label{vwsu(2)}
    {\widehat Z}_\text{VW}^{SU(2)}\big[ b \big](\tau,\bar \tau)  =   \frac{\widehat g_b(\tau,\bar \tau)}{\eta^3(\tau)}\ .
\end{align}
Note that \eqref{vwsu(2)} 
is normalized to be of weight $0$ under the modular
transformation that differs  from the conventional VW partition 
function \cite{Vafa:1994tf,Manschot:2021qqe} by $1/\eta^3(\tau)$ factor, i.e., ${\widehat Z}_\text{VW}(\tau,\bar \tau) = \eta^3(\tau) Z_\text{VW}(\tau,\bar \tau)$.

Based on the SymTFT picture, after gauging the one-form center symmetry, only the topological boundary state is changed and we obtain for free the partition function of the gauged theory. As discussed in Section \ref{sec:4}, depending on whether we perform a $T$ transformation on topological boundary states,  
we result in $SO(3)_+$ or $SO(3)_-$ gauge group. 
In particular, the formula \eqref{DWSO(3)} becomes 
\begin{align}
\begin{split}
    {\widehat Z}_\text{VW}^{SO(3)_+}\big[ b' \big](\tau,\bar \tau)  &= - \frac{1}{\sqrt2}\sum_{b=0,1} (-1)^{ b' b  } {\widehat Z}_\text{VW}^{SU(2)} \big[ b\big](\tau,\bar\tau),\\
    & = - \frac{1}{\sqrt 2} \left( 
    \frac{\widehat g_0(\tau,\bar \tau) + (-1)^{b'} \widehat g_1(\tau,\bar \tau) }{\eta^3(\tau)}\right) \ ,
\end{split}
\end{align}
and 
\begin{align}
\begin{split}
    {\widehat Z}_\text{VW}^{SO(3)_-}\big[ b' \big](\tau,\bar \tau) &= - \frac{e^{i\pi/4}}{\sqrt2}\sum_{b=0,1}  (-1)^{ b' b  }  e^{ \frac{\pi i}{2} b^2}  {\widehat Z}_\text{VW}^{SU(2)}  \big[ b\big](\tau,\bar \tau), \\
    & =- \frac{e^{i \pi/4}}{\sqrt 2} \left( 
    \frac{\widehat g_0(\tau,\bar \tau) + i (-1)^{b'} \widehat g_1(\tau,\bar \tau) }{\eta^3(\tau)}\right)\ . 
\end{split}
\end{align}
Here we emphasize again that \eqref{CP2VTVS} is used to have
consistent $SL(2,\mathbb{Z})$ transformations 
on the non-spin manifold $\mathbb{CP}^2$.

Indeed, using the modular properties of 
$\widehat{g}_b(\tau,\bar \tau)$, one can show the 
$SL(2,\mathbb{Z})$ duality symmetry of ${\cal N}=4$
super Yang-Mills theory:
\begin{align}
\begin{split}
    {\widehat Z}_\text{VW}^{SU(2)}\big[ b\big](\tau,\bar \tau)
    & = {\widehat Z}_\text{VW}^{SO(3)_+}\big[b](-\frac{1}{\tau}, -\frac{1}{\bar\tau})
    \\
    {\widehat Z}_\text{VW}^{SO(3)_+}\big[ b\big](\tau,\bar \tau) 
    & = {\widehat Z}_\text{VW}^{SO(3)_-}\big[ b\big](\tau+1,\bar \tau+1)\ . 
\end{split}
\end{align}
In terms of the original VW partition functions, one has 
\begin{align}
\begin{split}
    Z_\text{VW}^{SU(2)}\big[ b\big](\tau,\bar \tau)
    & = ( -i\tau )^{3/2} ~ Z_\text{VW}^{SO(3)_+}\big[ b\big](-\frac{1}{\tau}, -\frac{1}{\bar\tau})\ , 
    \\
    Z_\text{VW}^{SO(3)_+}\big[ b\big](\tau,\bar \tau) 
    & = e^{\pi i /4 }~Z_\text{VW}^{SO(3)_-}\big[ b\big](\tau+1,\bar \tau+1)\ , 
\end{split}
\end{align}
which agree perfectly with the transformation rules in 
\cite{Vafa:1994tf}.

\subsection*{Acknowledgements}
We would like to thank Matthew Buican, Masazumi Honda, Kimyeong Lee, Jan Manschot, David Tong, Piljin Yi, and Yunqin Zheng for helpful discussions. SL would like to thank the organizers of ``KAIST Workshop on Aspects of Quantum Field Theory" 
for the opportunity to present the current work at KAIST.  The work of ZD is supported by an STFC Consolidated Grant, ST$\backslash$T000686$\backslash$1 ``Amplitudes, strings \& duality". No new data were generated or analysed during this study. SL is supported by KIAS Grant PG056502. QJ is supported by KIAS Grant PG080802 and the National Research Foundation of Korea (NRF) Grant No. RS-2024-00405629.

\newpage

\appendix

\section{Condensation Defects}\label{app:A}
The $SL(2,\mathbb{Z})$ symmetry in the bulk SymTFT can be described by codimension-one symmetry defects.  It was shown in~\cite{Fuchs:2012dt,Roumpedakis:2022aik,Kaidi:2022cpf} that in a $(d+1)$-dimensional TFT, such kind of symmetry defects $\mathcal{D}$ extending along a co-dimension one hypersurface $M_{d}$ are built by condensing certain types of topological defects $\mathcal{L}$ along $M_{d}$. If the topological defects $\mathcal{L}$ generate a $q$-form symmetry inside $M_{d}$, the condensation defect $\mathcal{D}$ is equivalently understood as gauging the $q$-form symmetry inside $M_{d}$ which is referred to as $1$-gauging of the $q$-form symmetry \cite{Roumpedakis:2022aik}. In this appendix, we will present the detailed construction of the $SL(2,\mathbb{Z})$ symmetry defect using topological boundary states.

We will then consider the consequence of a generic $SL(2,\mathbb{Z}_N)$ action on the topological boundary states $|(t,s)\rangle$ of $T^4$.
Consider an $SL(2,\mathbb{Z}_N)$ transformation $\Lambda = \begin{pmatrix}
   a & b \\
    c & d 
\end{pmatrix}$
with $ad-bc=1$. It is defined such that the two-form fields $B$ and $\widetilde{B}$ are transformed according to
    \begin{equation}
        \left(\begin{array}{c}
            \widetilde{B}\\B
        \end{array} \right) \rightarrow \left( \begin{array}{cc}
            a & b\\c & d
        \end{array} \right) \left(\begin{array}{c}
           \widetilde{B}\\B
        \end{array} \right)\ .
    \end{equation}
Recall that a generic surface operator $S_{(e,m)}$ is defined as
    \begin{equation}
        S_{(e,m)}[\Gamma] \equiv \exp \left[i \oint_{\Gamma} \left(\begin{array}{cc}
            m & e
        \end{array} \right) \left(\begin{array}{c}
            \widetilde{B}\\B
        \end{array} \right) \right] = \exp \left[i \oint_{\Gamma} e B + m \widetilde{B} \right]\ ,
    \end{equation}
therefore the electric/magnetic charges are transformed effectively as
    \begin{equation}
        \left( \begin{array}{c}
             e\\m  \end{array} \right) \rightarrow  \left(\begin{array}{cc}
            d & b \\
            c & a
        \end{array} \right) \left( \begin{array}{c}
            e \\m  \end{array} \right)\ .
    \end{equation}

This means a general surface operator $S_{(e,m)}$ will be transformed by $V_{\Lambda}$ to $S_{(de+bm,ce+am)}$. One can also check the transformation is consistent with the decomposition
    \begin{equation}
        S_{(e,m)}[\Gamma] = \exp\left[ -em\frac{\pi i}{N} \mathfrak{P}(\Gamma) \right] S_{(0,m)} [\Gamma] S_{(e,0)}[\Gamma]\ .
    \end{equation}

Beginning with the "position" states $|(t,s)\rangle$, a general $SL(2,\mathbb{Z}_N)$ transformation will map it to other topological boundary states as
    \begin{equation}
        |(t,s)\rangle \xrightarrow{V_{\Lambda}} V_{\Lambda}|(t,s)\rangle\ .
    \end{equation}
They are diagonalized by operators $S_{(d,c)}[\Gamma]$
    \begin{equation}\label{5DBF-generic-eigenstate-1}
        S_{(d,c)}[\Gamma]  V_{\Lambda} |(t,s)\rangle = V_{\Lambda} S_{(1,0)}[\Gamma] |(t,s)\rangle = \omega^{p\cdot t + q\cdot s}V_{\Lambda} |(t,s)\rangle\ ,
    \end{equation}
and $S_{(b,a)}[\Gamma]$ will become the raising operators
    \begin{equation}\label{5DBF-generic-eigenstate-2}
        S_{(b,a)}[\Gamma]V_{\Lambda}|(t,s)\rangle = V_{\Lambda}S_{(1,0)}[\Gamma]|(t,s)\rangle = V_{\Lambda}|(t-q,s-p)\rangle\ .
    \end{equation}
For example, we have seen the $S$-transformation maps the "position" states $|(t,s)\rangle$ into the "momentum" states $|\tilde{t},\tilde{s}\rangle$,
and the $T$-transformation will stack a phase $\omega^{-t \cdot s}$.
Note that the Pontryagin square on $T^4$ is simply the cup product.

\subsubsection*{$SL(2,\mathbb{Z}_N)$ orbits on $|(t,s)\rangle$}
In the following, we will construct the topological boundary states $V_{\Lambda} |(t,s)\rangle$ explicitly by specifying the action of $V_{\Lambda}$ on boundary states. First, let us focus on the vacuum $V_{\Lambda}|(0,0)\rangle_{I}$ of the dual state. Consider the operator
    \begin{equation}
        P_{\Lambda} = \frac{1}{N^6} \sum_{\Gamma \in H_2(T^4,\mathbb{Z}_N)} S_{(d,c)}[\Gamma]
    \end{equation}
which satisfies $P_{\Lambda}^{\dagger} = P_{\Lambda}$, $P_{\Lambda}^2 = P_{\Lambda}$ and $S_{(d,c)}[\Gamma] P_{\Lambda} = P_{\Lambda} S_{(d,c)} [\Gamma] = P_{\Lambda}$. By definition, the state $P_{\Lambda}|(0,0)\rangle$ satisfies\footnote{In principle, beginning with any state $|b_0\rangle$ one can apply the projection operator $P_{\Lambda}$ on it. However, one might have $P_{\Lambda} |b_0\rangle = 0$ for some $b_0$ and it satisfies the eigenstate equation trivially. To avoid that, it is sufficient to simply choose $b_0=(0,0)$.}
    \begin{equation}
        S_{(d,c)}[\Gamma] P_{\Lambda}|(0,0)\rangle = P_{\Lambda}|(0,0)\rangle\ ,
    \end{equation}
therefore $P_{\Lambda}|(0,0)\rangle$ is proportional to the vacuum $V_{\Lambda} |(0,0)\rangle$ and we write
    \begin{equation}
        V_{\Lambda} |(0,0)\rangle = \mathcal{N}_{\Lambda} P_{\Lambda}|(0,0)\rangle\ ,\quad \mathcal{N}_{\Lambda} = \frac{N^3}{\gcd(N,c)^3}\ ,
    \end{equation}
where $\mathcal{N}_{\Lambda}$ is chosen to ensure $V_{\Lambda}|(0,0)\rangle|$ have unit norm. Other states $V_{\Lambda}|(t,s)\rangle$ are obtained by acting the raising operator $S_{(b,a)}[\Gamma]$
\begin{align}\label{5DBF-t-s-Lambda-expression}
        V_{\Lambda}|(t,s)\rangle \equiv & S_{(b,a)}[\Gamma_{-s,-t}] \mathcal{N}_{\Lambda} P_{\Lambda} |(0,0)\rangle\nonumber\\
        =& S_{(b,a)}[\Gamma_{-s,-t}] \frac{\mathcal{N}_{\Lambda}}{N^6} \sum_{t',s'} S_{(d,c)}[\Gamma_{t',s'}] |(0,0)\rangle\nonumber\\
        =& S_{(b,a)}[\Gamma_{-s,-t}] \frac{\mathcal{N}_{\Lambda}}{N^6} \omega^{-dc t'\cdot s'}  \sum_{t',s'} \widetilde{U}^c[\Gamma_{t',s'}] U^d[\Gamma_{t',s'}]|(0,0)\rangle\nonumber\\
        =&\frac{\mathcal{N}_{\Lambda}}{N^6}\sum_{t',s'} \omega^{-dc t'\cdot s' -ba t\cdot s} \widetilde{U}^a[\Gamma_{-s,-t}] U^b[\Gamma_{-s,-t}] |(c t', c s')\rangle\nonumber\\
        =&\frac{\mathcal{N}_{\Lambda}}{N^6} \sum_{t',s'} \omega^{-dc t'\cdot s' -ba t\cdot s - bc (s\cdot t'+s'\cdot t)}|(c t'+at,cs'+as)\rangle\ .
    \end{align}
It is easy to deduce that different states $V_{\Lambda}|(t,s)\rangle$ are orthogonal to each other, which also implies $V_{\Lambda}$ is unitary. 

Let us consider the special cases $\Lambda=S,T$ to illustrate the construction. For $S$-transformation, we have $a=d=0,b=-1,c=1$ so the projection operator and the normalization factor are
    \begin{equation}
        P_{S} = \frac{1}{N^6} \sum_{p,q} \widetilde{U}[\Gamma_{p,q}]\ ,\quad \mathcal{N}_S = N^3\ .
    \end{equation}
Therefore we have the vacuum
    \begin{equation}
        V_S|0\rangle = \frac{1}{N^3} \sum_{p,q}\widetilde{U}[\Gamma_{p,q}]|0,0\rangle = \frac{1}{N^3} \sum_{p,q}|(-q,-p)\rangle = \frac{1}{N^3} \sum_{t,s}|(t,s)\rangle\ . 
    \end{equation}
This can also be understood as a condensation of $\widetilde{U}$-defects along the boundary. Other states are raised by acting $S_{(-1,0)}[\Gamma] = U^{-1}[\Gamma]$ and the generic state $V_S|b\rangle$ can be read from \eqref{5DBF-t-s-Lambda-expression} as
    \begin{equation}
        V_S|(t,s)\rangle = \frac{1}{N^3} \sum_{t',s'} \omega^{s\cdot t' + t \cdot s'} |(t',s')\rangle
    \end{equation}
which is exactly the same as the "momentum" basis $|(\tilde{t},\tilde{s})\rangle$ discussed before with $\tilde{t}=t,\tilde{s}=s$.

For $T$-transformation we have $a=b=d=1,c=0$, so the projection operator and the normalization factor are
\begin{equation}
    P_T = \frac{1}{N^6}\sum_{p,q}U[\Gamma_{p,q}]\ , \quad \mathcal{N}_T=1\ .
\end{equation}
The operator $P_{T}$ acts trivially on $|0\rangle$ so that $V_T|0\rangle = |0\rangle$. Other states are raised using $S_{1,1}[\Gamma] = \omega^{-\frac{1}{2}\int \gamma \wedge \gamma} \widetilde{U}[\Gamma] U[\Gamma]$ and it implies
    \begin{equation}
        V_T|(t,s)\rangle = \omega^{-t\cdot s} |(t,s)\rangle\ .
    \end{equation}
Indeed we obtain the phase factor stacked on $|(t,s)\rangle$.

\subsubsection*{$SL(2,\mathbb{Z}_N)$ defects from topological boundary states}
In this section, we will focus on the unitary operator $V_{\Lambda}$ and give a concrete expression of $V_{\Lambda}$ in terms of condensation of surface operators $S_{(e,m)}$. By definition, $V_{\Lambda}$ transforms the field $(\widetilde{B},B)^T$ to $\Lambda (\widetilde{B},B)^T$. Acting by another operators $V_{\Lambda'}$ will generate $\Lambda \Lambda' (\widetilde{B},B)^T$, namely
    \begin{equation}
        \left( \begin{array}{c}
            \widetilde{B}\\B \end{array} \right) \xrightarrow{V_{\Lambda}} \Lambda\left( \begin{array}{c}
           \widetilde{B}\\B \end{array} \right)\xrightarrow{V_{\Lambda'}} \Lambda \Lambda' \left( \begin{array}{c}
            \widetilde{B}\\B \end{array} \right)
    \end{equation}
such that the fusion rule is
    \begin{equation}
        V_{\Lambda'} \times V_{\Lambda} = V_{\Lambda \Lambda'}\ .
    \end{equation}
In the previous section, we defined $V_{\Lambda}$ via its action on the boundary states $|(t,s)\rangle$
    \begin{equation}
        V_{\Lambda} |(t,s)\rangle = S_{(b,a)}[\Gamma_{-s,-t}] \mathcal{N}_{\Lambda} P_{\Lambda} |(0,0)\rangle\ ,
    \end{equation}
and one can check this definition of $V_{\Lambda}$ also satisfies the fusion rule. We leave the proof to Appendix \ref{App:fuson}.
Now we would like to express $V_{\Lambda}$ in a more explicit form via surface operators.
We pick a basis of surface operators $S_{(1,0)}[\Gamma_{p,q}] = U[\Gamma_{p,q}], S_{(0,1)}[\Gamma_{\tilde{p},\tilde{q}}] = \widetilde{U}[\Gamma_{\tilde{p},\tilde{q}}]$ and write down a general ansatz
    \begin{equation}
        V_{\Lambda} = \sum_{p,q,\tilde{p},\tilde{q}} \Theta(\Lambda)_{p,q}^{\tilde{p},\tilde{q}} \widetilde{U}[\Gamma_{\tilde{p},\tilde{q}}] U[\Gamma_{p,q}]\ ,   \end{equation}
where the surface operator $\widetilde{U}[\Gamma_{\tilde{p},\tilde{q}}] U[\Gamma_{p,q}] $ satisfies the following orthogonality relation 
\begin{align}
    &\frac{1}{N^6}\textrm{Tr}\left((\widetilde{U}
    [\Gamma_{\tilde{p}',\tilde{q}'}]U[\Gamma_{p',q'}] )^{\dagger} \widetilde{U}[\Gamma_{\tilde{p},\tilde{q}}] U[\Gamma_{p,q}] \right)\nonumber
       = \delta_{p-p',0} \delta_{q-q',0} \delta_{\tilde{p}-\tilde{p}',0} \delta_{\tilde{q}-\tilde{q}',0}\ ,
\end{align}
hence we are able to determine the coefficient $\Theta(\Lambda)_{p,q}^{\tilde{p},\tilde{q}}$,
\begin{align}
    \Theta(\Lambda)_{p,q}^{\tilde{p},\tilde{q}} =& \frac{1}{N^6} \textrm{Tr}\left( (\widetilde{U}[\Gamma_{\tilde{p},\tilde{q}}]U[\Gamma_{p,q}] )^{\dagger}  V_{\Lambda} \right)\nonumber\\
    =& \frac{1}{N^6} \sum_{t,s} \langle (t,s)| U[\Gamma_{-p,-q}] \widetilde{U}[\Gamma_{-\tilde{p},-\tilde{q}}]  V_{\Lambda} |(t,s)\rangle \nonumber\\
        =&\frac{1}{N^6}\sum_{t,s} \omega^{-p \cdot t - q\cdot s} K_{\Lambda}(t,s;\tilde{p},\tilde{q})\ .
    \end{align}
In the equation above, we have define a kernel function $K_{\Lambda}(t,s;\tilde{p},\tilde{q}) \equiv \langle (t-\tilde{q},s-\tilde{p})| V_{\Lambda}|(t,s)\rangle$. We will explicitly work out its expression in \eqref{Appendix-identityproof-inner-product} so we just quote the result below,
    \begin{equation}
        K_{\Lambda}(t,s;\tilde{p},\tilde{q}) = \frac{\mathcal{N}_{\Lambda}}{N^6} \sum_{t',s'} \omega^{-dc t'\cdot s' -ba t\cdot s - bc (s\cdot t'+s'\cdot t)}\delta_{t-\tilde{q},ct'+at}\delta_{s-\tilde{p},cs'+as}\ .
    \end{equation}
Then let us consider some examples.
\subsubsection*{T-defects}
For $T$-transformation we have $a=b=d=1,c=0$ and $\mathcal{N}_T=1$. The kernel is
    \begin{equation}
        K_{T}(t,s;\tilde{p},\tilde{q}) = \omega^{-t\cdot s}\delta_{\tilde{q},0}\delta_{\tilde{p},0}\ .
   \end{equation}
And we have
    \begin{equation}
        \Theta(T)_{p,q}^{\tilde{p},\tilde{q}} = \frac{1}{N^6}\sum_{t,s} \omega^{-t\cdot s -p \cdot t - q\cdot s}\delta_{\tilde{q},0}\delta_{\tilde{p},0}
        =\frac{1}{N^3} \omega^{p\cdot q} \delta_{\tilde{q},0}\delta_{\tilde{p},0}\ .
    \end{equation}

Therefore the condensation defect $T$ can be written as a condensation of surface operator $U[\Gamma]$
    \begin{equation}
        V_T = \frac{1}{N^3} \sum_{p,q} \omega^{p \cdot q} U[\Gamma_{p,q}]\ .
    \end{equation}
\subsubsection*{S-defects}
For $S$-transformation we have $a=d=0,b=-1,c=1$ and $\mathcal{N}_{S} = N^3$. The kernel is
    \begin{equation}
        K_{S}(t,s;\tilde{p},\tilde{q}) = \frac{1}{N^3}  \omega^{2 t\cdot s - \tilde{q}\cdot s - \tilde{p} \cdot t}\ ,
    \end{equation}
and we have
    \begin{align}
        \Theta(S)_{p,q}^{\tilde{p},\tilde{q}} =& \frac{1}{N^9} \sum_{t,s} \omega^{2 t\cdot s - (p+\tilde{p})\cdot t - (q+\tilde{q})\cdot s}\ .
    \end{align}
The condensation defect $S$ can be written as
    \begin{equation}
        V_S = \frac{1}{N^9} \sum_{p,q,\tilde{p},\tilde{q}} \sum_{t,s} \omega^{2 t\cdot s - (p+\tilde{p})\cdot t - (q+\tilde{q})\cdot s} \widetilde{U}[\Gamma_{\tilde{p},\tilde{q}}] U[\Gamma_{p,q}]\ .
    \end{equation}

\section{Proof of the Fusion Rule}
\label{App:fuson}
In this section, we will prove the fusion rule for the topological boundary states defined by
    \begin{equation}
        V_{\Lambda} |(t,s)\rangle = S_{(b,a)}[\Gamma_{-s,-t}] \mathcal{N}_{\Lambda} P_{\Lambda} |(0,0)\rangle\ ,
    \end{equation}
respects the fusion rule
    \begin{equation}
        V_{\Lambda} \times V_{\Lambda'} = V_{\Lambda'\Lambda}\ .
    \end{equation}
It is sufficient to consider $\Lambda'=T$ or $S$ since generic $\Lambda'$ can be obtained inductively.

Let us define the prime version $V'_{\Lambda} \equiv V_{\Lambda'\Lambda} V^{\dagger}_{\Lambda'}$, and the components of $\Lambda'\Lambda$ are
    \begin{equation}
        \Lambda' \Lambda = \left(\begin{array}{cc}
            \hat{a} & \hat{b} \\
            \hat{c} & \hat{d}
        \end{array} \right) = \left(\begin{array}{cc}
            aa'+b'c & a'b+b'd \\
            c'a+d'c & c'b+d'd
        \end{array} \right).
    \end{equation}
We will prove $V'_{\Lambda} = V_{\Lambda}$ by comparing the matrix elements,
    \begin{equation}
        \langle (t,s)| V_{\Lambda} |(t',s')\rangle,\quad \langle (t,s)| V'_{\Lambda} |(t',s')\rangle\ .
    \end{equation}
Using the algebra of $V'_{\Lambda}$ we can show that
    \begin{equation}
        S_{(d,c)}[\Gamma_{p,q}] V'_{\Lambda} |(t',s')\rangle = V'_{\Lambda} S_{(\delta,\gamma)}[\Gamma_{p,q}]|(t',s')\rangle = \omega^{p\cdot t' + q \cdot s'} V'_{\Lambda}|(t',s')\rangle
    \end{equation}
and also
    \begin{equation}
        S_{(b,a)}[\Gamma_{p,q}] V'_{\Lambda} |(t',s')\rangle = V'_{\Lambda} S_{(\beta,\alpha)}[\Gamma_{p,q}]|(t',s')\rangle =  V'_{\Lambda}|(t'-q,s'-p)\rangle\ .
    \end{equation}
Therefore the boundary states $V'_{\Lambda}|(t,s)\rangle$ satisfy the same operator equations compared to $V_{\Lambda}|(t,s)\rangle$. Since the irreducible representation of the algebra is unique, they must equal to each other up to some overall constant. We can write
\begin{equation}
    V'_{\Lambda}|(t,s)\rangle = \langle (0,0)|V^{\dagger}_{\Lambda} V'_{\Lambda}|(0,0)\rangle V_{\Lambda}|(t,s)\rangle\ ,
\end{equation}    
where the overall constant can be obtained by setting $t=s=0$. The matrix elements of $V_{\Lambda}$ and $V_{\Lambda'}$ are related by
    \begin{equation}
        \langle (t,s)| V'_{\Lambda} |(t',s')\rangle = \langle (0,0)|V^{\dagger}_{\Lambda} V'_{\Lambda}|(0,0)\rangle \langle (t,s)| V_{\Lambda} |(t',s')\rangle\ ,
    \end{equation}
and we can deduce that
    \begin{equation}
        V'_{\Lambda} =  \langle (0,0)|V^{\dagger}_{\Lambda} V'_{\Lambda}|(0,0)\rangle  V_{\Lambda}\ ,
    \end{equation}
and $V_{\Lambda}$ and $V'_{\Lambda}$ are identified up to some constant factor $\langle (0,0)|V^{\dagger}_{\Lambda} V'_{\Lambda}|(0,0)\rangle$\ .

Let us fix the normalization factor $\langle (0,0)|V^{\dagger}_{\Lambda} V'_{\Lambda}|(0,0)\rangle$ carefully. Remember $P_{\Lambda}$ is
    \begin{equation}
        P_{\Lambda} = \frac{1}{N^6} \sum_{\Gamma \in H_2(T^4,\mathbb{Z}_N)} S_{(d,c)}[\Gamma]\ .
    \end{equation}
Using $V_{\Lambda} |(0,0)\rangle = \mathcal{N}_{\Lambda} P_{\Lambda} |(0,0)\rangle$ we have
    \begin{align}
        &\langle (0,0)|V^{\dagger}_{\Lambda} V'_{\Lambda} |(0,0)\rangle \nonumber \\
        =& \mathcal{N}_{\Lambda}  \langle (0,0)| P_{\Lambda} V'_{\Lambda}   |(0,0)\rangle \nonumber\\
        =& \mathcal{N}_{\Lambda} \langle (0,0)| V'_{\Lambda}   P_{I}|(0,0)\rangle \nonumber\\
        =& \mathcal{N}_{\Lambda} \langle (0,0)| V'_{\Lambda}  |(0,0)\rangle\ ,
    \end{align}
where we use $P^{\dagger}_{\Lambda} = P_{\Lambda}$ and $P_{\Lambda}V'_{\Lambda} =V'_{\Lambda} P_{I}$. Using $V'_{\Lambda} = V_{\Lambda' \Lambda} V^{\dagger}_{\Lambda'}$ we can write
    \begin{align}
        &\langle (0,0)| V'_{\Lambda}  |(0,0)\rangle \nonumber\\
        =& \langle (0,0)| V_{\Lambda' \Lambda} V^{\dagger}_{\Lambda'}  |(0,0)\rangle \nonumber\\
        =&\sum_{t,s} \langle (0,0)| V_{\Lambda' \Lambda}|(t,s)\rangle \langle (t,s)| V^{\dagger}_{\Lambda'}  |(0,0)\rangle 
    \end{align}
where we insert a pair of complete basis $\sum_{t,s} |(t,s)\rangle \langle (t,s)|$
and we need to evaluate the two factors $\langle (0,0)| V_{\Lambda' \Lambda}|(t,s)\rangle$ and $  \langle (t,s)| V^{\dagger}_{\Lambda'}|(0,0)\rangle$ separately. Recall that
    \begin{align}
        V_{\Lambda}|(t,s)\rangle =\frac{\mathcal{N}_{\Lambda}}{N^6} \sum_{t',s'} \omega^{-dc t'\cdot s' -ba t\cdot s - bc (s\cdot t'+s'\cdot t)}|(c t'+at,cs'+as)\rangle\ ,
    \end{align}
and one has
    \begin{equation}\label{Appendix-identityproof-inner-product}
        \langle (t'',s'')| V_{\Lambda}|(t,s)\rangle = \frac{\mathcal{N}_{\Lambda}}{N^6} \sum_{t',s'} \omega^{-dc t'\cdot s' -ba t\cdot s - bc (s\cdot t'+s'\cdot t)}\delta_{t'',ct'+at}\delta_{s'',cs'+as}\ .
    \end{equation}
Set $t''=s''=0$, we have
    \begin{align}
        &\langle (0,0)| V_{\Lambda}|(t,s)\rangle \nonumber\\ =& \frac{\mathcal{N}_{\Lambda}}{N^6} \sum_{t',s'} \omega^{-dc t'\cdot s' -ba t\cdot s - bc (s\cdot t'+s'\cdot t)}\delta_{ct'+at,0}\delta_{cs'+as,0}  \nonumber\\
        =& \frac{\mathcal{N}_{\Lambda}}{N^6} \sum_{t',s'} \omega^{t\cdot s'}\delta_{ct'+at,0}\delta_{cs'+as,0}
    \end{align}
where we used the delta function to simplify the expressions. The normalization factor is then
    \begin{align}
        &\langle (0,0)|V^{\dagger}_{\Lambda} V'_{\Lambda}|(0,0)\rangle \nonumber\\
        =& \mathcal{N}_{\Lambda} \sum_{t,s}\langle (0,0)| V_{\Lambda' \Lambda}|(t,s)\rangle \langle (t,s)| V^{\dagger}_{\Lambda'}  |(0,0)\rangle \nonumber\\
        =& \frac{\mathcal{N}_{\Lambda}\mathcal{N}_{\Lambda'}\mathcal{N}_{\Lambda \Lambda'}}{N^{12}} \sum_{t,s,t',s',t'',s''} \delta_{\hat{c}t'+\hat{d}t,0}\delta_{\hat{c}s'+\hat{d}s,0}\delta_{c't''+d't,0}\delta_{c's''+d's,0} \omega^{t\cdot s'-t\cdot s''}\ .
    \end{align}

As mentioned before, it is sufficient to focus on $\Lambda'=T$ or $S$ and the generic $\Lambda'$ will be constructed inductively later. First, let us consider $\Lambda'=T$ which means $a'=b'=d'=1,c'=0$ and $\mathcal{N}_{\Lambda'}=1$. The normalization factor is then
    \begin{align}
        &\frac{\mathcal{N}_{\Lambda} \mathcal{N}_{\Lambda T}}{N^{12}} \sum_{t,s,t',s',t'',s''} \delta_{\hat{c}t'+\hat{d}t,0}\delta_{\hat{c}s'+\hat{d}s,0}\delta_{t,0}\delta_{s,0} \omega^{t\cdot s'-t\cdot s''}\nonumber\\
        =& \frac{\mathcal{N}_{\Lambda} \mathcal{N}_{\Lambda T}}{N^{12}}\sum_{t',s',t'',s''} \delta_{\hat{c}t',0}\delta_{\hat{c}s',0} = \frac{\mathcal{N}_{\Lambda} \mathcal{N}_{\Lambda T}}{N^{6}}\sum_{t',s'} \delta_{\hat{c}t',0}\delta_{\hat{c}s',0}\ .
    \end{align}
Since we have $\hat{c}=c$ when $\Lambda'=T$, the two factors $\mathcal{N}_{\Lambda} $ and $\mathcal{N}_{\Lambda T}$ are the same and we have
    \begin{equation}
        \mathcal{N}_{\Lambda} = \mathcal{N}_{\Lambda T} = \frac{N^3}{\gcd(c,N)^3}\ ,
    \end{equation}
therefore we get
    \begin{equation}
        \langle (0,0)|V^{\dagger}_{\Lambda} V'_{\Lambda}|(0,0)\rangle = \frac{1}{\gcd(c,N)^6} \sum_{t',s'} \delta_{ct',0}\delta_{cs',0} = 1\ .
    \end{equation}

On the other hand, when $\Lambda'=S$, which means $a'=d'=0,b'=-1,c'=1$ and $\mathcal{N}_{\Lambda'}=N^3$. The normalization factor is then
    \begin{align}
        &\frac{\mathcal{N}_{\Lambda}\mathcal{N}_{\Lambda S}}{N^9} \sum_{t,s,t',s',t'',s''} \delta_{\hat{c}t'+\hat{d}t,0}\delta_{\hat{c}s'+\hat{d}s,0}\delta_{t'',0}\delta_{s'',0} \omega^{t\cdot s'-t\cdot s''}\nonumber\\
        =&\frac{\mathcal{N}_{\Lambda}\mathcal{N}_{\Lambda S}}{N^9}  \sum_{t,s,t',s'} \delta_{\hat{c}t'+\hat{d}t,0}\delta_{\hat{c}s'+ds,0} \omega^{t\cdot s'}\ .
    \end{align}
Further one has $\hat{c}=d,\hat{d}=-c$ when $\Lambda'=S$, the factor can be written as
    \begin{equation}
        \frac{1}{N^3 \gcd(c,N)^3 \gcd(d,N)^3} \sum_{t,s,t',s'} \delta_{d t'-ct,0}\delta_{ds'-c s,0} \omega^{t\cdot s'}\ .
    \end{equation}
In order to evaluate it, we can consider doing an $SL(2,\mathbb{Z})$ transformation and defined,
    \begin{equation}
        \left( \begin{array}{c}
             u_i  \\
             u'_i 
        \end{array}\right) = \left( \begin{array}{cc}
            -c & d \\
            -a & b
        \end{array}\right) \left( \begin{array}{c}
             t_i  \\
             t'_i 
        \end{array}\right)\ ,\quad \left( \begin{array}{c}
             v_i  \\
             v'_i 
        \end{array}\right) = \left( \begin{array}{cc}
            -c & d \\
            -a & b
        \end{array}\right) \left( \begin{array}{c}
             s_i  \\
             s'_i 
        \end{array}\right)\ ,
    \end{equation}
where $i=1,2,3$ runs over all indices. Since the determinant of the $SL(2,\mathbb{Z})$ matrix is one, we do not have any Jacobian factor and the factor is then
\begin{align}
    &\frac{1}{N^3 \gcd(c,N)^3 \gcd(d,N)^3}  \sum_{u,v,u',v'} \delta_{u,0}\delta_{v,0} \omega^{cd u'\cdot v'}\nonumber\\
    =& \frac{1}{N^3 \gcd(c,N)^3 \gcd(d,N)^3}  \sum_{u',v'} \omega^{cd u'\cdot v'}\nonumber\\
    =& \frac{1}{\gcd(c,N)^3 \gcd(d,N)^3}  \sum_{u'} \delta_{cdu',0}\ .
\end{align}
Notice that we cannot have both $c=d=0$ for an $SL(2,\mathbb{Z}_N)$ matrix. Further, if one of them is zero, for example, $c=0$ and $d\neq 0$, then $d$ must satisfy $ad = 1 \ \textrm{mod}\ N$ and both $a,d$ are coprime with $N$. In this case it is easy to see the factor is one. On the other hand, if both $c$ and $d$ are not equal to zero, 
notice that for any $SL(2,\mathbb{Z}_N)$ matrix, we can make $\gcd(c,d)=1$ by shifting $c\rightarrow c+N, d\rightarrow d+N$ such that $\Lambda$ becomes and $SL(2,\mathbb{Z})$ matrix with $ad-bc=1$. If that is the case, there exists a multiplicative property which says if $c,d$ are relatively coprime, one has
    \begin{equation}
        \gcd(c,N) \gcd(d,N) = \gcd(cd,N)\quad \forall N\in \mathbb{Z}\ ,
    \end{equation}
and we have,
    \begin{equation}
        \langle (0,0)|V^{\dagger}_{\Lambda} V'_{\Lambda}|(0,0)\rangle = \frac{1}{\gcd(cd,N)^3}  \sum_{u'} \delta_{cdu',0}=1\ .
    \end{equation}

We have proven
    \begin{equation}
        V_{\Lambda} \times V_T = V_{T\Lambda}\ ,\quad V_{\Lambda} \times  V_S = V_{S\Lambda}
    \end{equation}
and since any $SL(2,\mathbb{Z}_N)$ matrix $\Lambda'$ can be generated by $S,T$ transformation, we can deduce the fusion rule
    \begin{equation}
        V_{\Lambda} \times V_{\Lambda'} = V_{\Lambda' \Lambda}\ .
    \end{equation}

\bibliographystyle{JHEP}
\bibliography{main}

\end{document}